\documentclass[11pt]{article}
\usepackage{amsmath, amsthm, amssymb}
\usepackage{fullpage}
\usepackage{graphicx}
\usepackage{url}

\newcommand{\mnras}{Mon. Not. R. Astron. Soc.}
\newcommand{\MLL}  {M_{\ell\ell}}
\newcommand{\ttbar} {t\bar{t}}
\newcommand{\qT}       {q_{T}}
\newcommand{\Pythia} {{\sc Pythia}}
\newcommand{\DYonly}{{\sl DY-only}}
\newcommand{\All}{{\sl all processes}}

\begin{document}
\begin{center}
{\large\bf The Use of Minimal Spanning Trees in Particle Physics}  \\

Jessica Lovelace Rainbolt and Michael Schmitt

Department of Physics and Astronomy, Northwestern University, Evanston, IL 60201

\today
\end{center}

\begin{abstract}
Minimal spanning trees (MSTs) have been used in cosmology and astronomy to distinguish distributions of points in a multi-dimensional space.  They are essentially unknown in particle physics, however.  We briefly define MSTs and illustrate their properties through a series of examples.  We show how they might be applied to study a typical event sample from a collider experiment and conclude that MSTs may prove useful in distinguishing different classes of events.
\end{abstract}

\section{Introduction}
The quest for physics beyond the standard model (SM) takes on two forms: 1) direct observation of non-SM events in some kinematic region, and 2) indirect observation of non-SM interactions through a minute deviation of an observable from its SM value.  In the former case, a deviation from the expected distribution of a kinematic variable may be localized, e.g. a narrow peak in an invariant mass distribution, or it may appear as a broad feature that can only be detected on a statistical basis, such as an enhancement of the transverse momentum distribution of vector bosons due to an anomalous trilinear coupling.  Very often the researcher makes a histogram for a kinematic quantity and applies a statistical test for compatibility with the distribution expected from the SM alone.  In this approach one hopes to know in advance which kinematic quantities will reveal new physics if it is present, and this requires insight and a fortunate choice of new physics model.  Different kinematic quantities are examined one at a time; the researcher is effectively marginalizing over the other quantities in each case.  Sometimes multiple kinematic quantities are combined in a so-called multivariate analysis, and a single indicator variable is constructed that allows the researcher to exploit several kinematic quantities simultaneously; this technique requires a model for the non-SM physics signal in order to optimize or train the indicator variable.   Multivariate techniques are nearly always tied to more or less specific models of new physics.

We propose a method that stands between the extremes of histogramming individual quantities and constructing a highly-directed indicator variable.  Our method makes use of minimal spanning trees~(MST) constructed with events as they populate phase space.\footnote{We discuss phase space here, but our techniques and studies can also be applied to a more general feature space.}  The distribution of events in phase space is determined by a production process followed by a decay process, which in turn are determined by the density of states and by matrix elements computed from an underlying fundamental theory.  For example, values for the invariant mass~($\MLL$) and transverse momentum~($\qT$) of muon pairs produced in the Drell-Yan process at a hadron collider are accurately generated based on SM calculations.  The LHC experiments have carefully measured differential cross sections $d\sigma / d\MLL$, $d\sigma / d\qT$, and $d\sigma / dY$ for the inclusive production of same-flavor~\cite{LHCbook}, opposite-charge lepton pairs in pp~collisions, and these measurements confirm precise SM predictions.  For example, a narrow peak for $\MLL \approx 91$~GeV reflects the Z~boson resonance, and the lepton pairs are produced more centrally as their invariant mass increases.  These measurements do not attempt to probe correlations among kinematic variables, except perhaps to measure the differential cross section for one quantity in ``slices'' of another.  Correlations or other structures that may be present in phase space may be overlooked and are certainly not the goal of these measurements.  Our method is based on a construct that can preserve and reflect correlations or other structures in phase space.  This method might complement traditional differential cross section measurements and ultimately provide a new tool for identifying deviations from the SM that indicate new physics processes.

This paper is organized as follows.  First we define MSTs and describe their construction.  Second we briefly recall results from applications of MSTs to cosmological structure.  Third we analyze a set of values for one stochastic quantity in preparation for multi-dimensional cases --- these we call one-dimensional ``trees.''  Fourth we present a number of artificial examples that demonstrate in a dramatic fashion how MSTs can be sensitive to the way stochastic variables are distributed in a multidimensional space.  Fifth we illustrate the efficacy of MSTs for simulated dilepton events that might be observed at an LHC experiment.  Finally we summarize our exploratory results and conclude.

\section{Overview of minimal spanning trees}
\begin{figure} \centering
\includegraphics[width=1.0\textwidth]{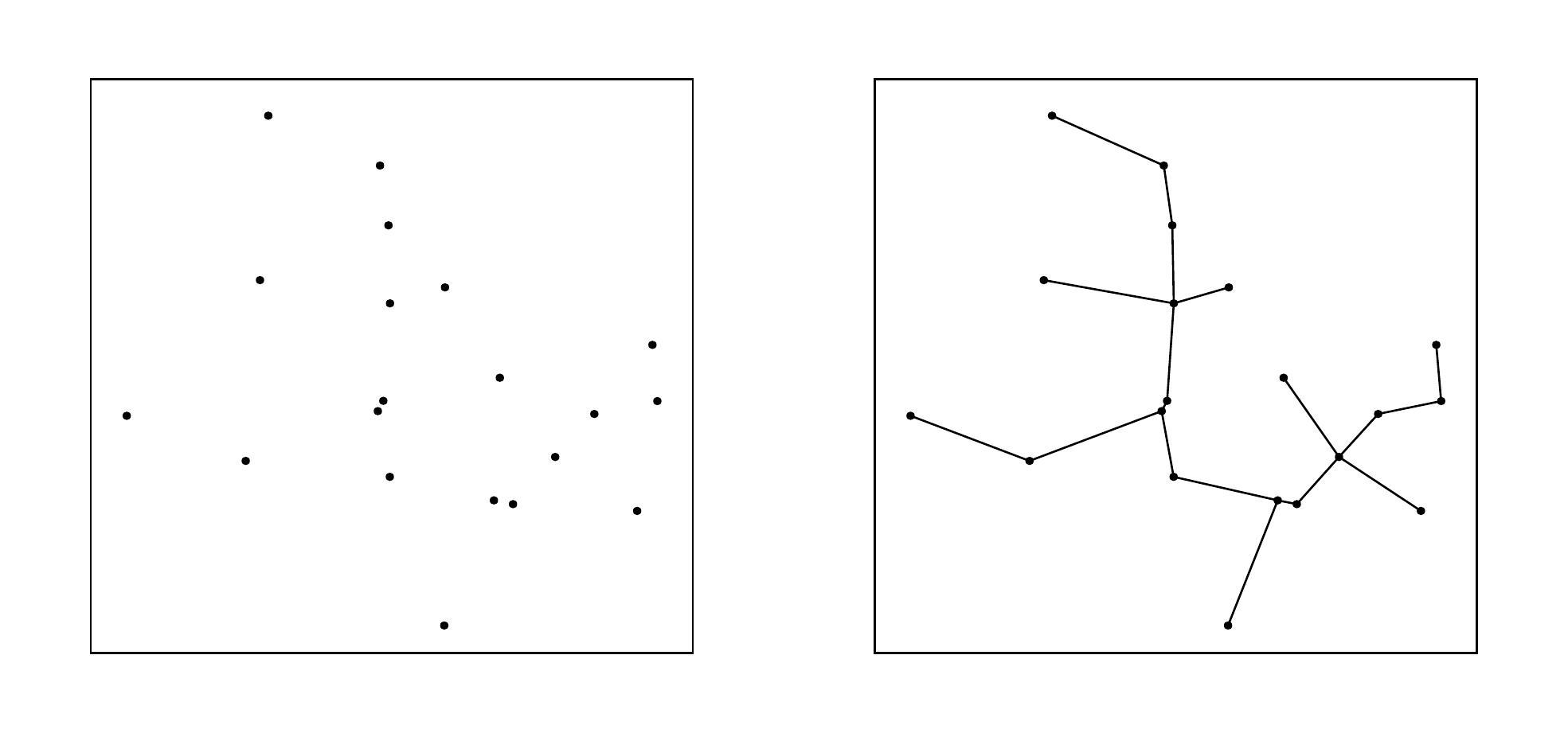}
\caption{(left): A collection of 20 randomly-distributed vertices in two dimensions.  (right): The MST constructed from those vertices.}
\label{fig:example}
\end{figure}

The Euclidean minimal spanning tree~(MST) is a graph connecting $m$ points in $n$ dimensions~\cite{Gower}.  Each point, called a ``vertex,'' is connected to at least one other via a line segment, called an ``edge,'' whose length is the Euclidean distance between the two vertices it joins.  The edges in the MST are chosen such that the sum of their lengths is minimized and that all $m$ vertices are connected; that is, the tree is both minimal and spanning.  By construction, such a graph will have no closed loops or ``circuits.''  For any given set of vertices, there is a unique MST so long as no two pairs of vertices are separated by exactly equal distances.  Figure~\ref{fig:example} shows a very simple MST with 20 vertices in two dimensions.

\subsection{Construction}
A number of algorithms to construct MSTs are available, the best known of which are Prim's algorithm~\cite{Prim} and Kruskal's algorithm~\cite{Kruskal}.  We implemented Kruskal's algorithm in C++ code written expressly for this study.  With this implementation, a tree with 6000 nodes takes a single computer about thirty seconds to build.

\subsection{Statistical quantities}
The structure of an MST can be analyzed on a statistical basis.  The lengths of the edges in the tree, $l$, is perhaps the most intuitive MST quantity.  The shape of the distribution of $l$ reflects the density and uniformity of the tree's vertices.  The more sparsely the vertices are distributed in phase space, the greater the mean of the distribution of $l$.  Furthermore, a tree with more variation in the density of its vertices will have a wider distribution of $l$.

To compare the relative shapes of two trees, the length of each edge divided by the average of its tree's edges is a more useful quantity.  This quantity is called ``normalized length'' $\bar{l}$, so $\bar{l} = l / \langle l \rangle$.  The distribution of the logarithm of this quantity, $\ln(\bar{l})$, is effective in distinguishing trees with different structures.

Another intuitive quantity is the number of edges attached to each vertex, called its ``degree'' or $d$.  A distribution of $d$ with a sharp peak at 2 signifies a tree with a filamentary structure, or longer ``branches.''  A ``branch'' is a chain of edges with a loose end, all connected via vertices of degree 2.  The length of a branch is $b$ and one can compare $\ln(b)$ among the trees.

The quantities $l,\ \bar{l},\ d$, and $b$ pertain to a single tree, and distributions of these quantities allow a comparison of two or more trees.  In addition, we have devised quantities that relate to how two MSTs overlap in phase space (see Section~\ref{sec: examples}).

\subsection{Data considerations}
The MST relies upon the spatial arrangement of events in phase space, which must be taken into account when choosing which features to include.  The first of these considerations is scaling.  The MST uses Euclidean distance, so the structure of the tree is sensitive to the units and range of each feature.  If a single tree makes use of kinematic quantities which take varying units or differ greatly in their distribution, all features can be rescaled to prevent any single quantity from inappropriately guiding the analysis.  In this paper we have constructed examples that do not require rescaling; however, it might be necessary in a more general application.

Another consideration is the continuity of our feature space.  As a distance-based method, the MST is not suited to categorical or discrete features such as an integer number of jets.  The inclusion of discretized features may have unexpected consequences, such as creating artificial filaments and branches in one of the dimensions.


\section{MST analysis of cosmological structure}
One of the earliest applications of MSTs in a scientific analysis was reported by Barrow, Bhavsar, and Sonoda in 1985~\cite{Barrow}.  At the time, the filamentary nature of the distribution of galaxies was becoming apparent and Barrow et al.~wanted to find a tool that was sensitive to that filamentary structure.  Traditional statistical measures were not effective in identifying the filamentary structure and were insufficient for distinguishing competing models for galaxy formation.

Barrow et al.~took the available data on galaxy locations (``Zwicky'') and constructed an MST.  For comparison, they constructed a random grid of vertices (``Poisson'').  Their analysis is based on the fraction of edges of a given length in the MST after scaling by the mean length.  The results show that the Zwicky and the Poisson MSTs have rather different structures.  The number of nodes is modest (slightly more than 1000) and the trees can be distinguished by inspection.  More importantly, the distributions of the normalized edge lengths $\bar{l}$ are quite different, proving that the distribution of galaxies through space has structure beyond that of a random distribution in space --- the galaxies are more clustered.

Other analyses in astrophysics have employed MSTs.  For example, Campana et al.~employ them to identify sources of high-energy $\gamma$~rays~\cite{Campana}.  Allison et al.~used MSTs to trace the segregation of mass (the fact that massive stars are not distributed in the galaxy the same way as other stars)~\cite{Allison}.  Finally, MacFarlane et al.~applied MSTs to chemical tagging of stellar debris~\cite{MacFarlane}.  These three applications are rather different, and do not have obvious 
analogs in particle physics.  Nonetheless, they show interesting advantages over more standard methods from which particle physics might profit.

\section{One-dimensional trees}
We begin with three simple examples for a one-dimensional ``tree.''  While a one-dimensional tree has very little structure, these examples nonetheless illustrate some statistical features that are important for our discussion.  Since the trees are almost trivial, it is easy to see the connection between the distribution of ``vertices'' and the distribution of edge lengths.

Suppose we have a set of $N$ values $\{ x_i \}$ drawn from a given probability density function~(pdf).   They can be visualized as a set of vertices spread out along a line ---  a one-dimensional ``tree.''  If we put the values in ascending order, then the MST is simply the set of edges from one vertex to the next, starting from one end and ending at the other.  The length of the ``tree''  is just the sum of the distances from one vertex to the next, which is the same as the length spanned by the tree:
$ D = | x_{\mathrm{last}} - x_{\mathrm{first}} | $.
There is only one branch, and that branch contains all vertices connected one to the next; all vertices have degree two except the first and the last.

The only quantity of interest is the edge length, $l_i = | x_{i+1} - x_i |$ ($ i = 1, 2, \dots, N-1 $), or quantities derived from it.  In this section we focus only on the distribution of $l_i$.

\begin{figure} \centering
\includegraphics[width=1\textwidth]{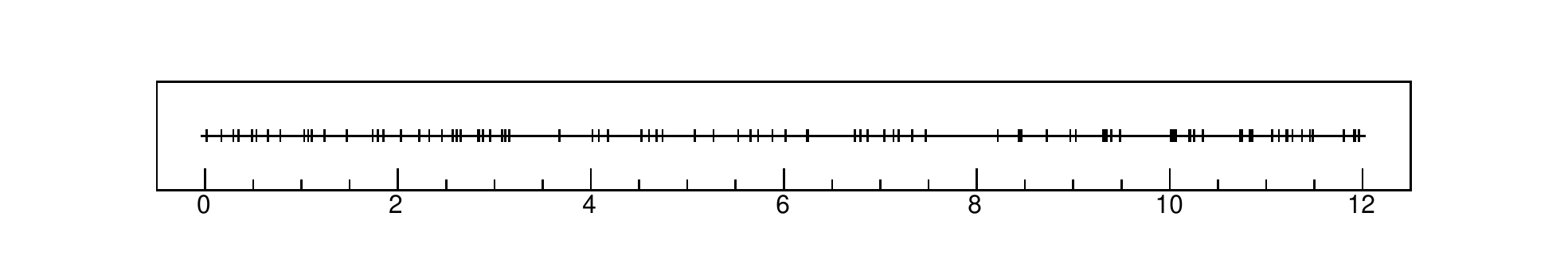}
\includegraphics[width=1\textwidth]{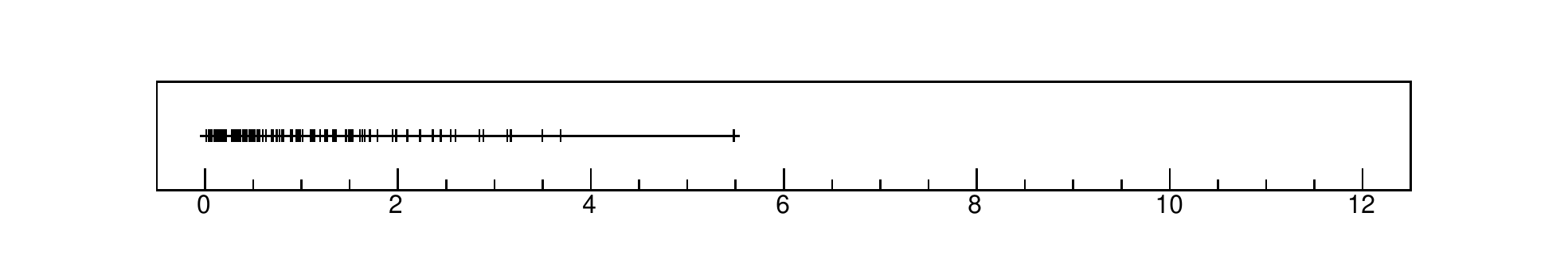}
\includegraphics[width=1\textwidth]{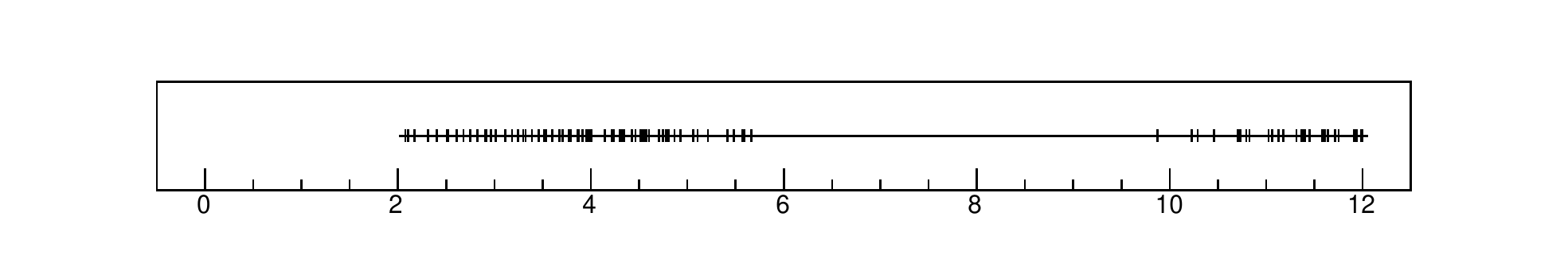}
\caption{Trees constructed from 100 vertices of the $f_u$ (top), $f_e$ (center), and $f_s$ (bottom) distributions.}
\label{fig: 1dtrees}
\end{figure}

We consider three illustrative cases, all of them defined for the interval $0 \le x \le 12$.  The first case is just the uniform distribution, with pdf $f_u(x) = 1/12$.  The second is sharply peaked toward the origin: $f_e(x) = e^{-x}$.  The third function is multi-modal with a region where vertices are suppressed: $f_s(x) = C \sin^2 (\pi x / 8)$ where $C$ is a normalization constant.  Example trees of these three distributions are shown in figure~\ref{fig: 1dtrees}.

\begin{figure} \centering
\includegraphics[width=0.495\textwidth]{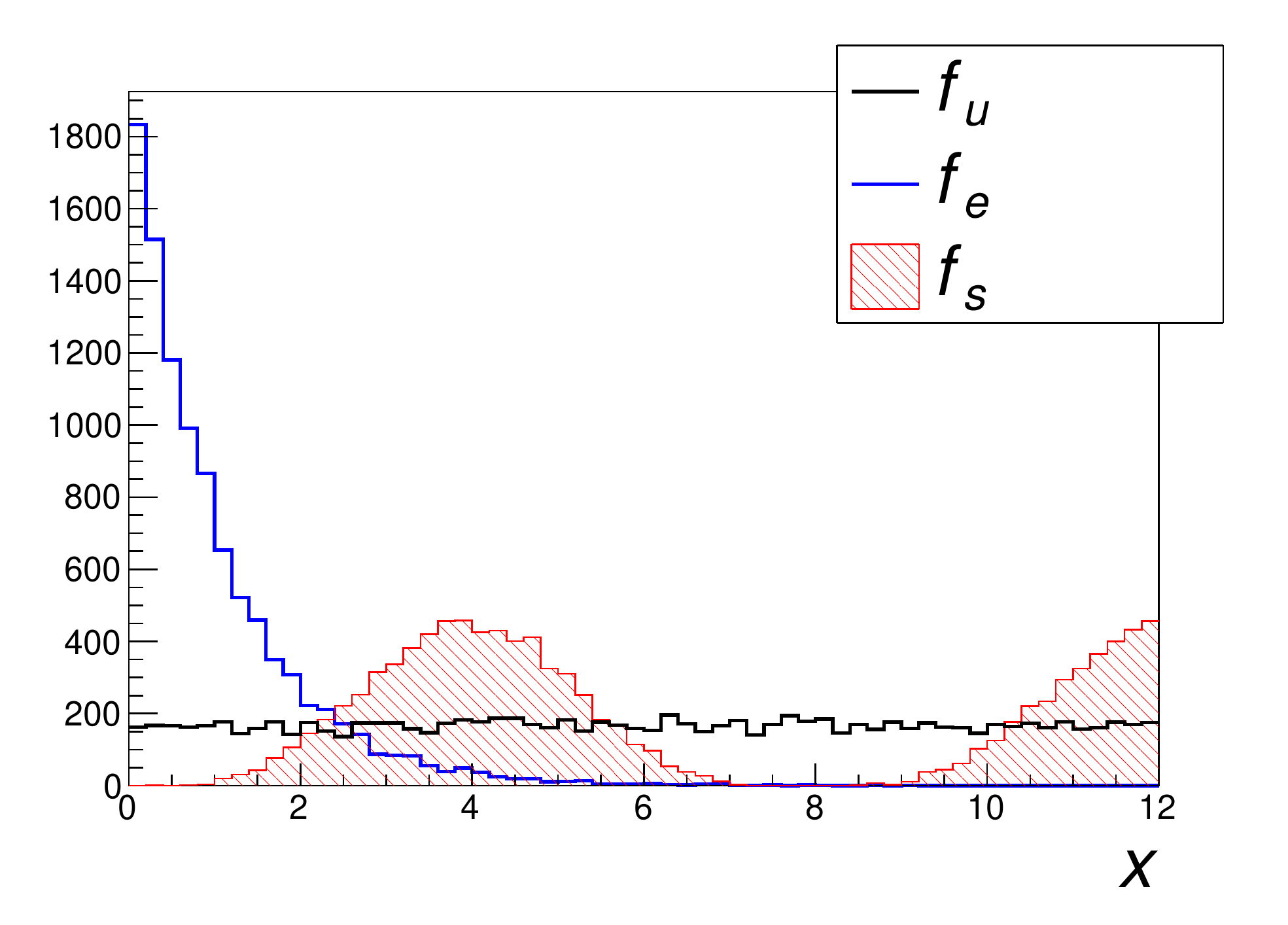}
\includegraphics[width=0.495\textwidth]{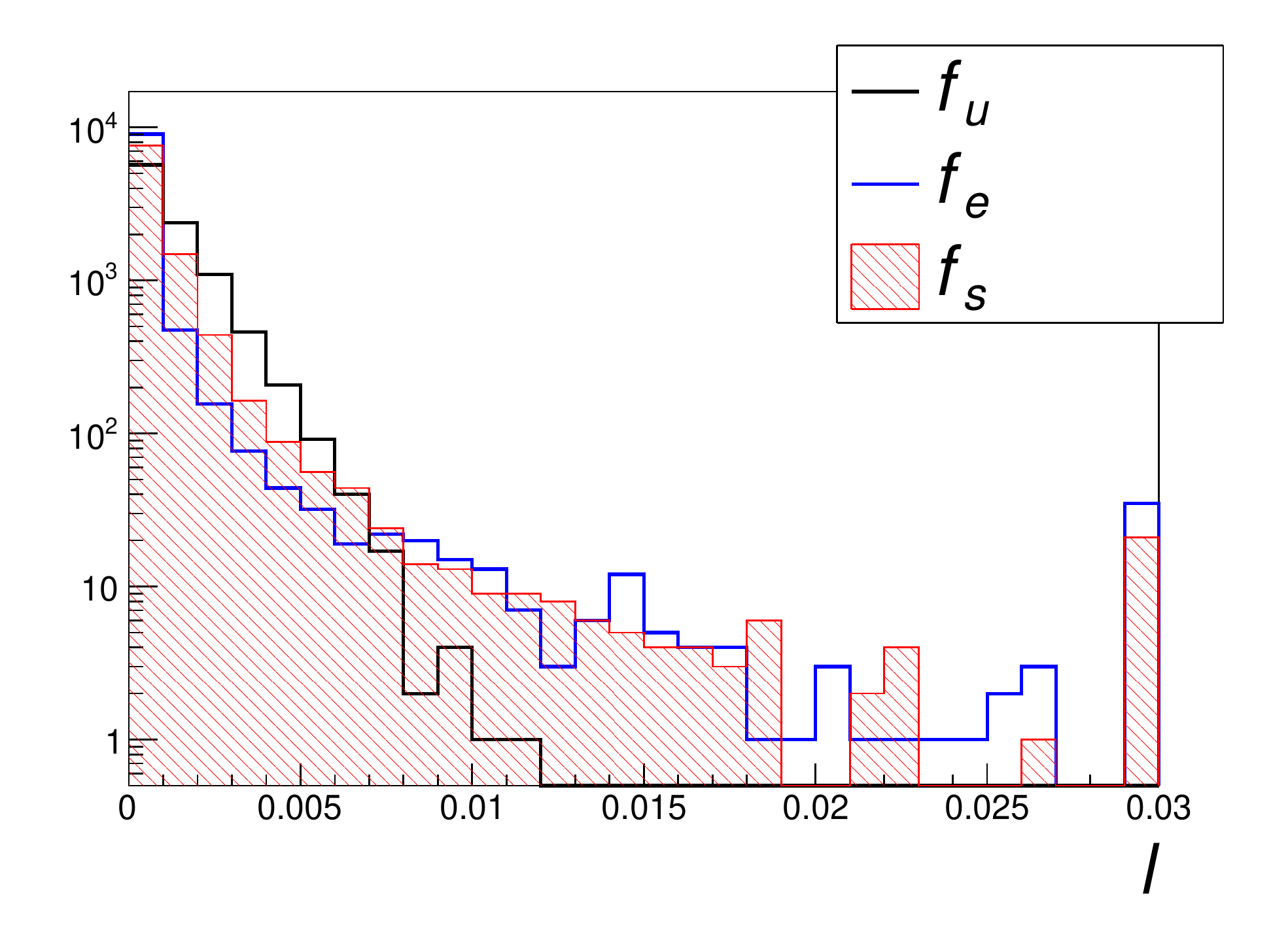}
\caption{(left): Three test distributions for one-dimensional ``trees.'' The black curve shows the uniform distribution, the blue curve shows the exponential distribution, and the shaded red region shows the $\sin^2(\pi x/8)$ distribution.  (right): The distributions for edge lengths for the three different trees.  The last bin is an overflow bin.}
\label{fig: oneD}
\end{figure}

We generated $10^5$ vertices ($x$ values) from each distribution, sorted them, and calculated the $10^5-1$ edge lengths.  Distributions of these edge lengths are shown in figure~\ref{fig: oneD}.    As one would expect, the exponential and $\sin^2$ distributions lead to long tails in the edge length distributions, in contrast to the uniform distribution which terminates far more quickly.  The three distributions differ greatly for smaller edge lengths, $l < 0.006$.

\section{Artificial examples} \label{sec: examples}
The examples from the previous section show clear discrimination power among the three pdfs, but for one dimension the analysis is rather simple and does not lead to deep insights.   In this section we examine four more complicated examples based on trees in more than one dimension.  While the examples are still quite artificial, they anticipate the features of the collider data we discuss later.

\subsection{Dense vs. sparse}
\begin{figure} \centering
\includegraphics[width=0.69\textwidth, trim={0 2cm 0 2cm}, clip]{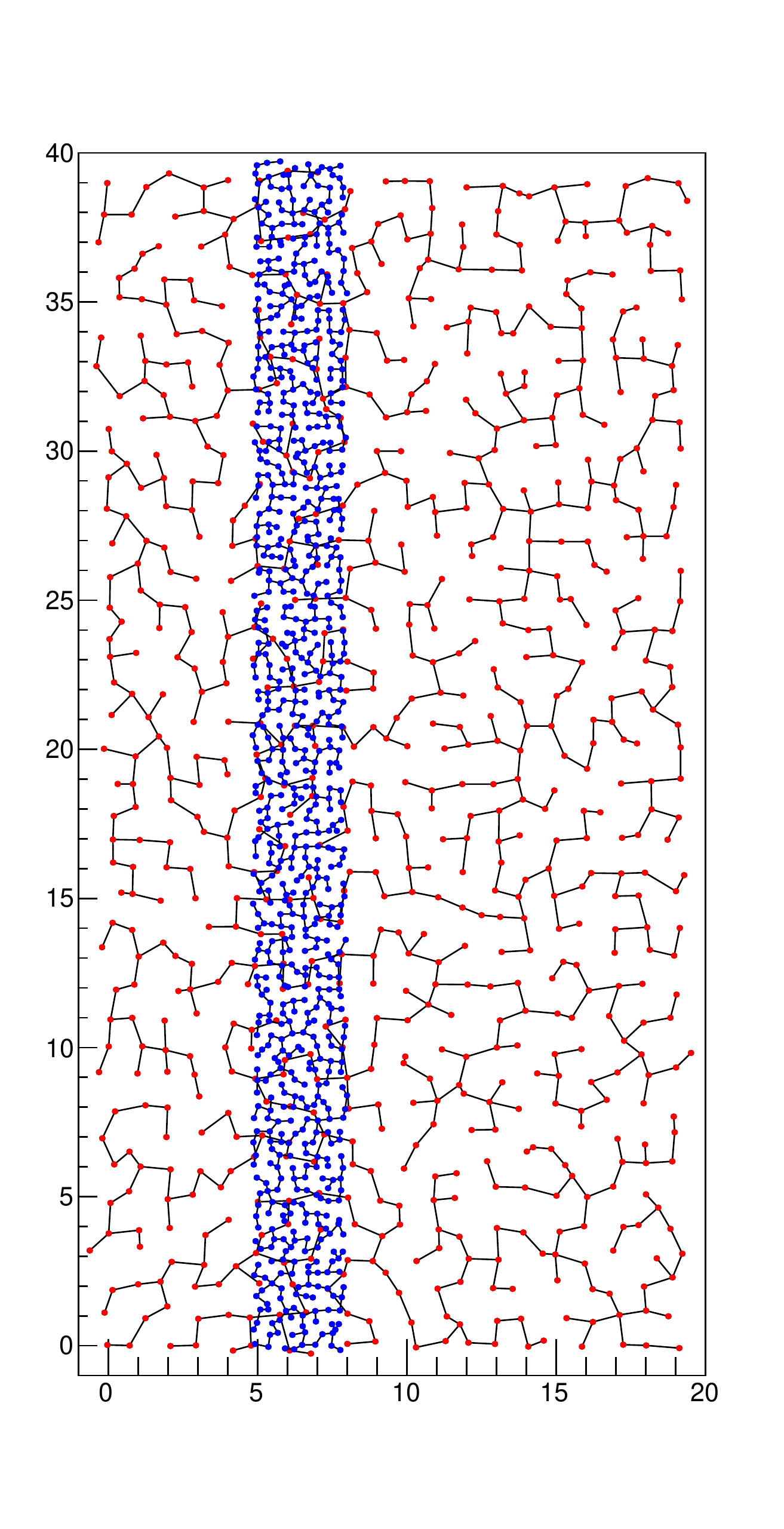}
\caption{The sparsely-populated tree (red) and the densely-populated tree (blue).}
\label{fig: ex1trees}
\end{figure}

For a first example, we placed 800 evenly-spaced vertices on two grids.  The first was sparsely populated with dimensions $20 \times 40$, and the second was densely populated with dimensions $3 \times 40$.  Small ($\sigma = 0.2$) random numbers were drawn from a Gaussian distribution to perturb the position of each vertex.  These perturbations created nonuniform edge lengths for each set of vertices, which is the necessary condition to build a unique MST.  Both trees are shown in figure~\ref{fig: ex1trees}.  These samples are easily distinguished by eye; our goal is to determine whether the MST provides a statistical basis for distinguishing them.

\begin{figure} \centering
\includegraphics[width=0.495\textwidth]{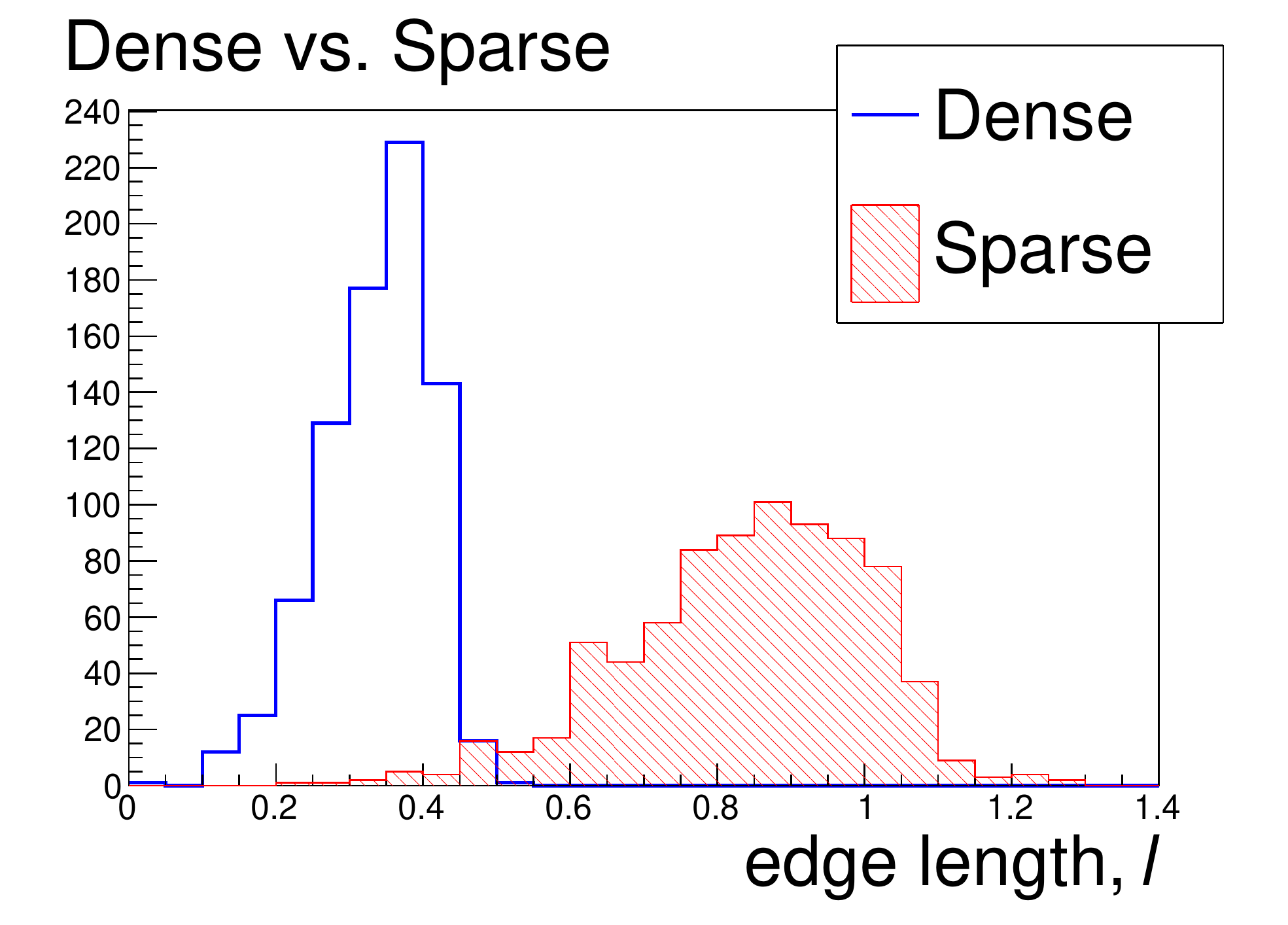}
\includegraphics[width=0.495\textwidth]{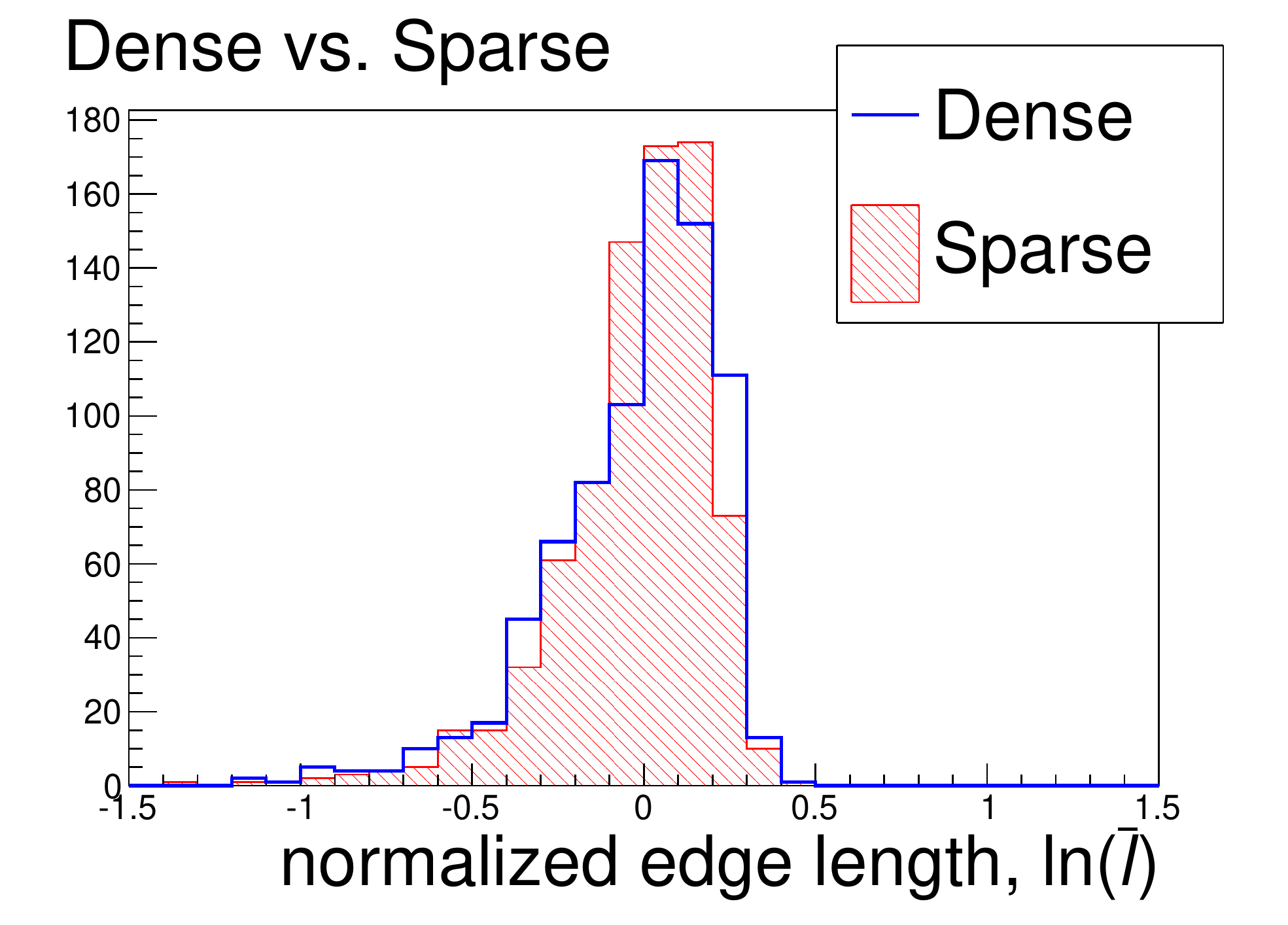}
\includegraphics[width=0.495\textwidth]{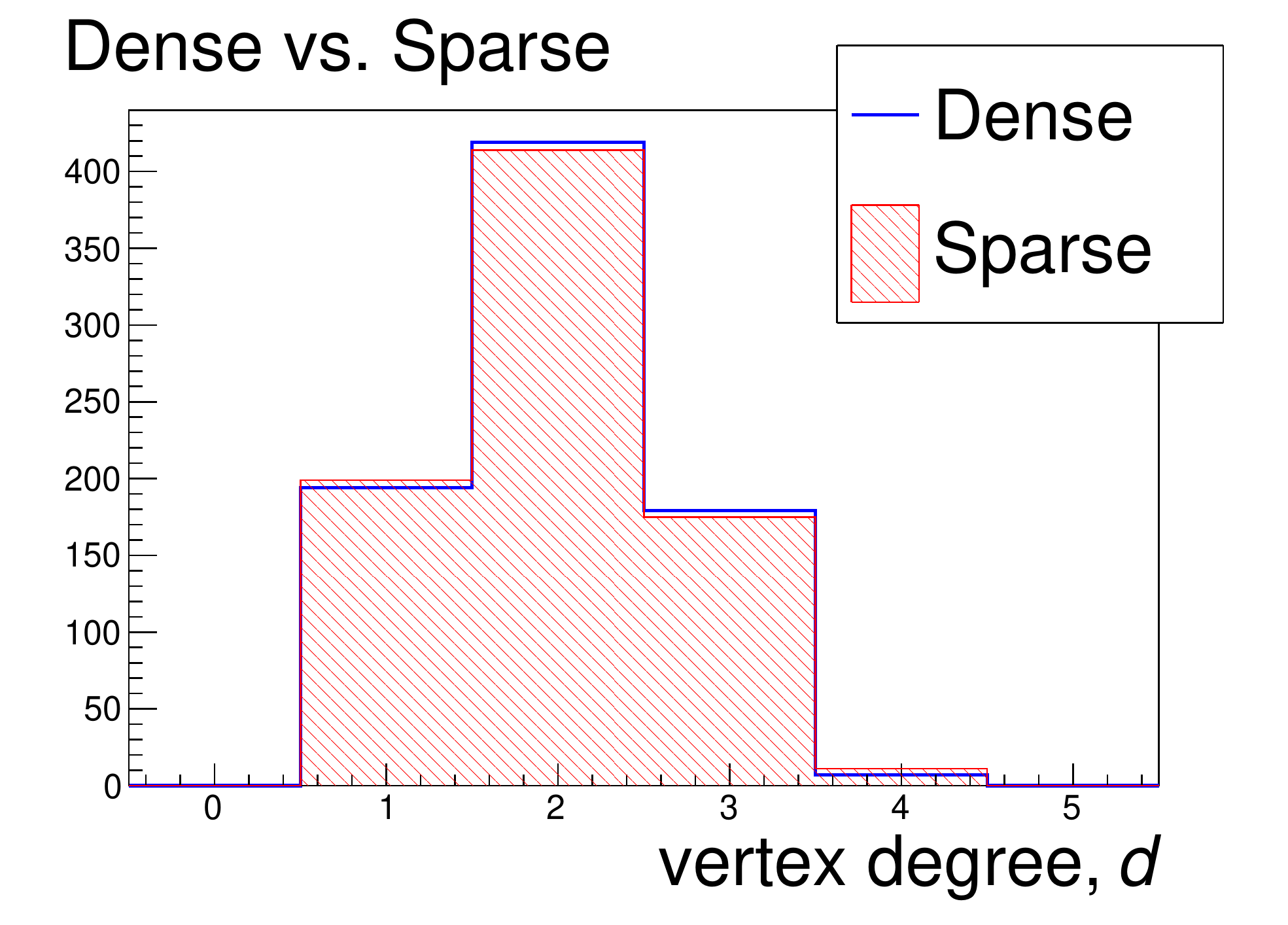}
\includegraphics[width=0.495\textwidth]{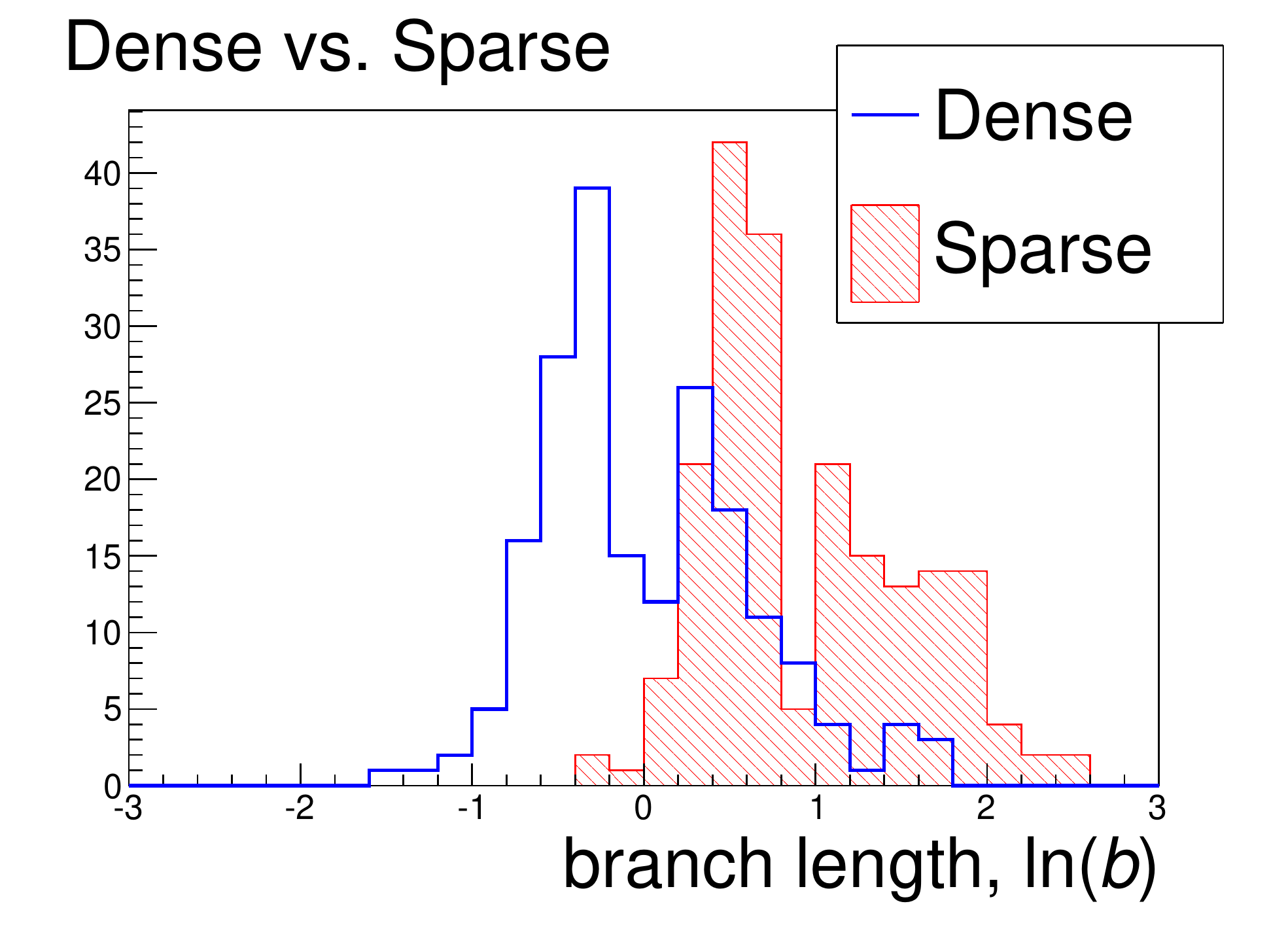}
\caption{The distributions of $l$ (top left), $\ln(\bar{l})$ (top right), $d$ (bottom left), and $\ln(b)$ (bottom right) for the dense and sparse trees shown in figure~\ref{fig: ex1trees}.}
\label{fig: ex1indiv}

\vskip 30pt

\includegraphics[width=0.495\textwidth]{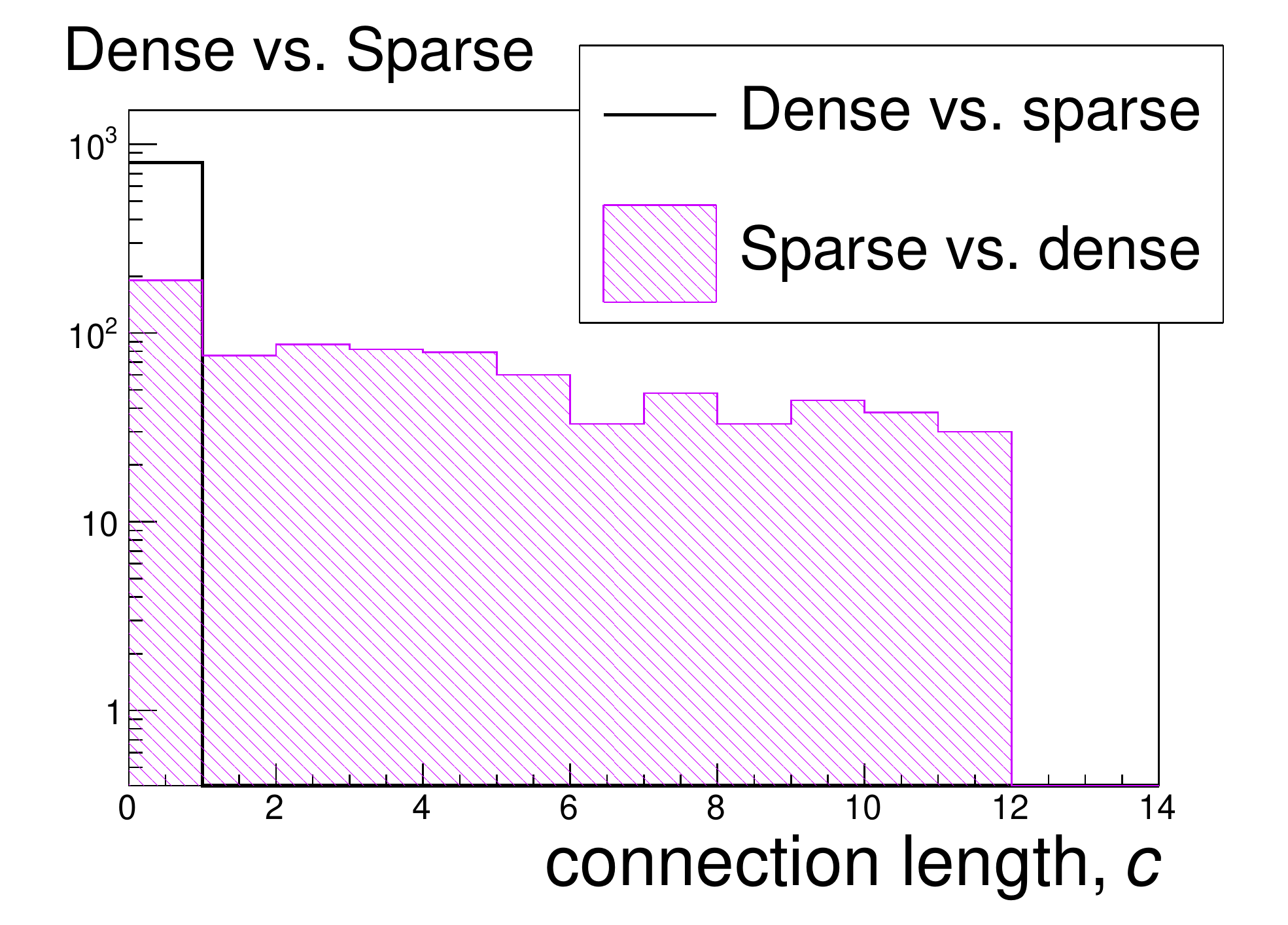}
\includegraphics[width=0.495\textwidth]{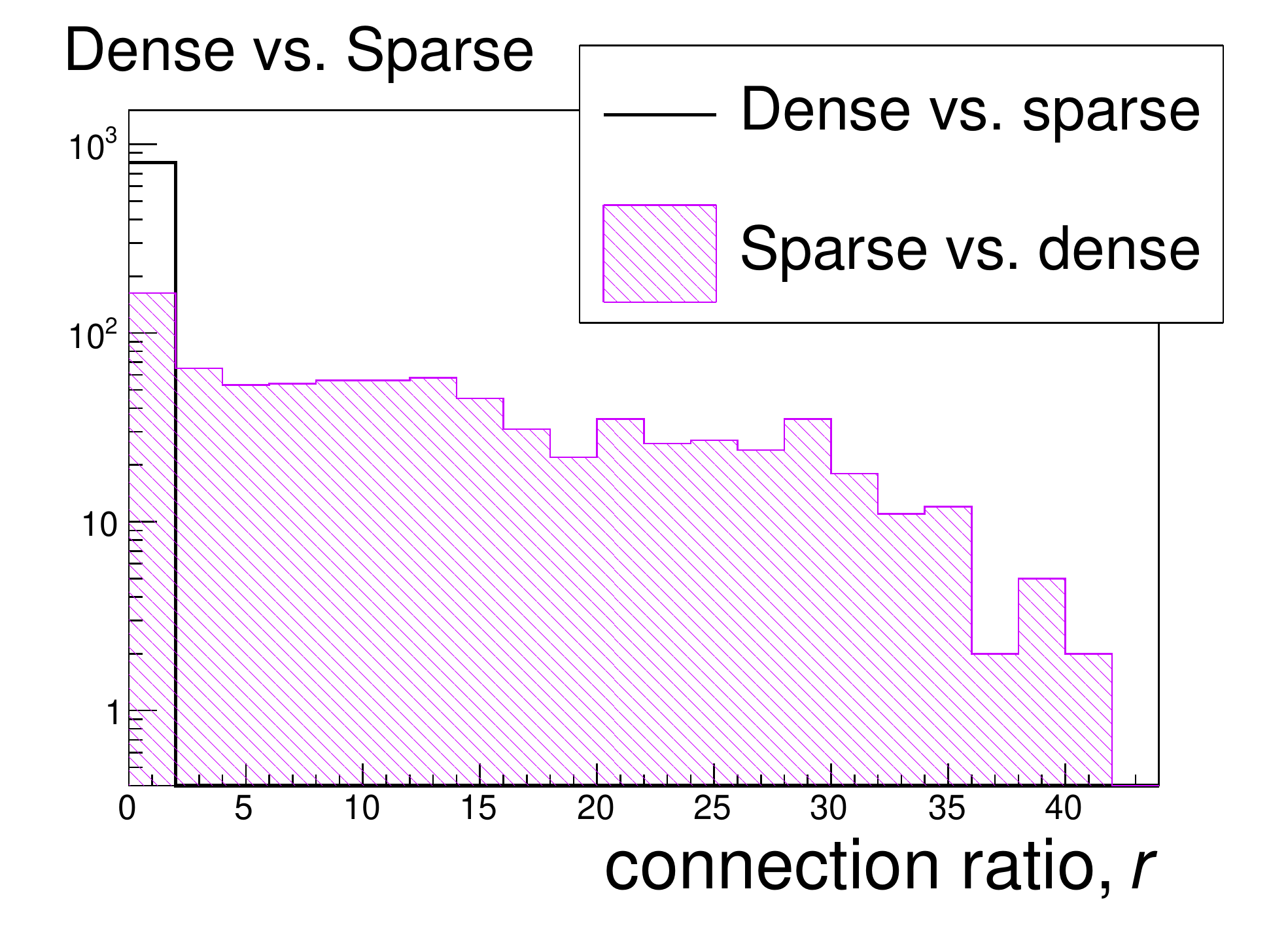}
\caption{(left): The distributions of $c$ for the dense and sparse trees.  The violet area represents the shortest distances from each vertex of the dense tree to the sparse tree.  For the black line, the trees are swapped.  (right): The distributions of $r$ for the dense and sparse trees.}
\label{fig: ex1comp}
\end{figure}

The statistical quantities of a given tree may be analyzed as shown in figure~\ref{fig: ex1indiv}.  It is illustrative to first look at the lengths of the edges in each tree, or $l$.  The trees in this example have visibly different densities, so one expects a clear difference in the means of the distributions of $l$.  Both trees are uniformly populated, however, so one does not necessarily expect a difference in their shape.  To determine whether there exists such a difference, we look at $\ln(\bar{l})$.  The shapes are very similar, but still distinguishable; the difference in shape comes from the difference in the aspect ratios of the trees.  For $d$, the symmetry of the peaks shows a relatively uniform distribution of vertices in 2-space, as expected for these samples.  The distributions of $d$ typically do not provide much information, at least in two dimensions.  The similarity in the distributions of $\ln(b)$ shows that the two trees are similar in shape but not in density.

In the context of this multidimensional example, we introduce the quantities that compare two trees directly.  One method of comparing two trees is finding the shortest distance between them.  For each vertex on one of the trees, we locate the nearest vertex on the other tree.  We will refer to the distance between them as the vertex's ``connection length,'' or $c$, which gives an idea of how far apart the trees are in our feature space.  In this example the dense tree is within the boundaries of the sparse tree, so small $c$ values are expected when measuring the dense tree against the sparse tree.  For the converse, one expects a wider but relatively uniform distribution of $c$.  It should be clear that the distribution of connection lengths for the dense tree with respect to the sparse tree is not the same as the distribution of connection lengths for the sparse tree with respect to the dense tree.  In figure~\ref{fig: ex1comp}, the peaks on the left show the area of overlap.  For the sparse tree, the bins on the right indicate the presence of events we would not expect to find in the dense tree.

If we compare the distance between trees in a way that accounts for their densities, we can pinpoint potential anomalies in a data sample.  Such anomalies could include, for example, a dense cluster of events in an area we expect to be sparsely populated.  One method of finding such anomalies is to compare the connection lengths found above to the lengths of the nearest edges of the tree in which we are interested.  Consider an arbitrary vertex in the dense tree.  We divide its $c$ by the average of the lengths of the $k$ edges nearest to this vertex, creating what we will call the ``connection ratio'' or $r$.  This $k$ is a tunable parameter, which we have chosen to be five.   The histograms for $r$ in figure~\ref{fig: ex1comp} look similar to those for $c$ because of the uniformity in the distribution of vertices, but the longer tail for the sparse tree shows the significance of its outlying vertices.

\subsection{Spacing in $x$}
\begin{figure} \centering
\includegraphics[width=0.69\textwidth, trim={0 2cm 0 2cm}, clip]{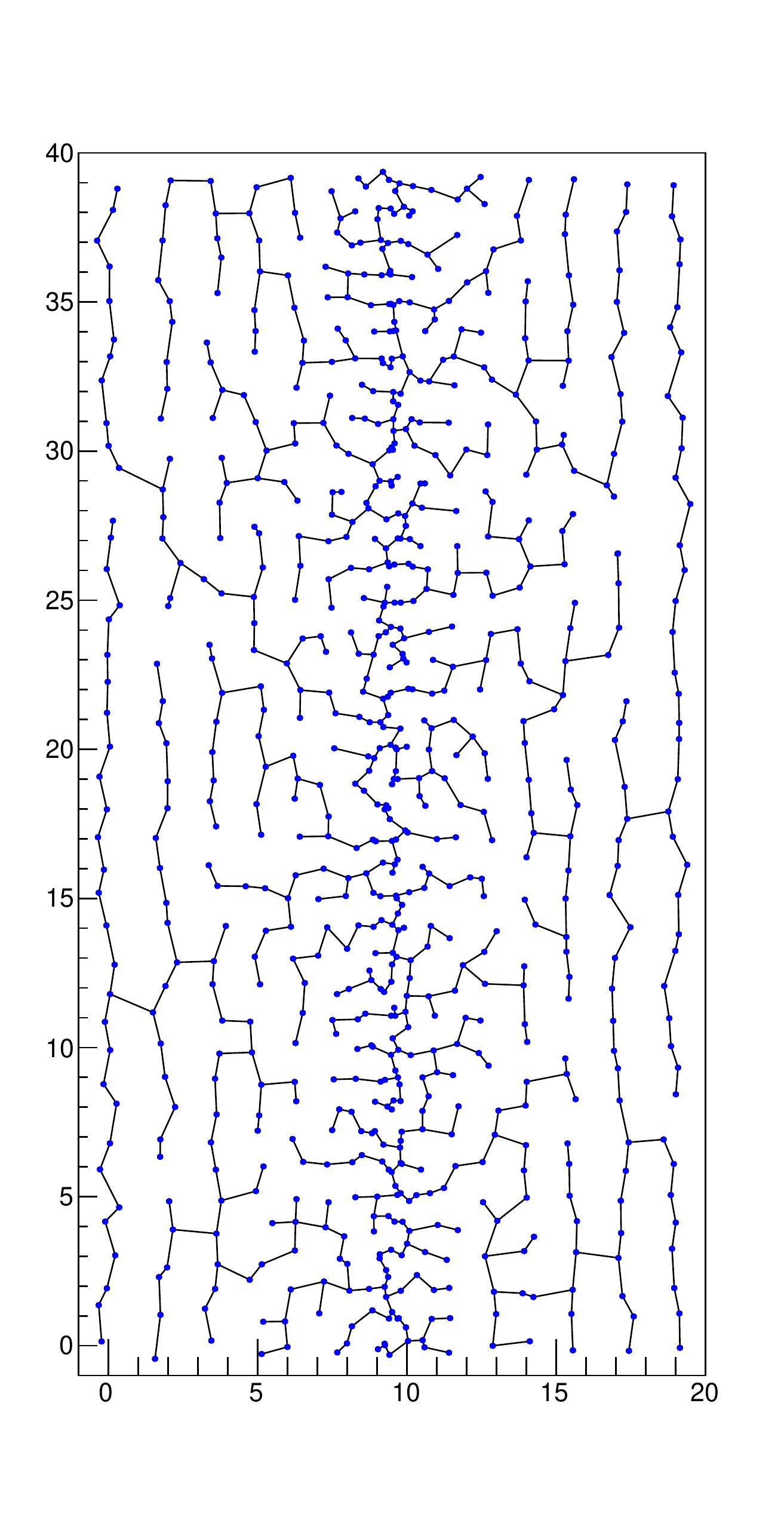}
\caption{The quadratically-distributed tree.  The uniformly-distributed tree is identical to the sparse (red) tree in figure~\ref{fig: ex1trees}.}
\label{fig: ex2trees}
\end{figure}

For a second example, we experimented with nonuniform spacing.  The first, ``uniform'' grid is identical to the sparse grid from the previous example.  The second, ``quadratic'' grid has the same dimensions and the same number of vertices, but they are distributed quadratically in the $x$ direction.  As is evident in figure~\ref{fig: ex2trees}, this creates long branches near the left and right boundaries of the quadratic MST and short branches near its horizontal center, which affect each measured quantity.

\begin{figure} \centering
\includegraphics[width=0.495\textwidth]{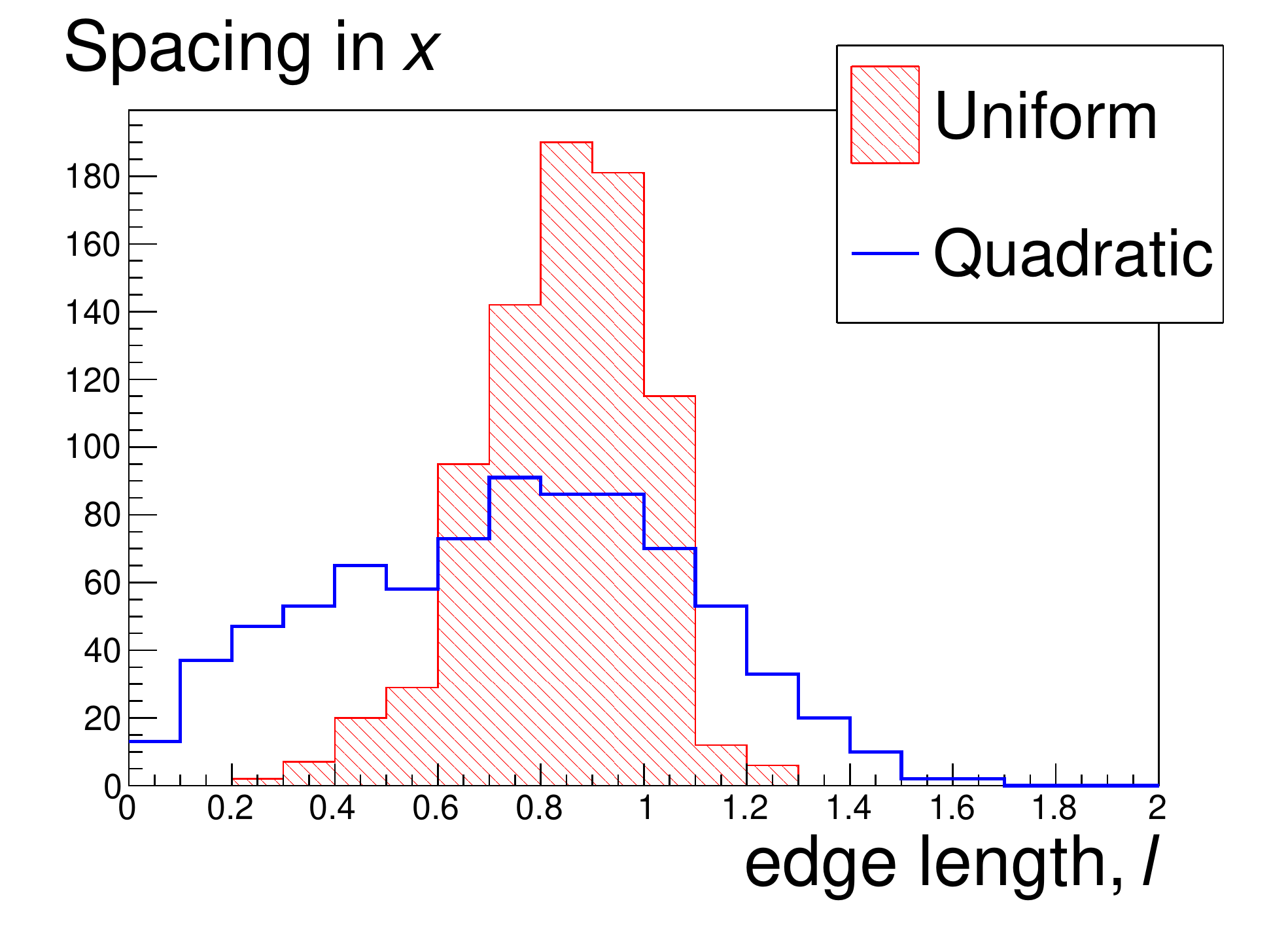}
\includegraphics[width=0.495\textwidth]{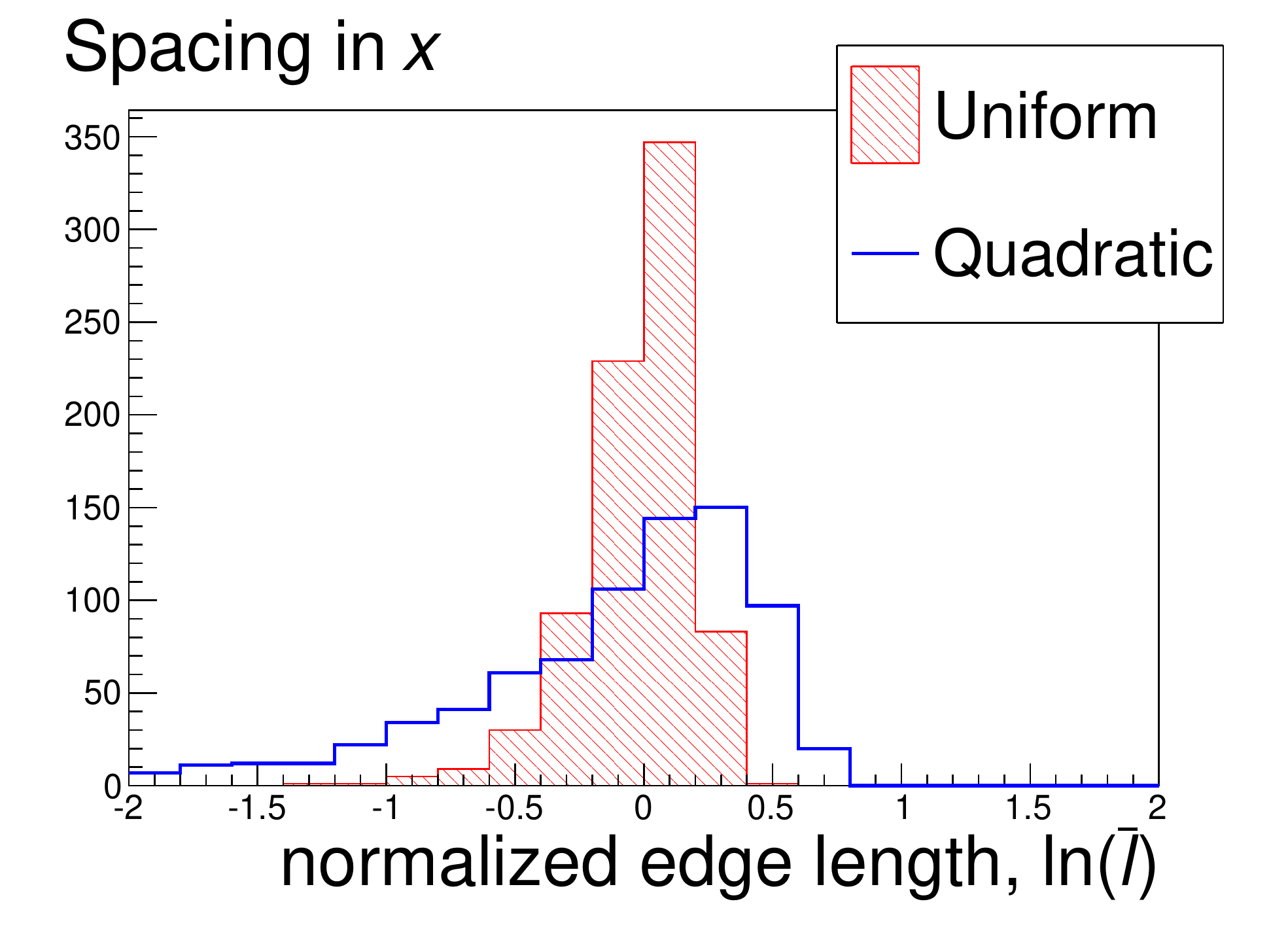}
\includegraphics[width=0.495\textwidth]{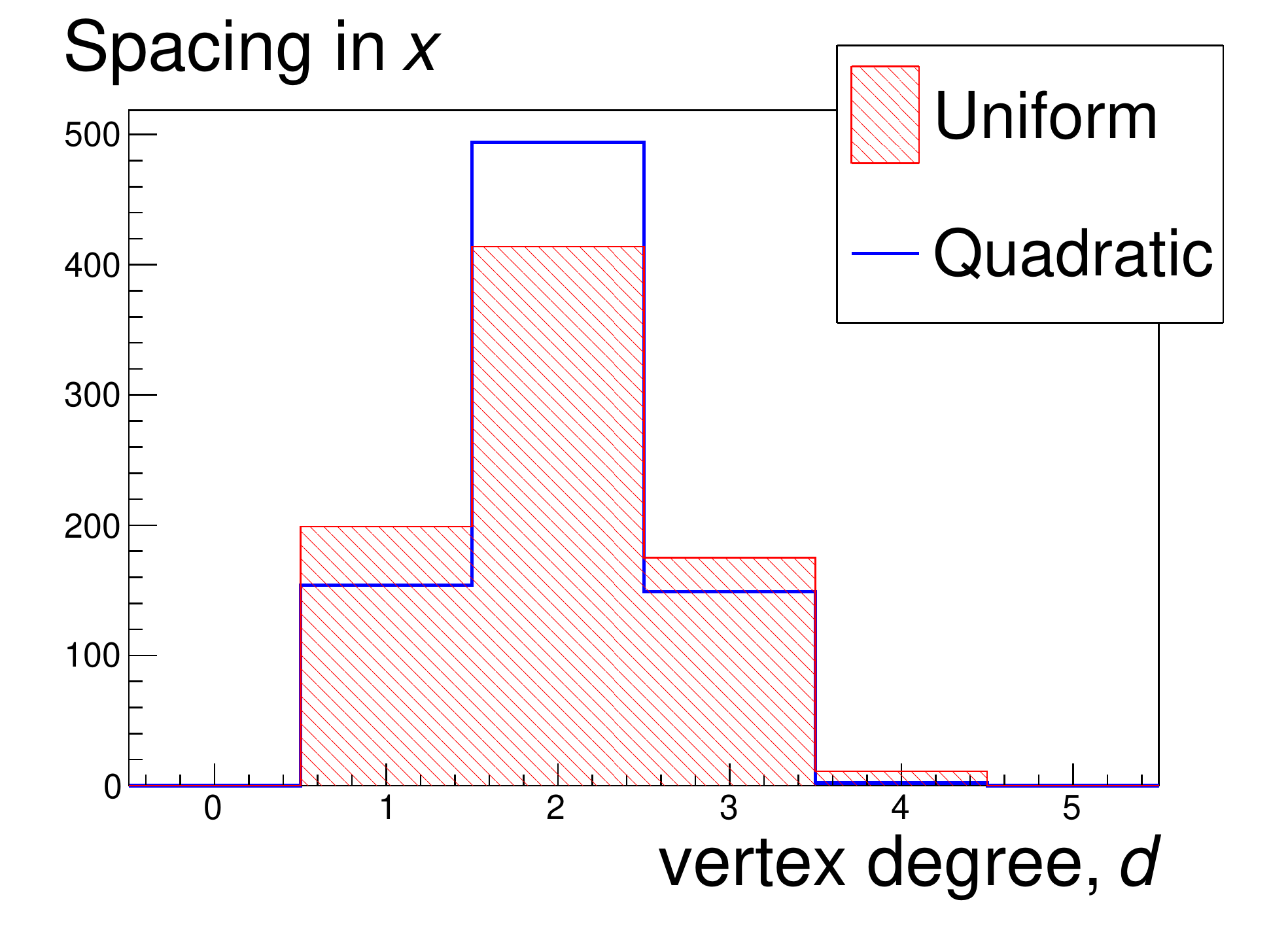}
\includegraphics[width=0.495\textwidth]{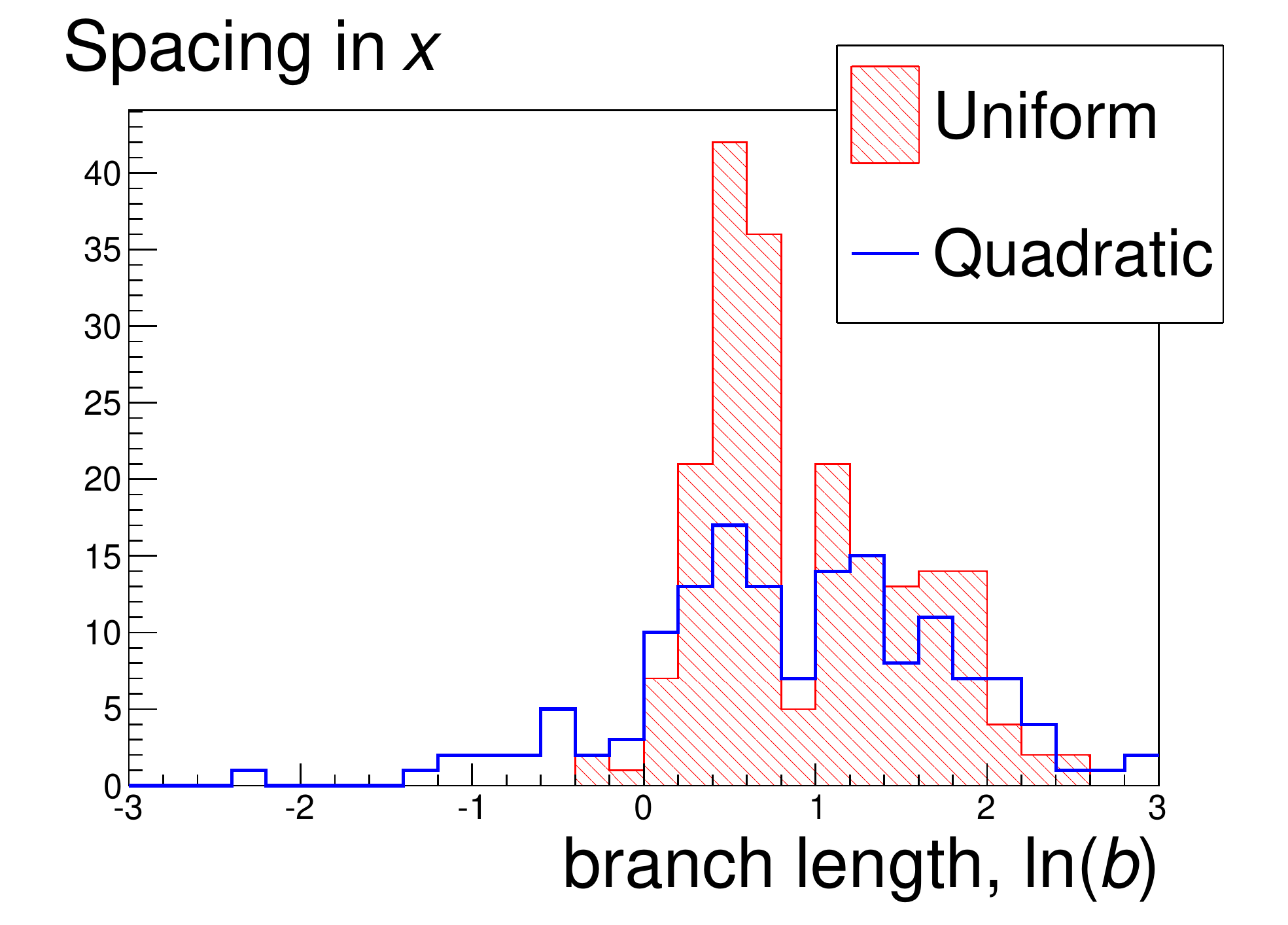}
\caption{The distributions of $l$ (top left), $\ln(\bar{l})$ (top right), $d$ (bottom left), and $\ln(b)$ (bottom right) for the quadratic and uniform trees.}
\label{fig: ex2indiv}

\vskip 30pt

\includegraphics[width=0.495\textwidth]{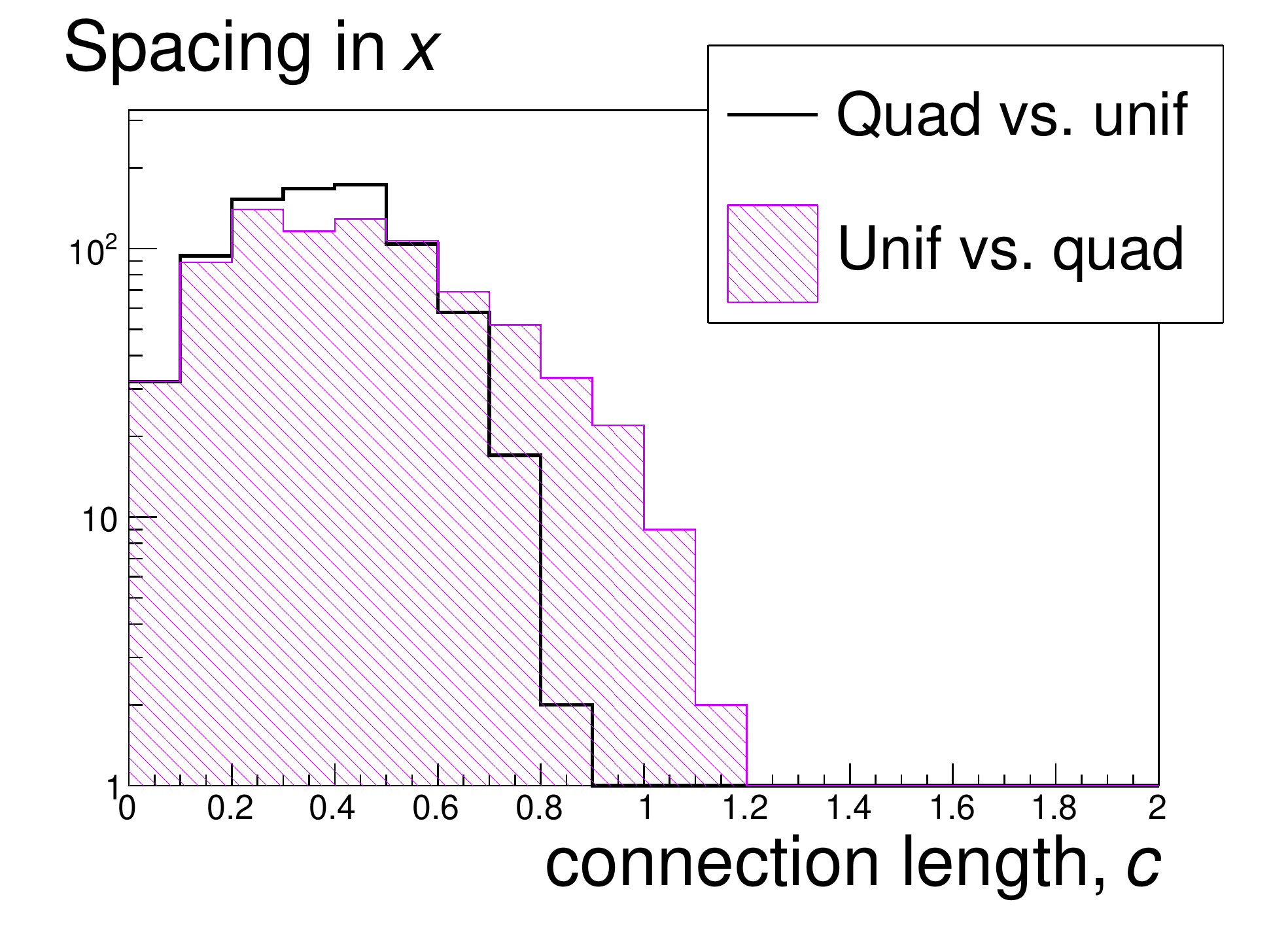}
\includegraphics[width=0.495\textwidth]{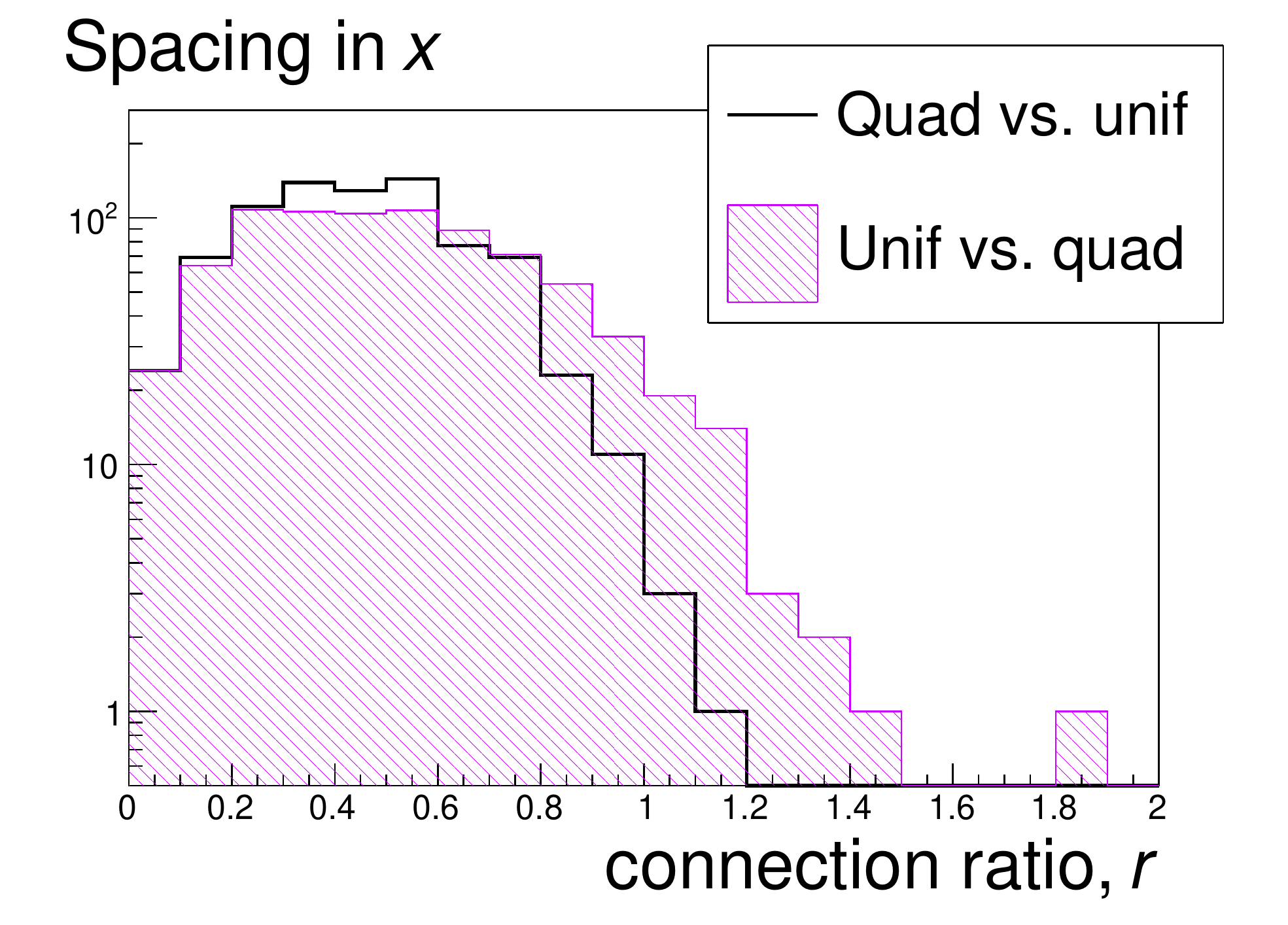}
\caption{The distributions of $c$ (left) and $r$ (right) for the quadratic and uniform trees.}
\label{fig: ex2comp}
\end{figure}

The distributions of the statistical quantities for these trees are shown in figure~\ref{fig: ex2indiv}.  The $l$ distributions have similar means, but the distribution for the quadratic tree is wider, indicating its variety of branch lengths.  This nonuniformity is also apparent in the wider, more skewed distribution of $\ln(\bar{l})$ for the quadratic tree.  The filamentary structure is further exemplified in the distributions of $d$, where the quadratic tree distribution peaks more sharply around 2.  The wider distribution of $\ln(b)$ for the quadratic tree signifies its shorter and longer branches.  Another effect of the sparse vertices on the sides of the quadratic tree is its narrower distribution of $c$, shown in figure~\ref{fig: ex2comp}.  The results for $r$ are similar because the trees in this example have similar average spacing as compared to figure~\ref{fig: ex1comp}.

\subsection{Disc vs. strip}
\begin{figure} \centering
\includegraphics[width=0.69\textwidth, trim={0 2cm 0 2cm}, clip]{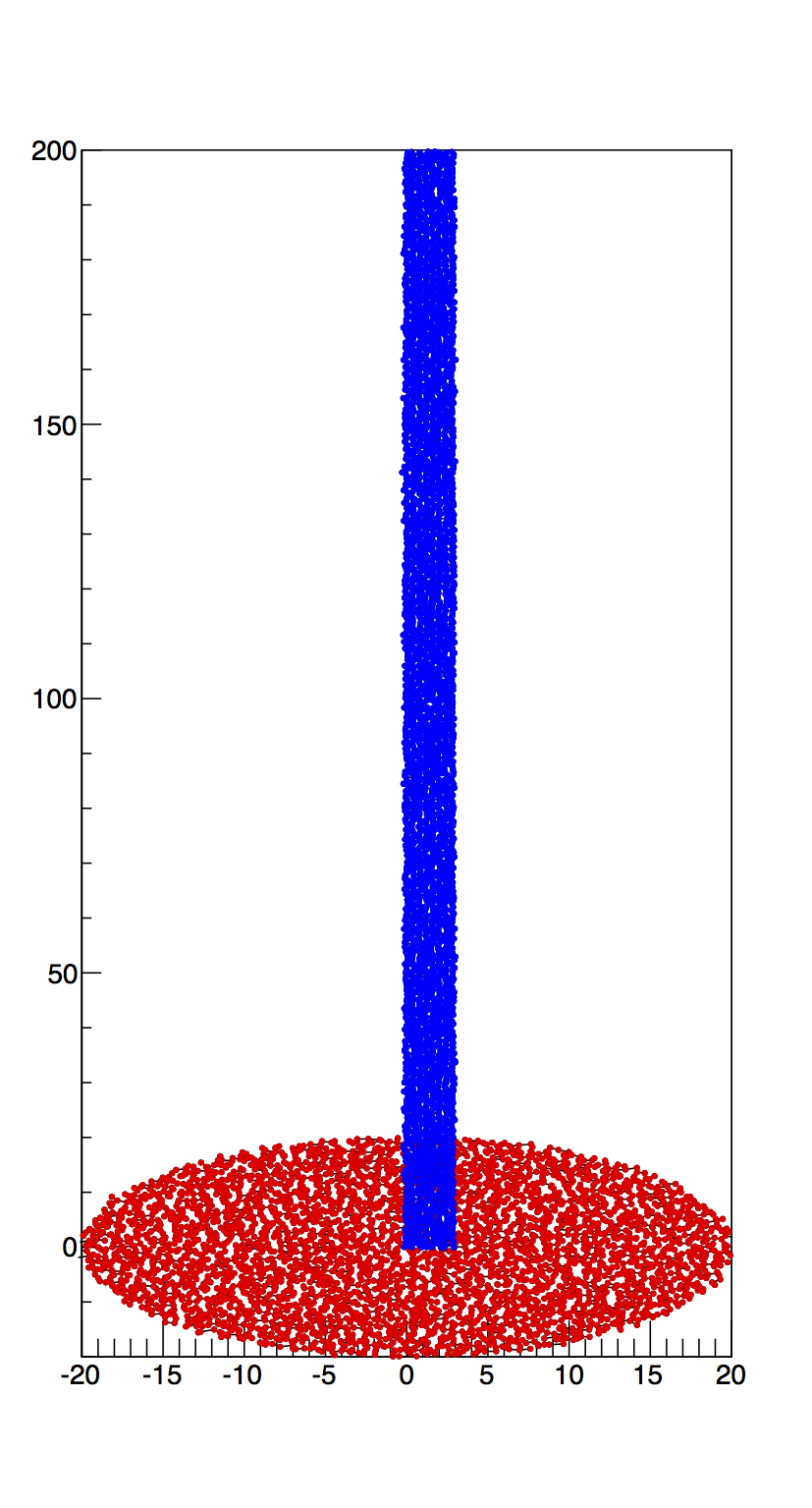}
\caption{The disc (red) and strip (blue) trees.}
\label{fig: ex3trees}
\end{figure}

Our third example compares two different tree geometries: one set of 4000 vertices uniformly distributed over a disc and one set of 4000 vertices uniformly distributed over a long, narrow rectangle which we will refer to as a ``strip.''  Both are shown in figure~\ref{fig: ex3trees}.  The strip geometry limits the length of branches that can be created in that tree as compared to the disc tree.

\begin{figure} \centering
\includegraphics[width=0.495\textwidth]{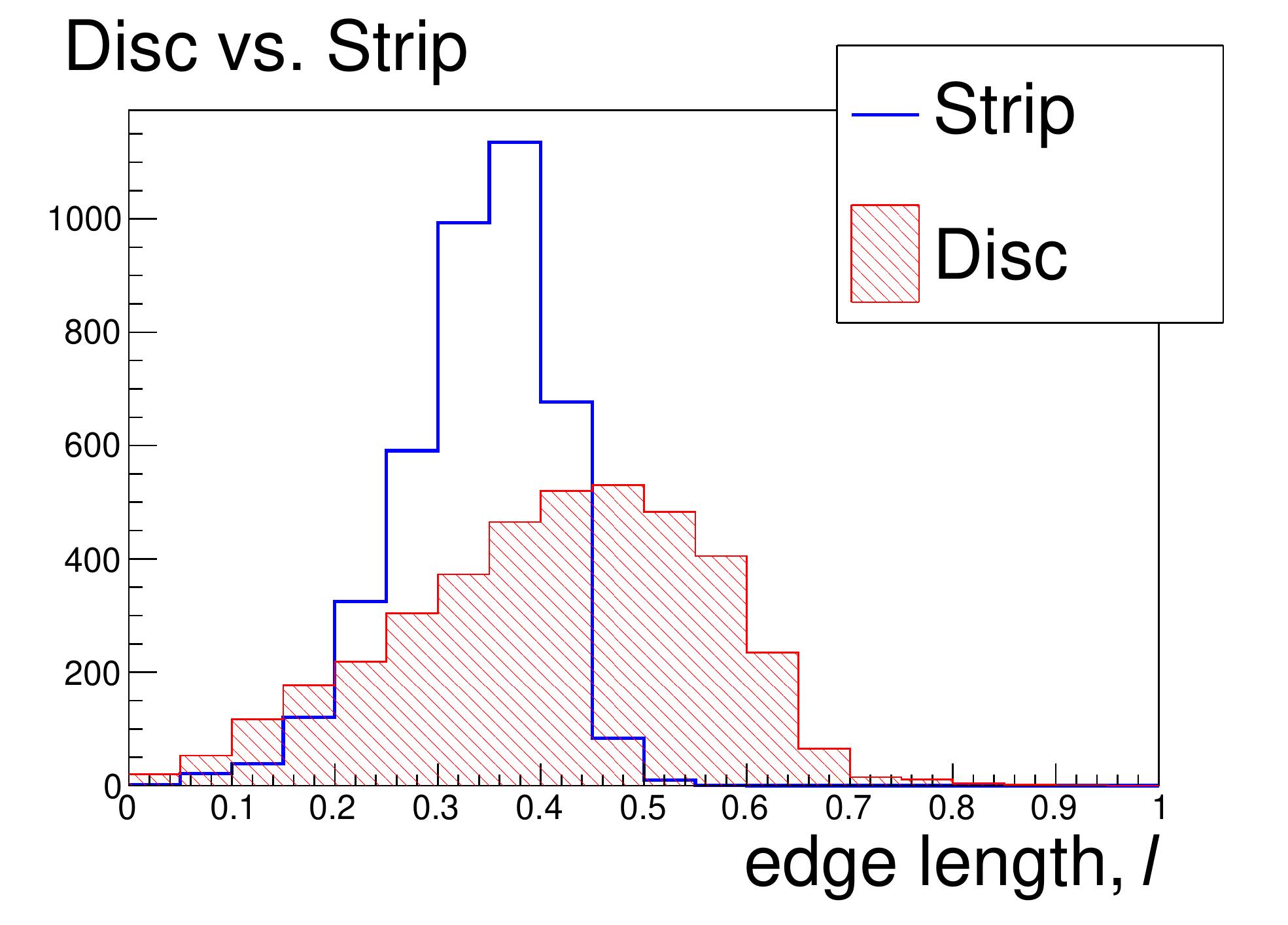}
\includegraphics[width=0.495\textwidth]{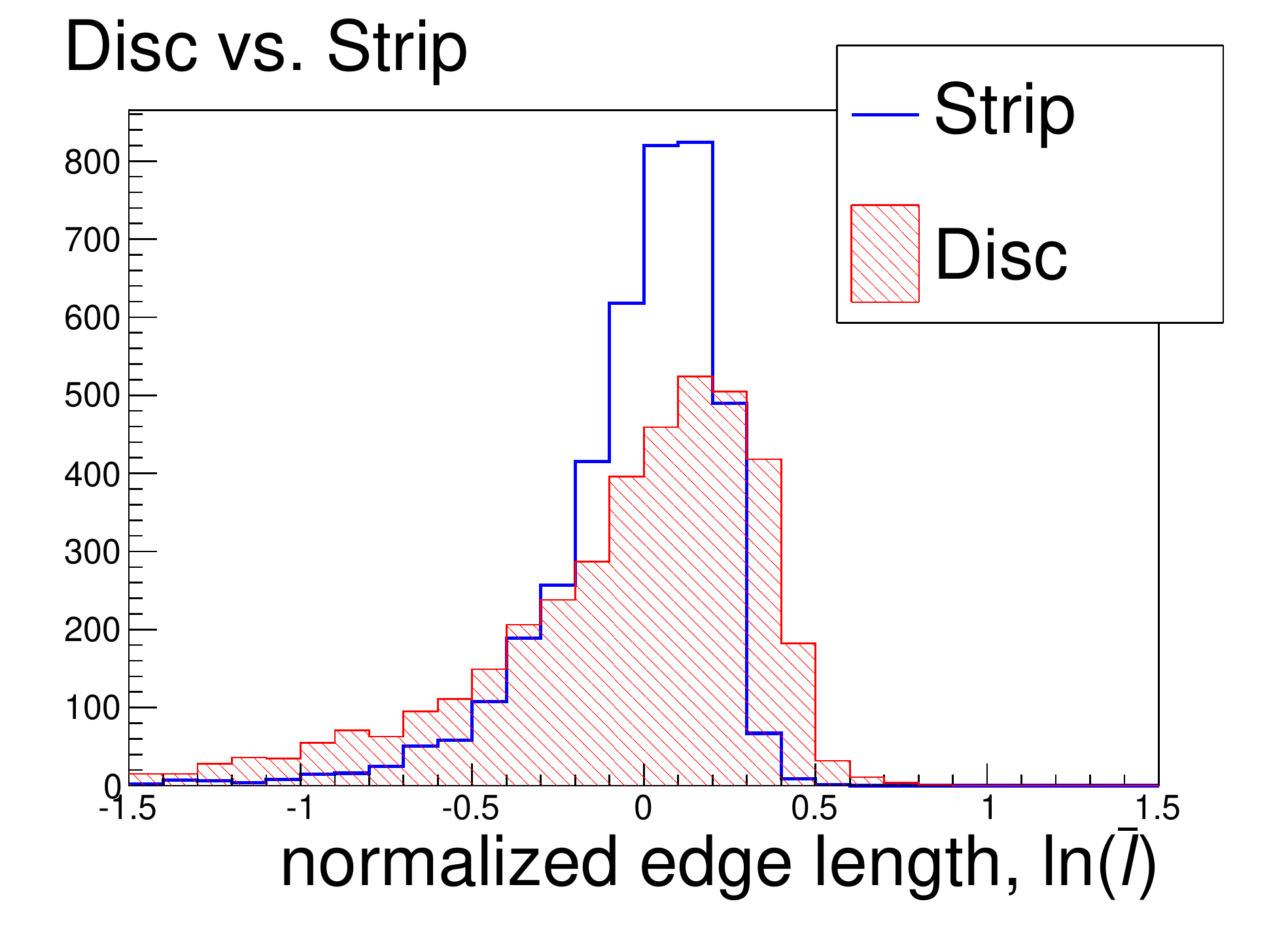}
\includegraphics[width=0.495\textwidth]{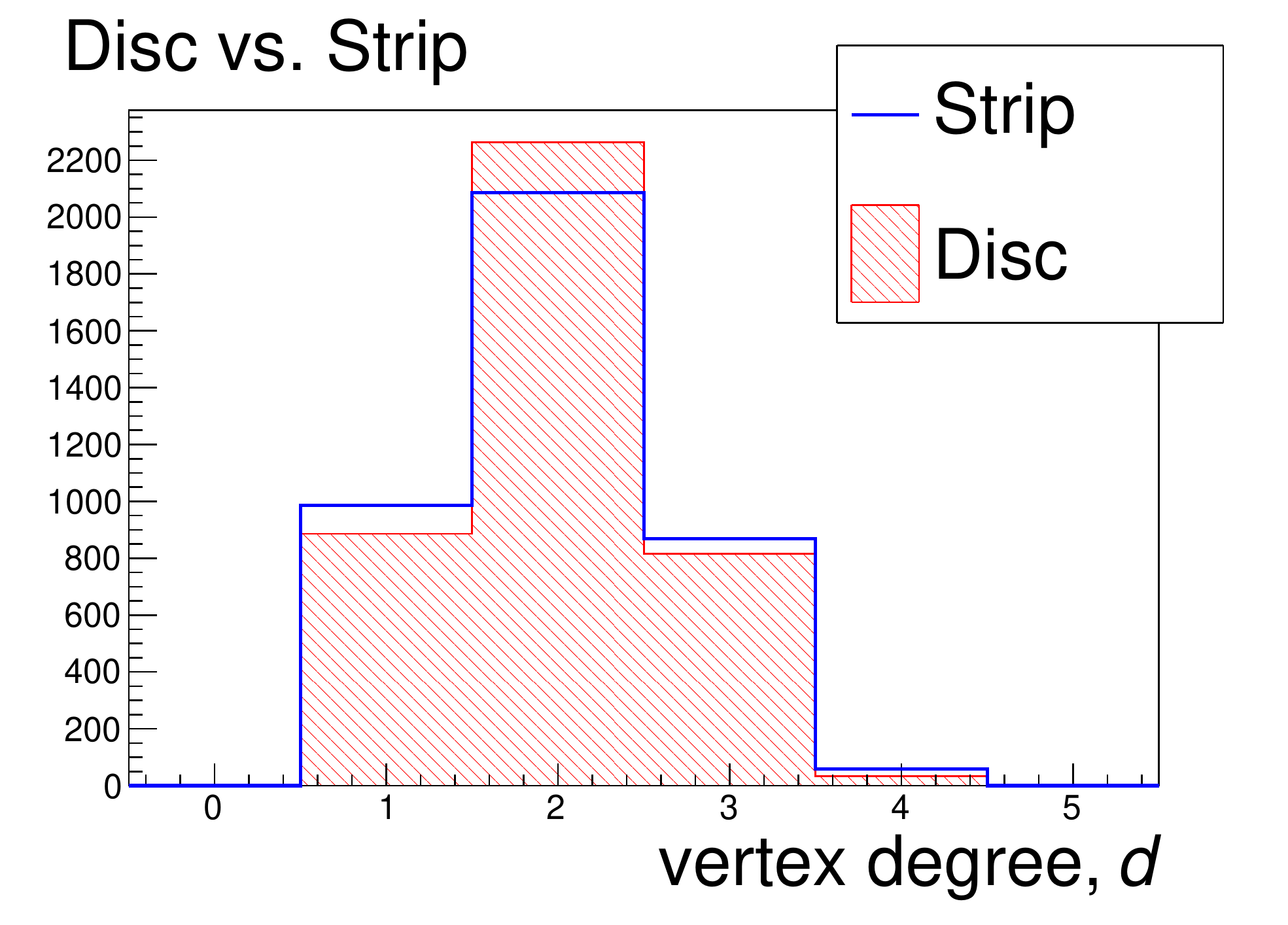}
\includegraphics[width=0.495\textwidth]{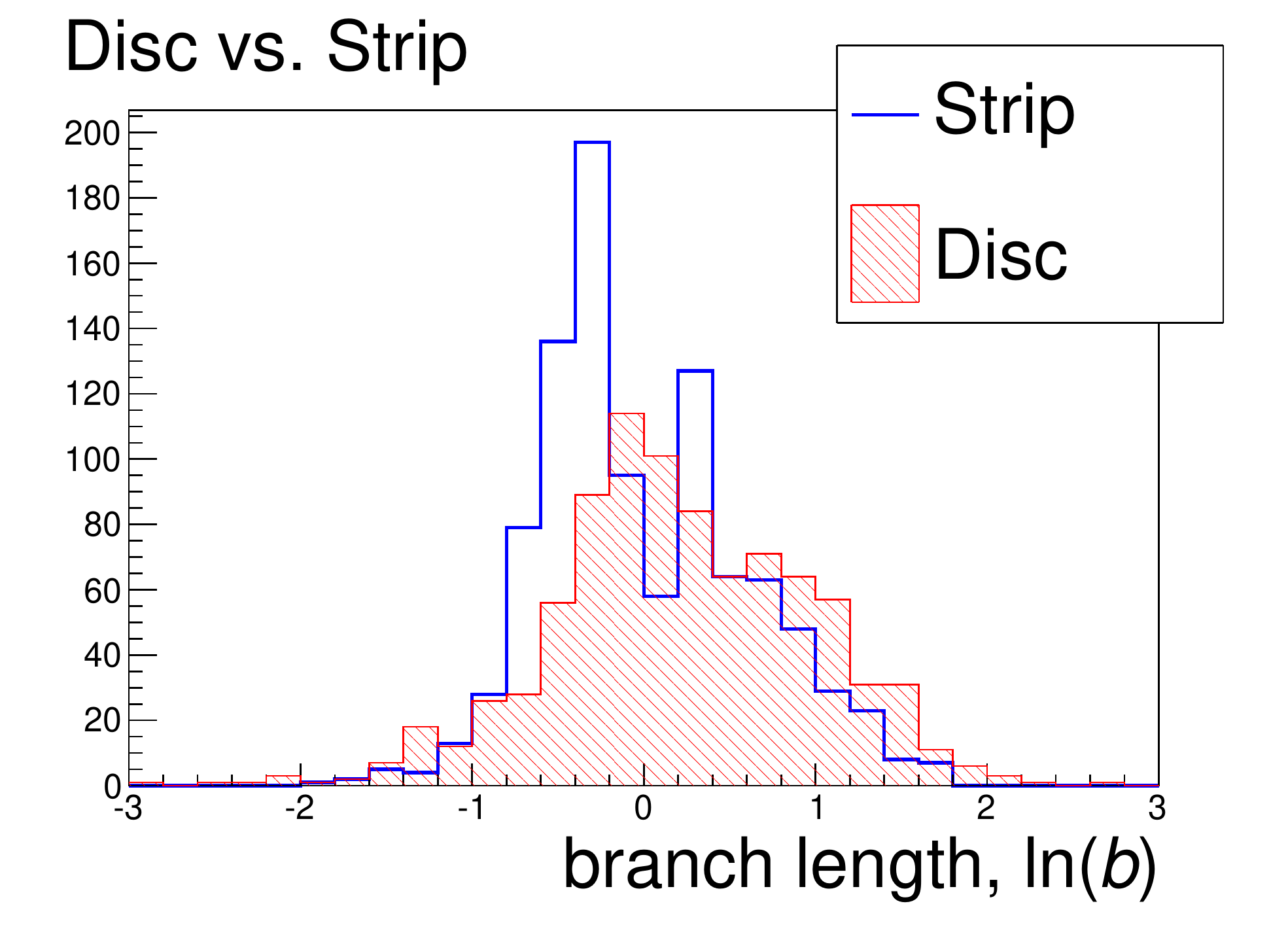}
\caption{The distributions of $l$ (top left), $\ln(\bar{l})$ (top right), $d$ (bottom left), and $\ln(b)$ (bottom right) for the disc and strip trees in figure~\ref{fig: ex3trees}.}
\label{fig: ex3indiv}

\vskip 30pt

\includegraphics[width=0.495\textwidth]{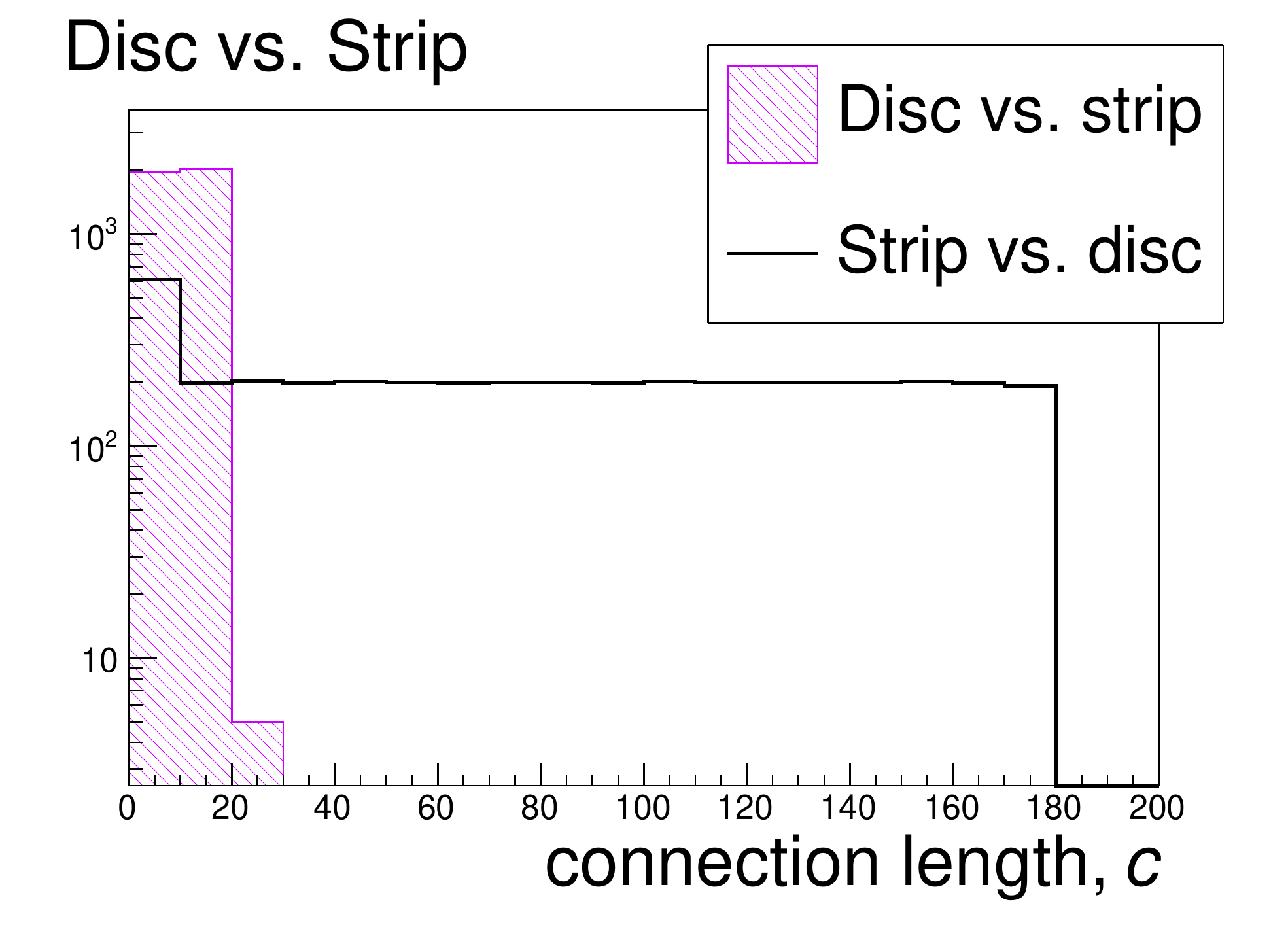}
\includegraphics[width=0.495\textwidth]{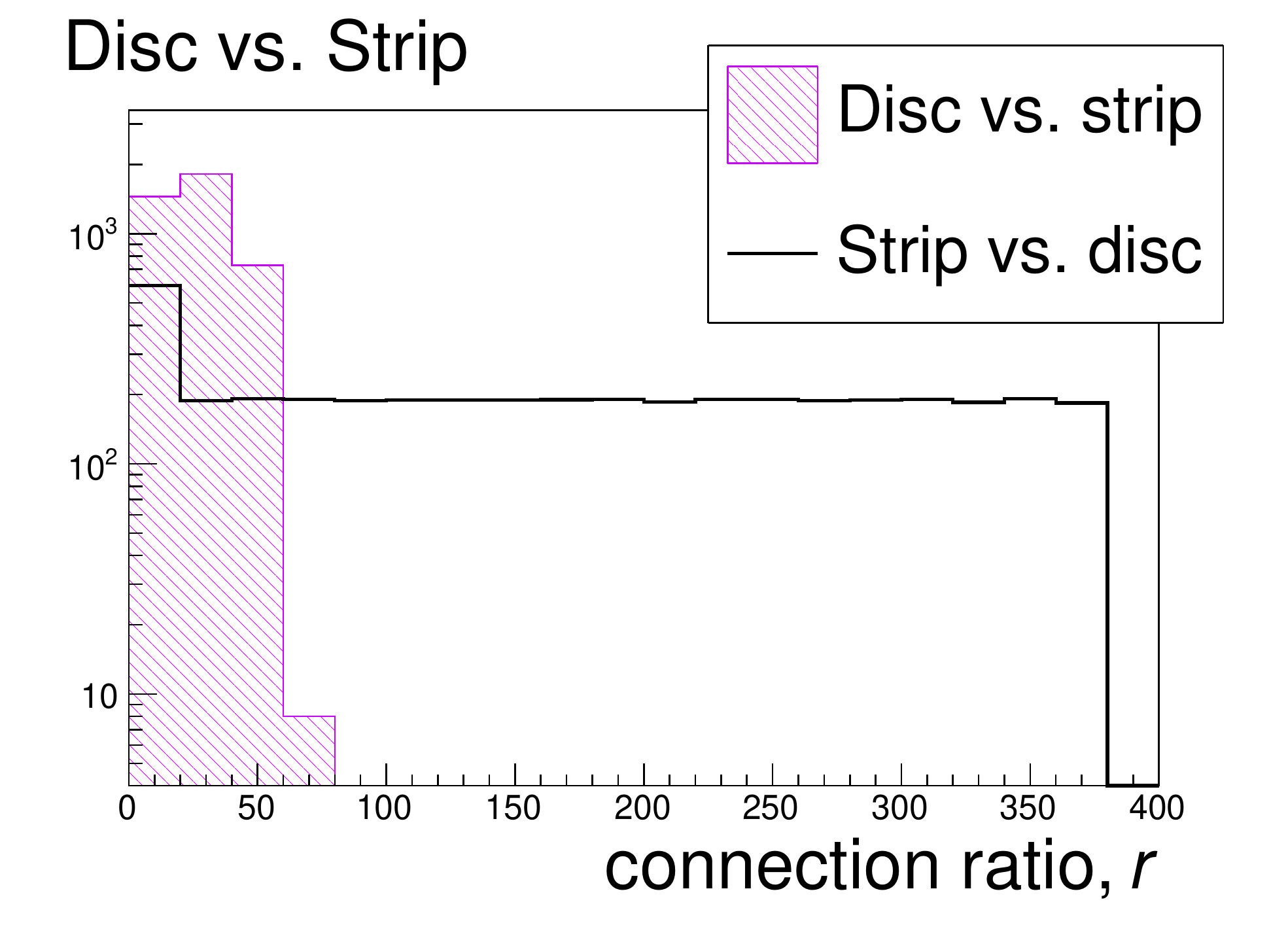}
\caption{The distributions of $c$ (left) and $r$ (right) for the disc and strip trees.}
\label{fig: ex3comp}
\end{figure}

The difference in the means of $l$ in figure~\ref{fig: ex3indiv} shows that the vertices of the strip tree are denser.  The broader, slightly shifted distribution for $\ln(\bar{l})$ of the disc tree is also due to its sparser distribution of vertices.  We see a wider distribution of $d$ for the strip tree because it has shorter branches.  These short branches are also visible in the $\ln(b)$ distribution.  The most noticeable difference in these trees aside from their shape is their differing locations in phase space.  The peaks on the left in the $c$ distributions in figure~\ref{fig: ex3comp} represent the area of overlap.  The bins on the right for the strip tree show a concentration of events that do not fit the disc distribution.  Rescaling the connection lengths to obtain $r$ gives us an even longer tail for the strip tree, magnifying the effect of its variant geometry.

\subsection{Hidden variable}
Our fourth example illustrates the multidimensional capabilities of the MST.  This example is important because it illustrates the potential of an MST to profit from structures or distributions in a multidimensional feature space that may be unanticipated by the researcher.  While this example plainly is contrived, it nonetheless illustrates how a multidimensional MST could provide non-obvious discriminating power.

\begin{figure} \centering
\includegraphics[width=0.69\textwidth, trim={0 1cm 0 1cm}, clip]{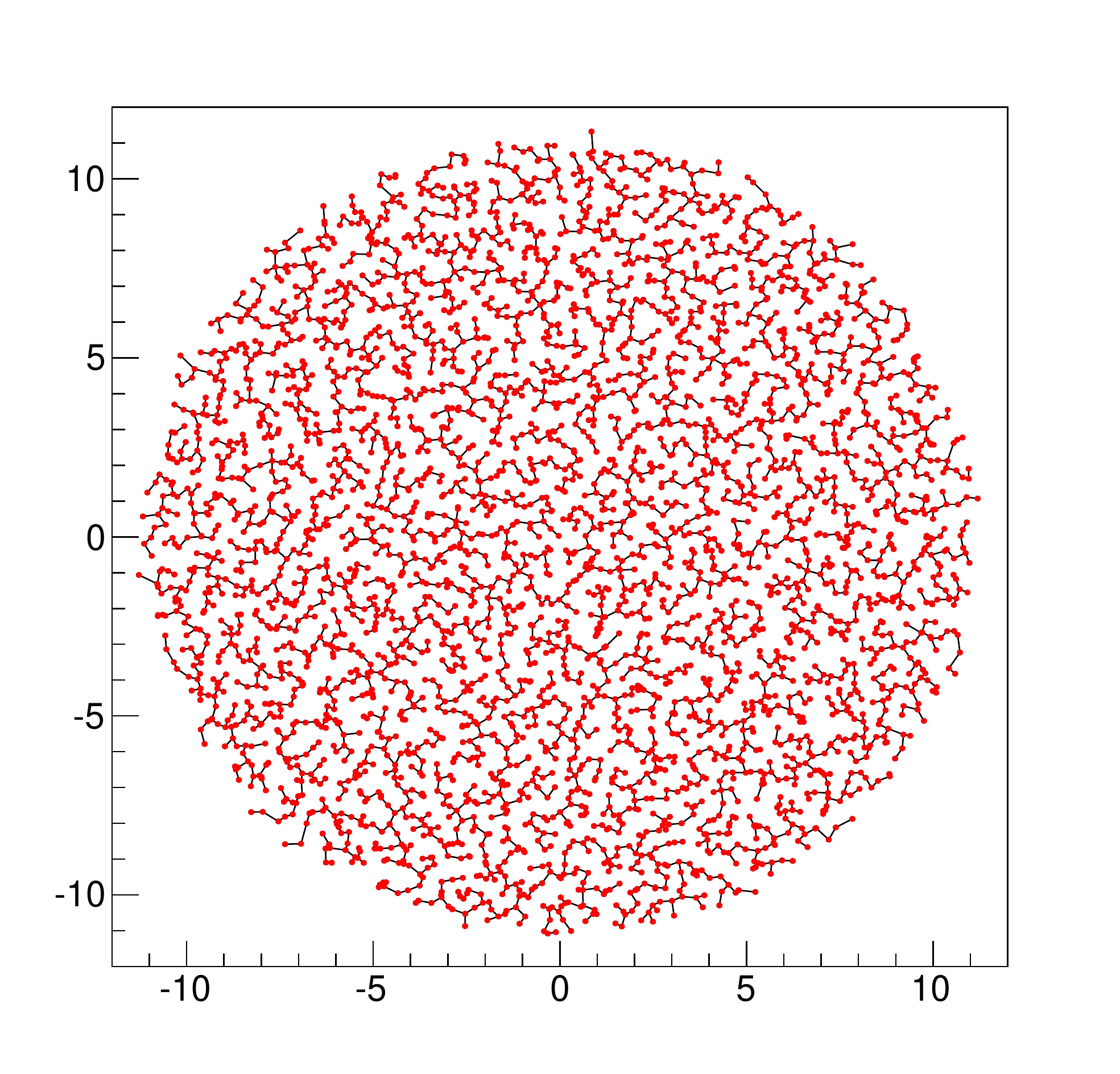}
\includegraphics[width=0.69\textwidth, trim={0 1cm 0 1cm}, clip]{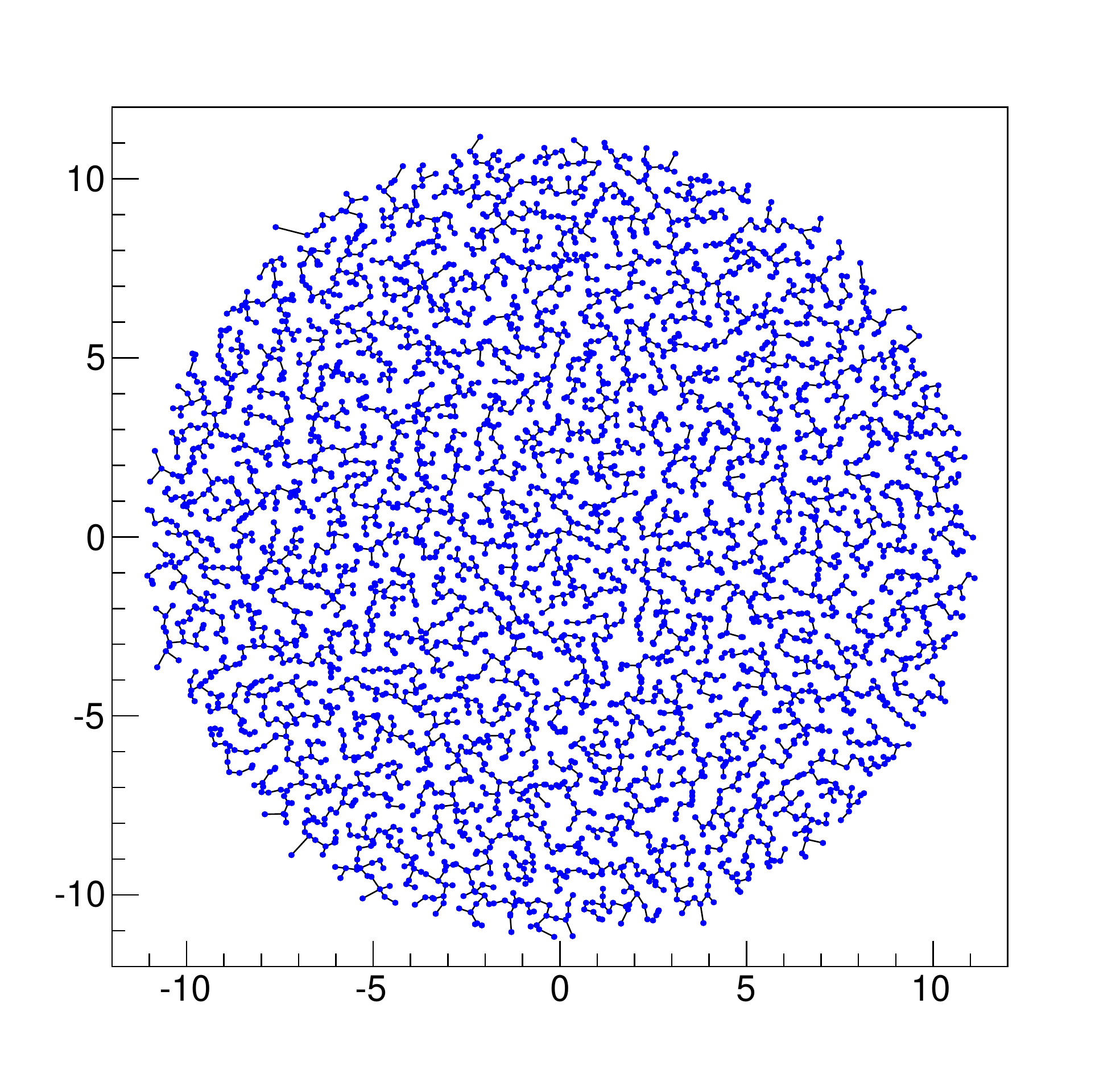}
\caption{The similar disc trees in two dimensions.}
\label{fig: ex4trees2d}
\end{figure}

\begin{figure} \centering
\includegraphics[width=0.495\textwidth]{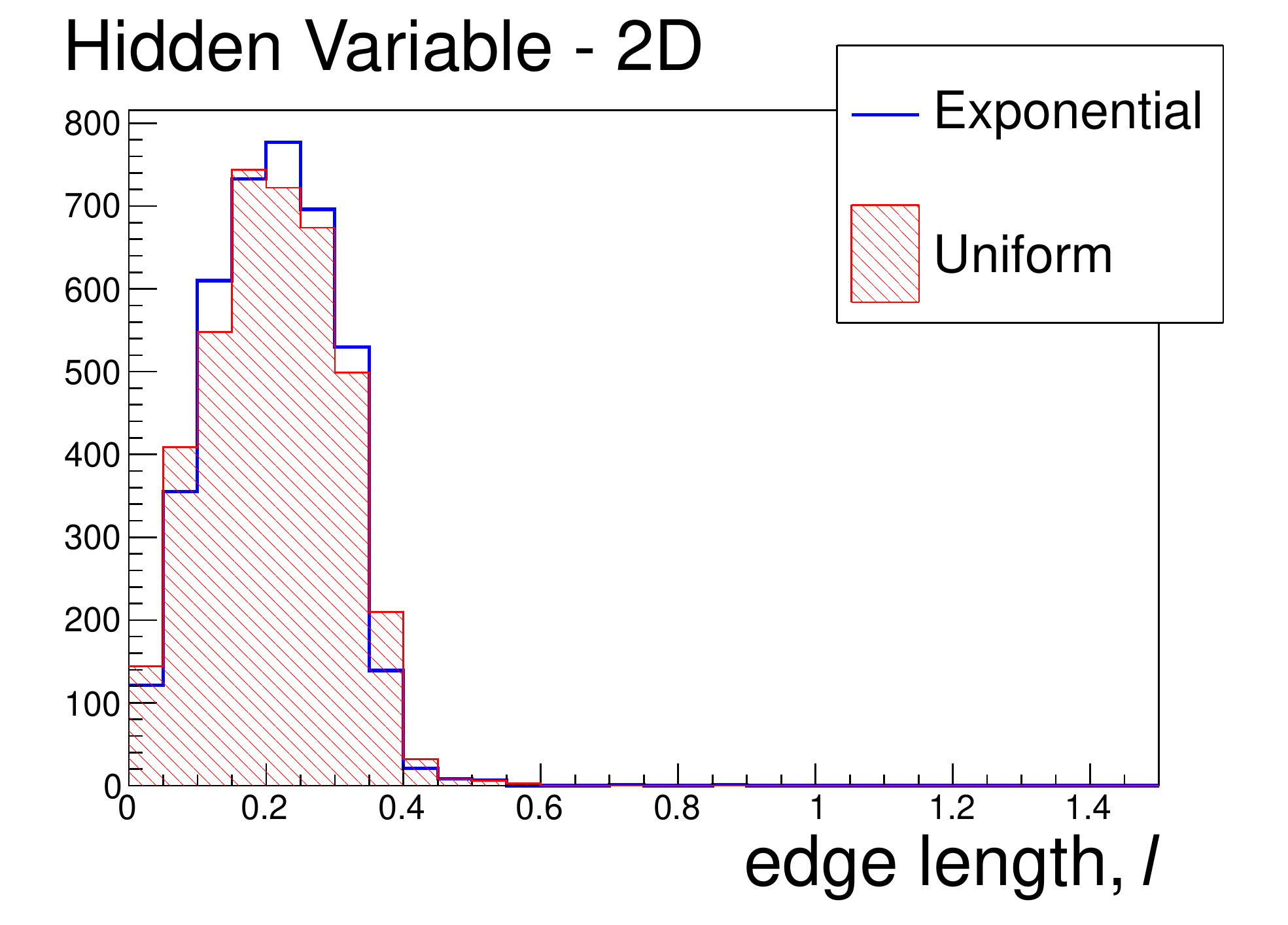}
\includegraphics[width=0.495\textwidth]{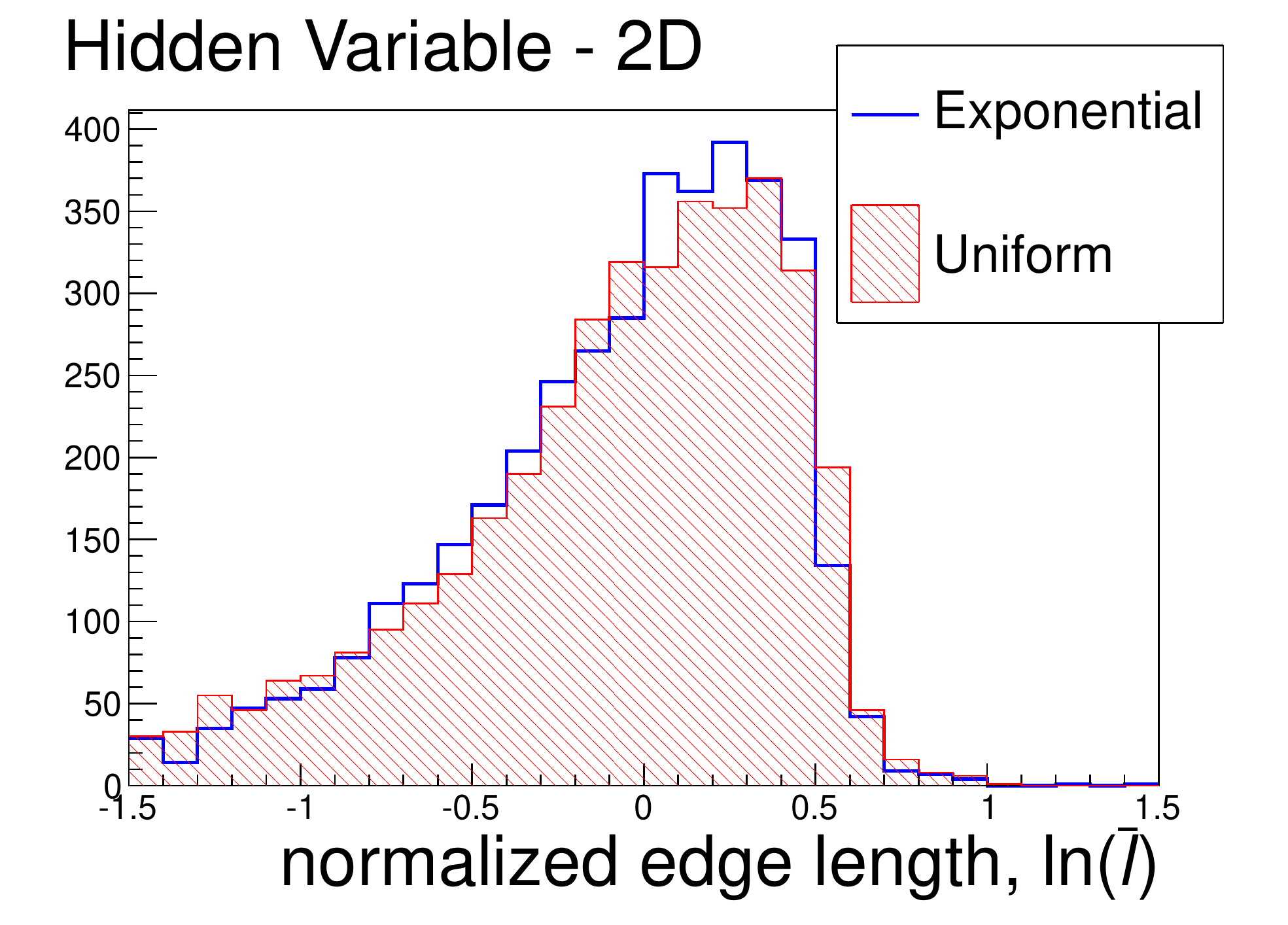}
\includegraphics[width=0.495\textwidth]{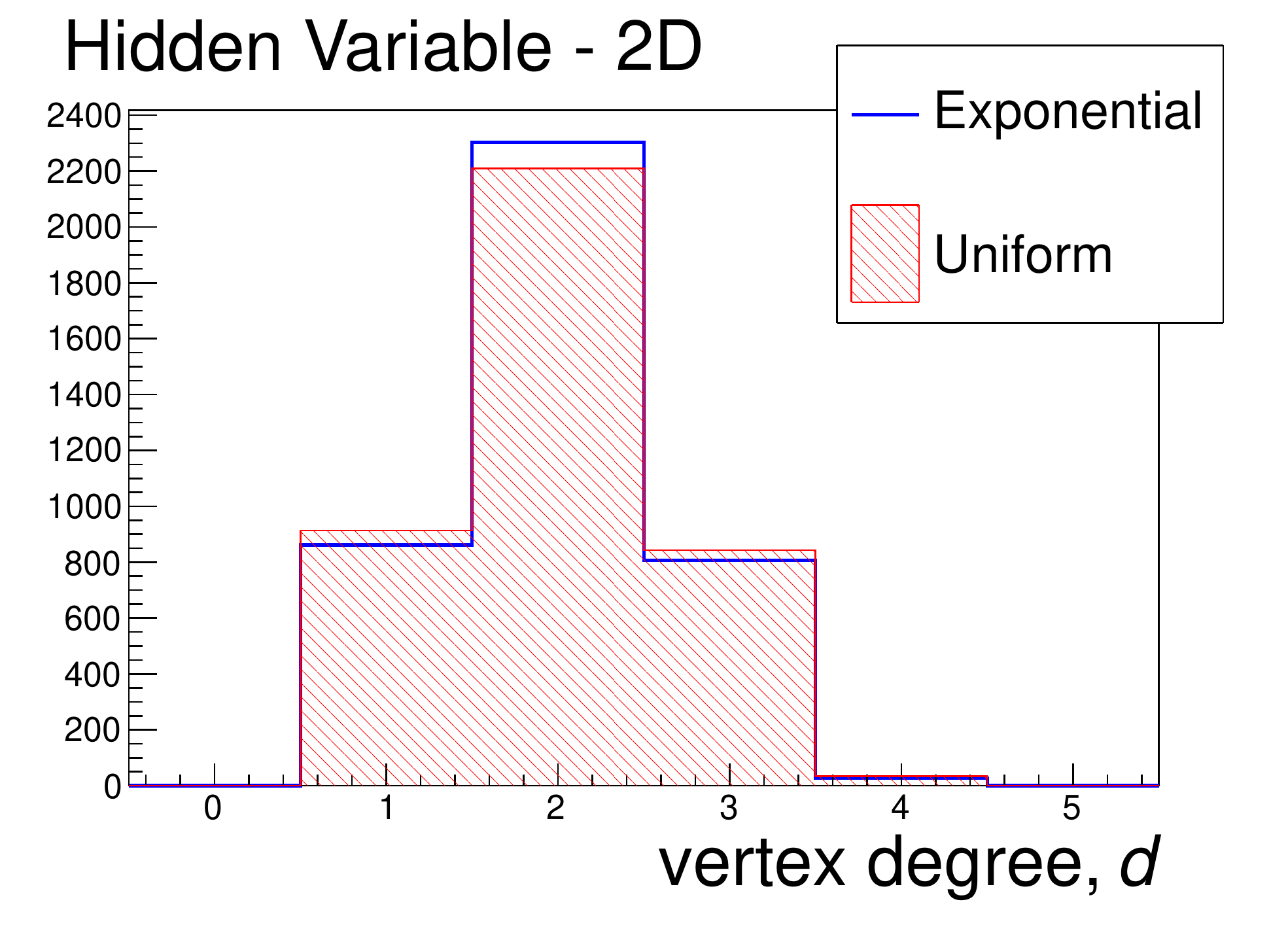}
\includegraphics[width=0.495\textwidth]{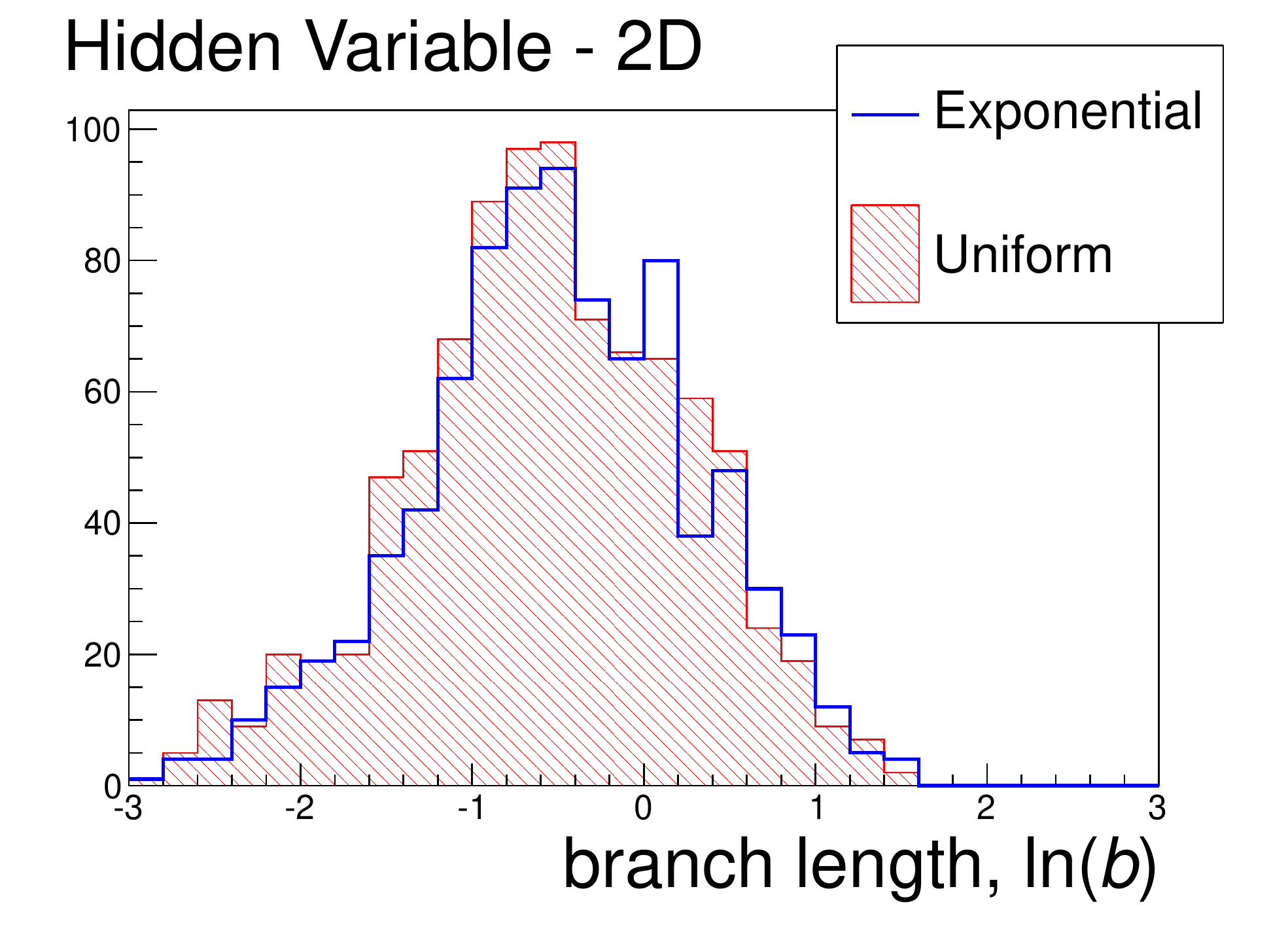}
\caption{The distributions of $l$ (top left), $\ln(\bar{l})$ (top right), $d$ (bottom left), and $\ln(b)$ (bottom right) for the similar disc trees in figure~\ref{fig: ex4trees2d}.}
\label{fig: ex4indiv2d}

\vskip 30pt

\includegraphics[width=0.495\textwidth]{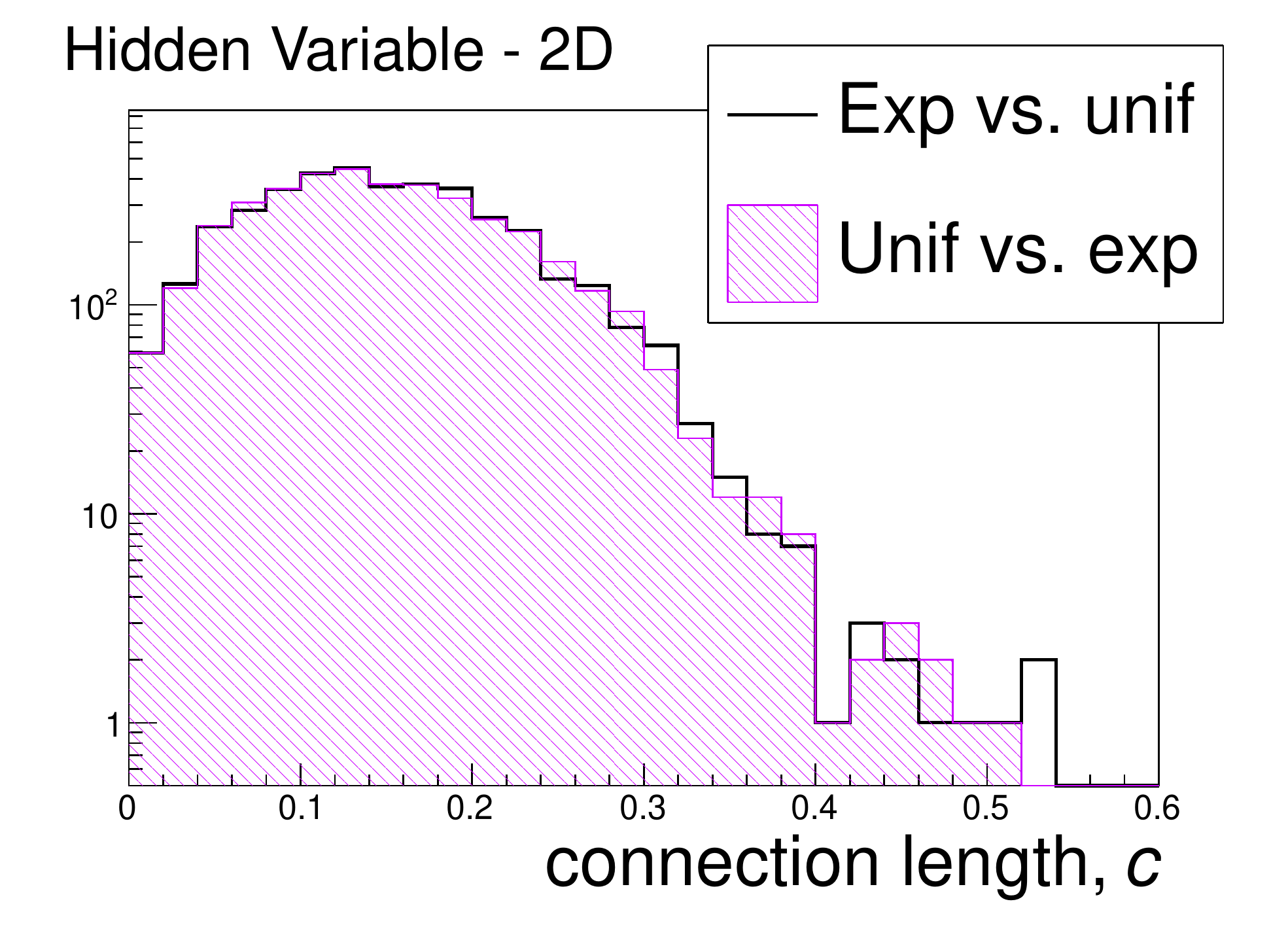}
\includegraphics[width=0.495\textwidth]{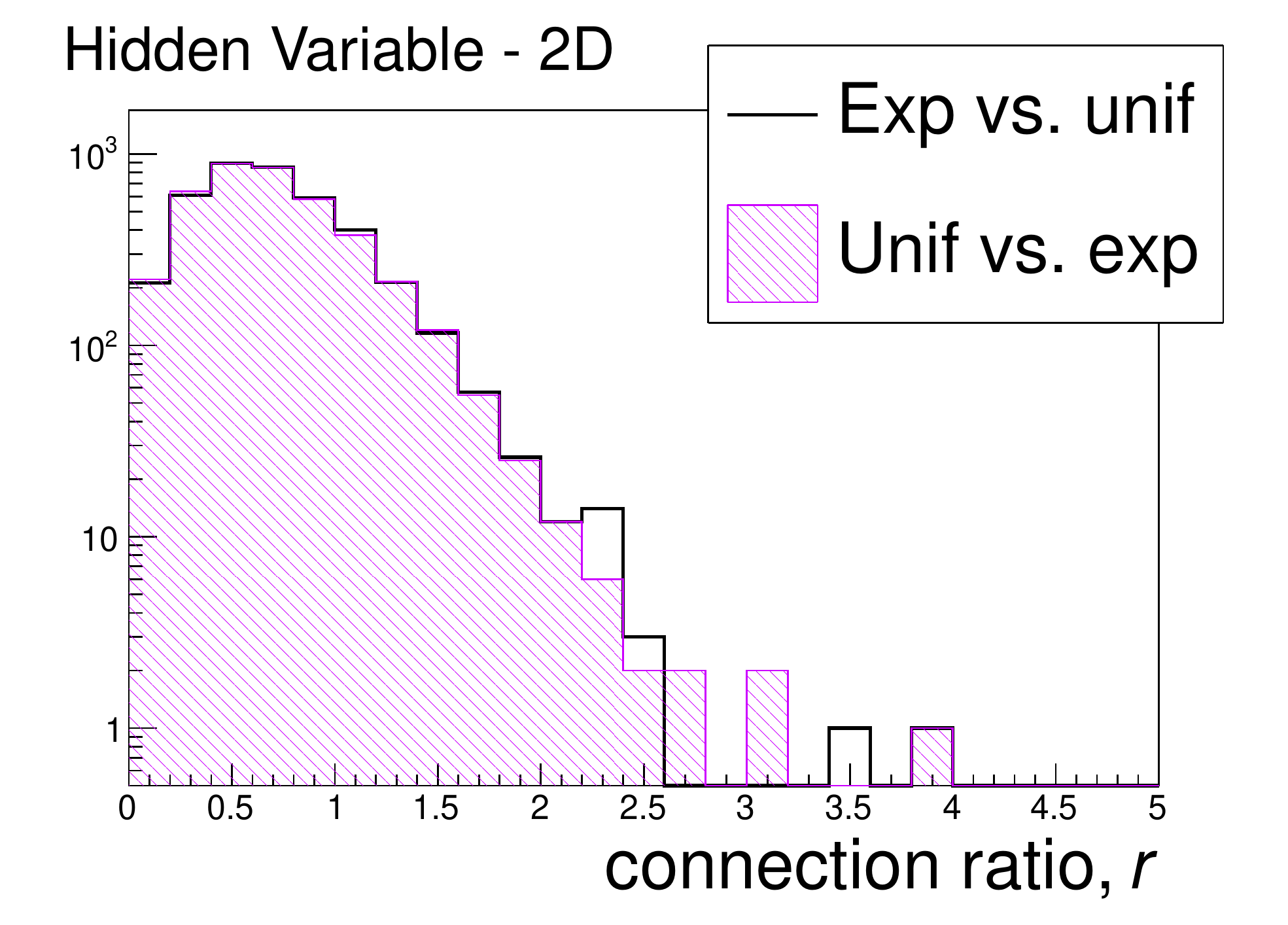}
\caption{The distributions of $c$ (left) and $r$ (right) for the similar disc trees.}
\label{fig: ex4comp2d}
\end{figure}

We uniformly distributed and perturbed two sets of 4000 vertices on the same disc in two-dimensional space, shown in figure~\ref{fig: ex4trees2d}.  Figures~\ref{fig: ex4indiv2d} and \ref{fig: ex4comp2d} show that the distributions of statistical quantities for the two discs are the same apart from statistical fluctuations.  In particular, the histograms of $c$ in figure~\ref{fig: ex4comp2d} are not only nearly identical, but also very narrow because the trees are so similar.

\begin{figure} \centering
\includegraphics[width=0.69\textwidth, trim={0 1cm 0 1cm}, clip]{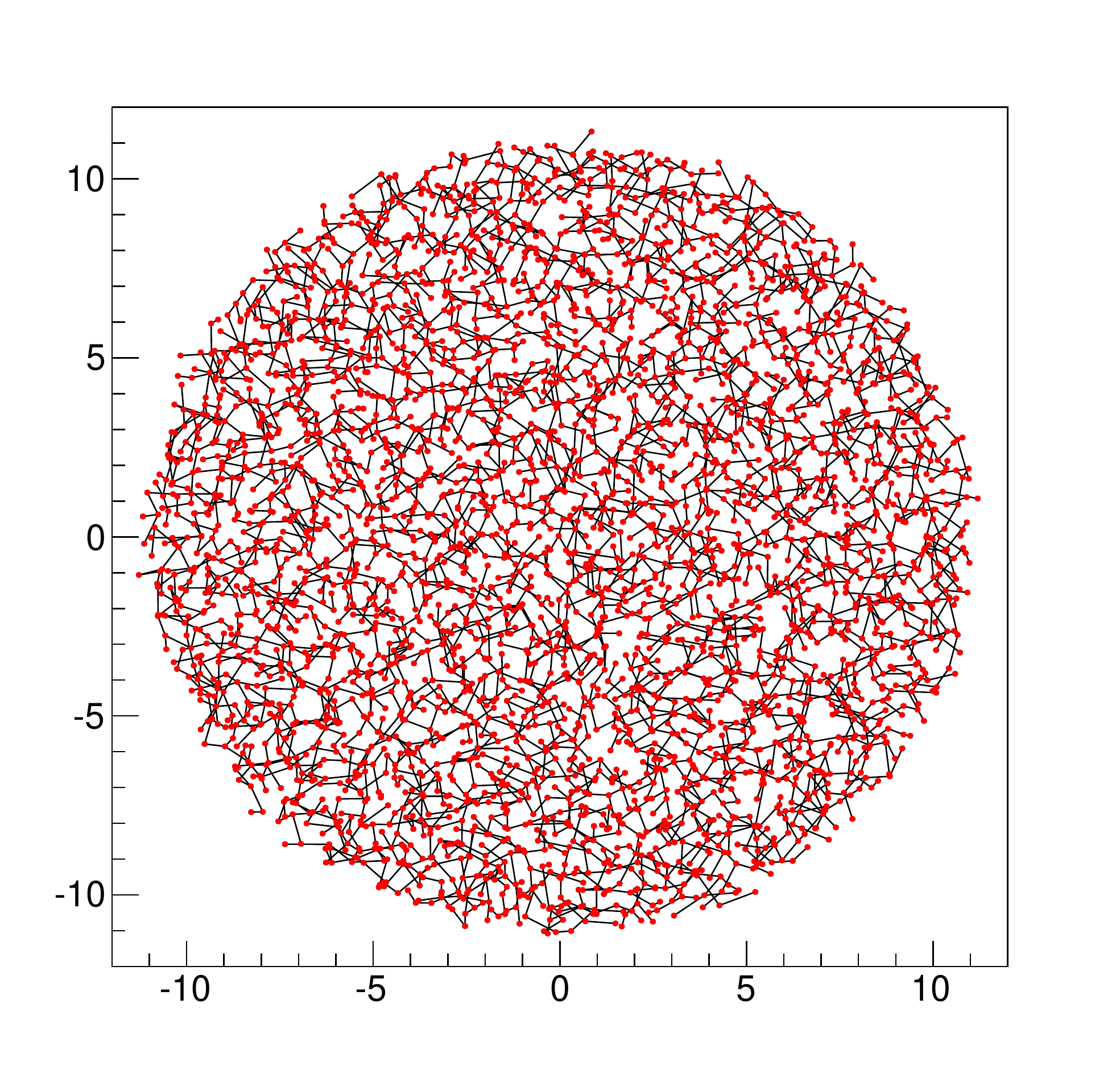}
\includegraphics[width=0.69\textwidth, trim={0 1cm 0 1cm}, clip]{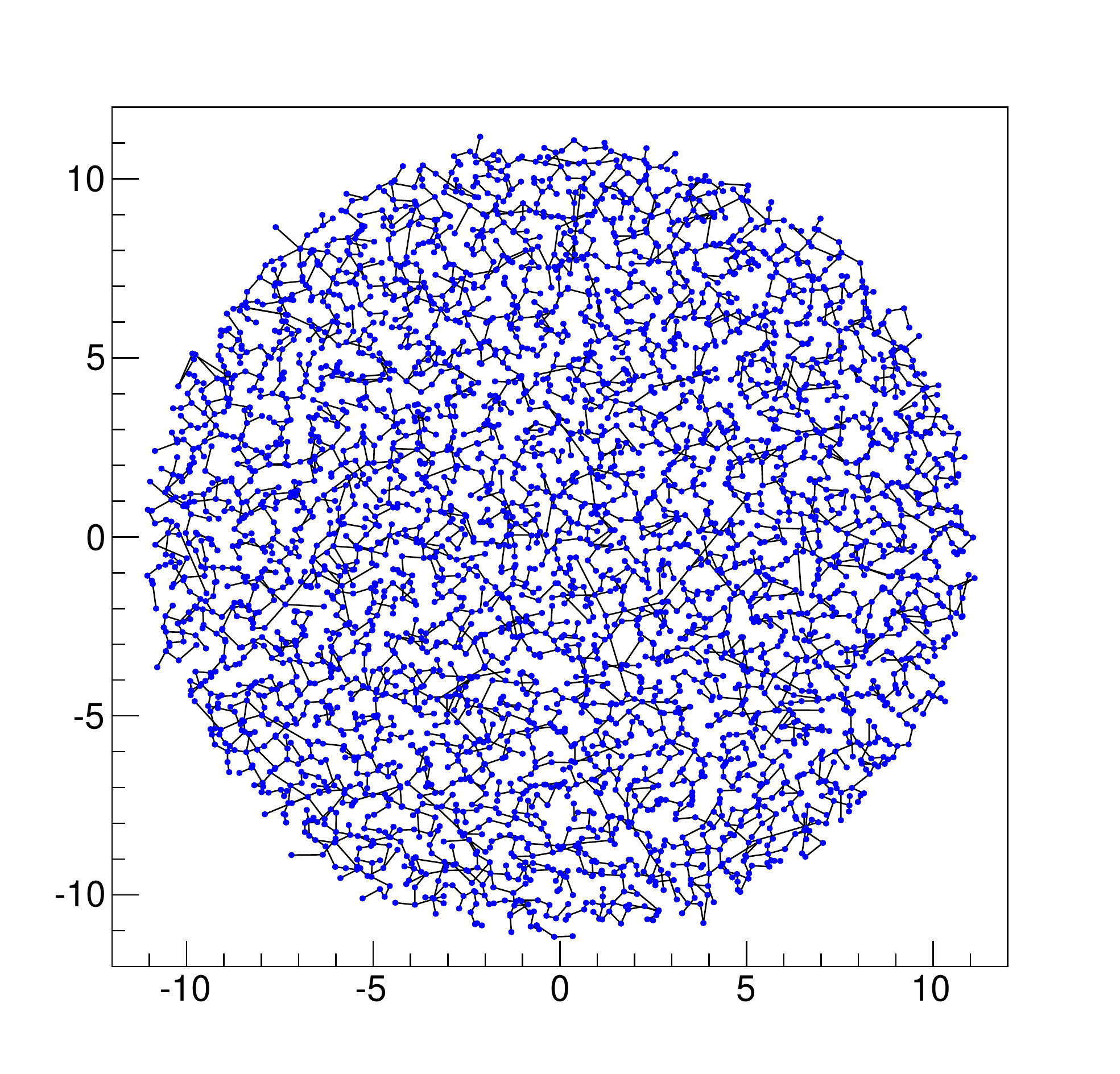}
\caption{Projections in $xy$ of the uniform tree (red) and the exponential tree (blue).}
\label{fig: ex4trees3d}
\end{figure}

\begin{figure} \centering
\includegraphics[width=0.69\textwidth]{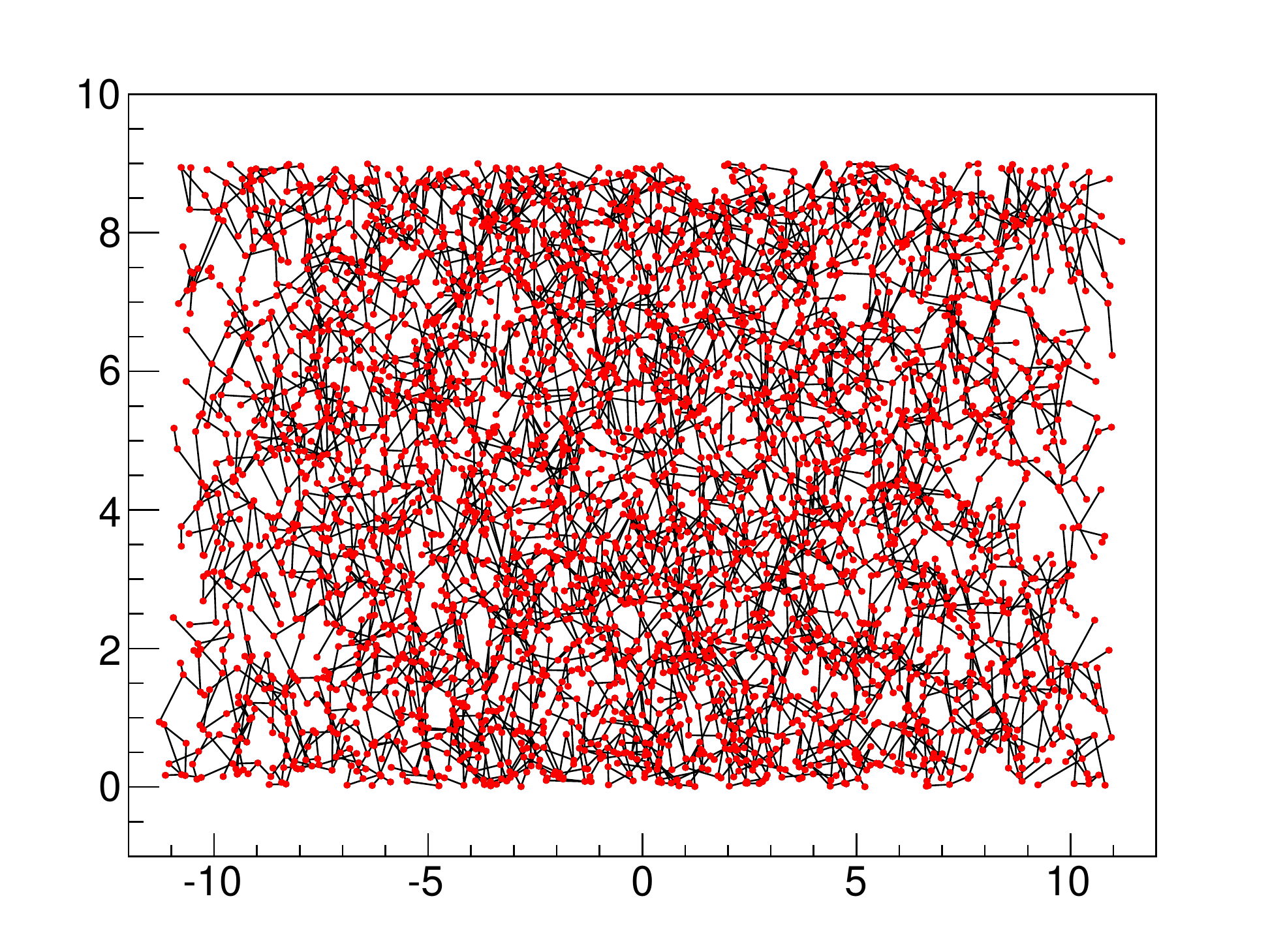}
\includegraphics[width=0.69\textwidth]{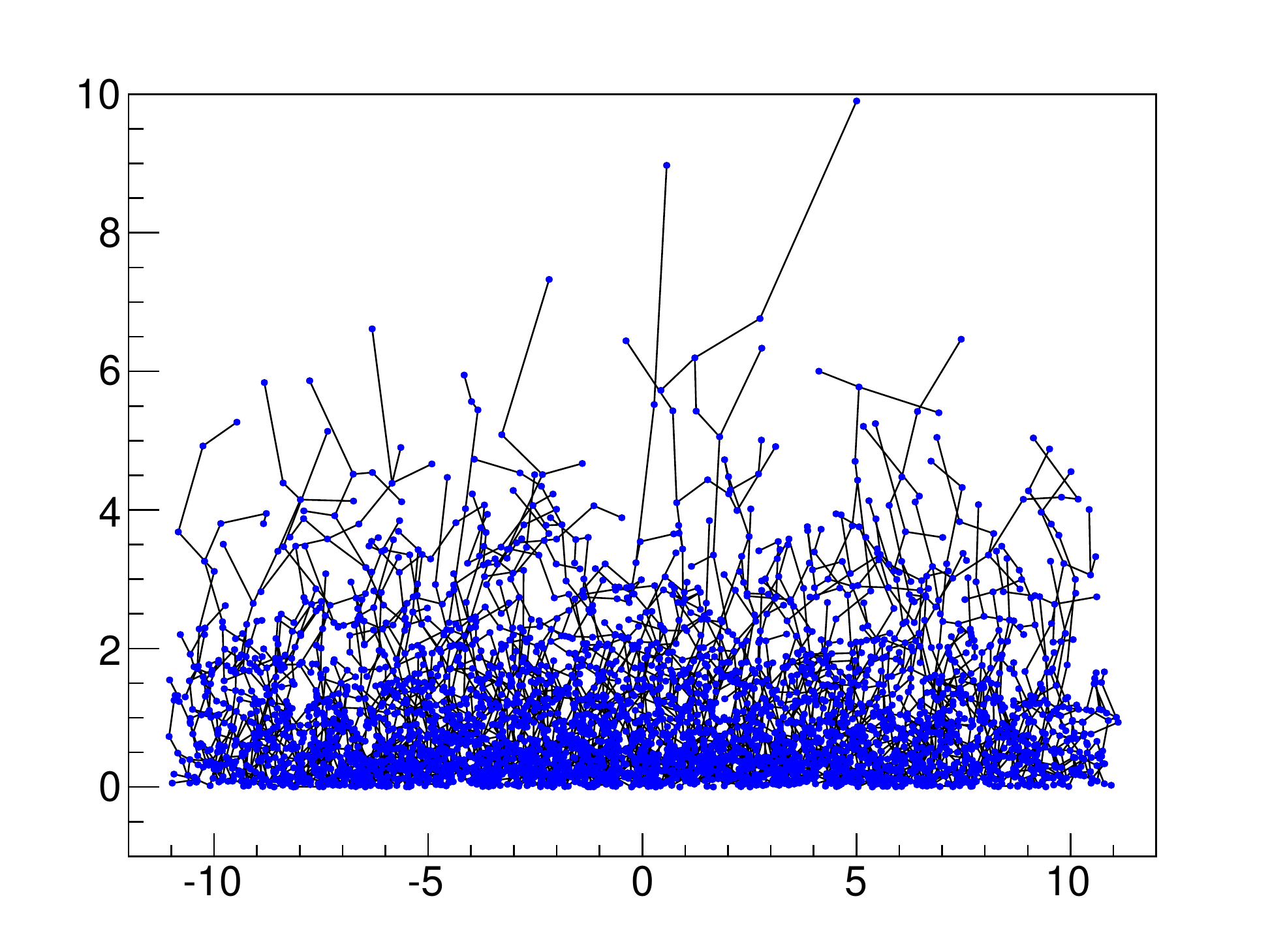}
\caption{Projections in $xz$ of the uniform tree (red) and the exponential tree (blue).}
\label{fig: ex4treesXZ3d}
\end{figure}

We then added a third variable and built the trees in all three dimensions.  For the first disc, we filled the third ($z$) dimension with vertices from a uniform distribution.  For the second, we used an exponential distribution.  The familiar $xy$ projections of these trees are shown in figure~\ref{fig: ex4trees3d}; they are essentially the same as figure~\ref{fig: ex4trees2d}.  Their $xz$ projections are shown in figure~\ref{fig: ex4treesXZ3d}.  The introduction of the new variable significantly alters the structure of each tree, making them qualitatively different even though  the $xy$ view in figure~\ref{fig: ex4trees3d} is essentially the same.  The difference becomes apparent, however, when the distributions of each of our statistical quantities are examined.

\begin{figure} \centering
\includegraphics[width=0.495\textwidth]{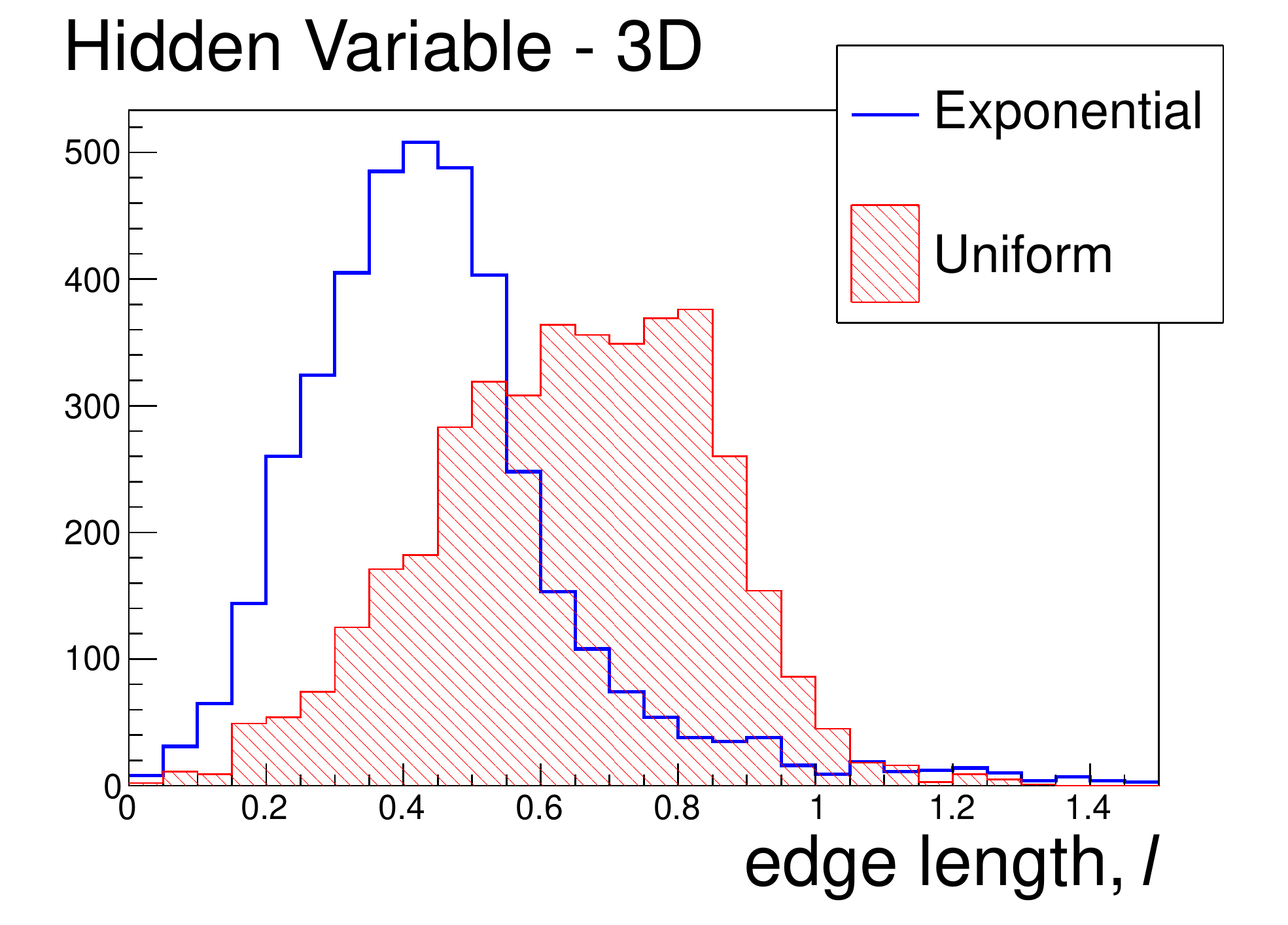}
\includegraphics[width=0.495\textwidth]{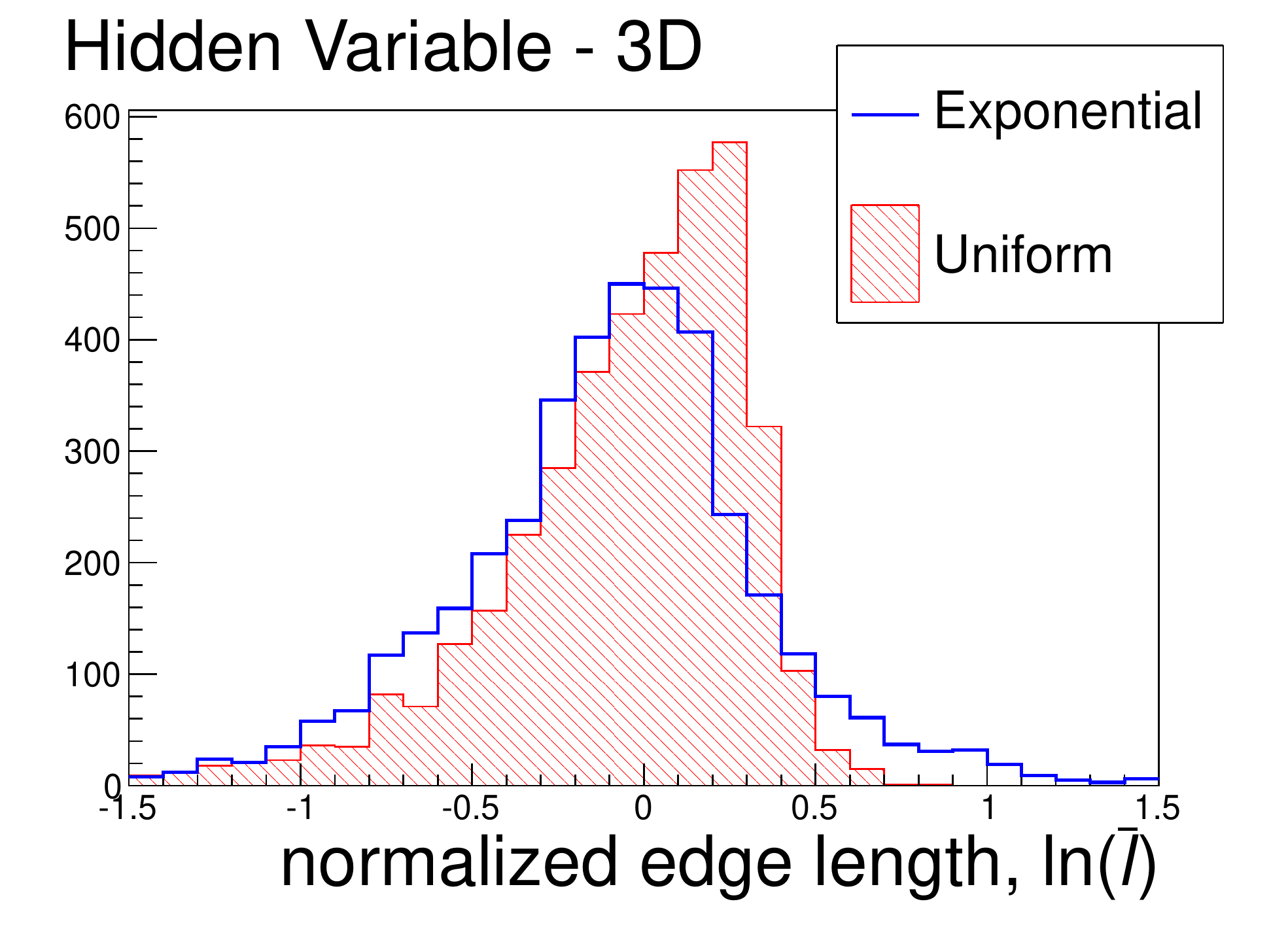}
\includegraphics[width=0.495\textwidth]{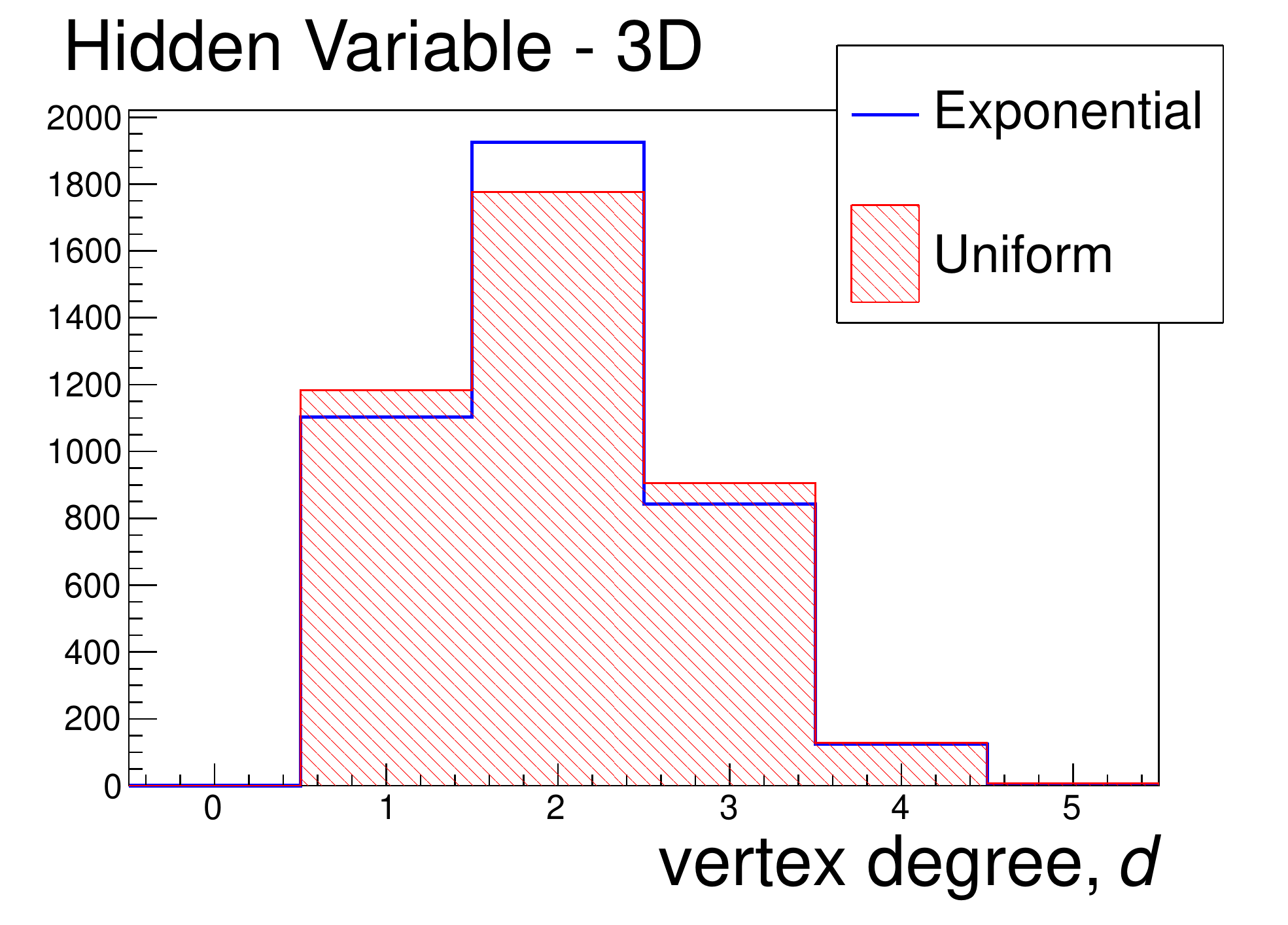}
\includegraphics[width=0.495\textwidth]{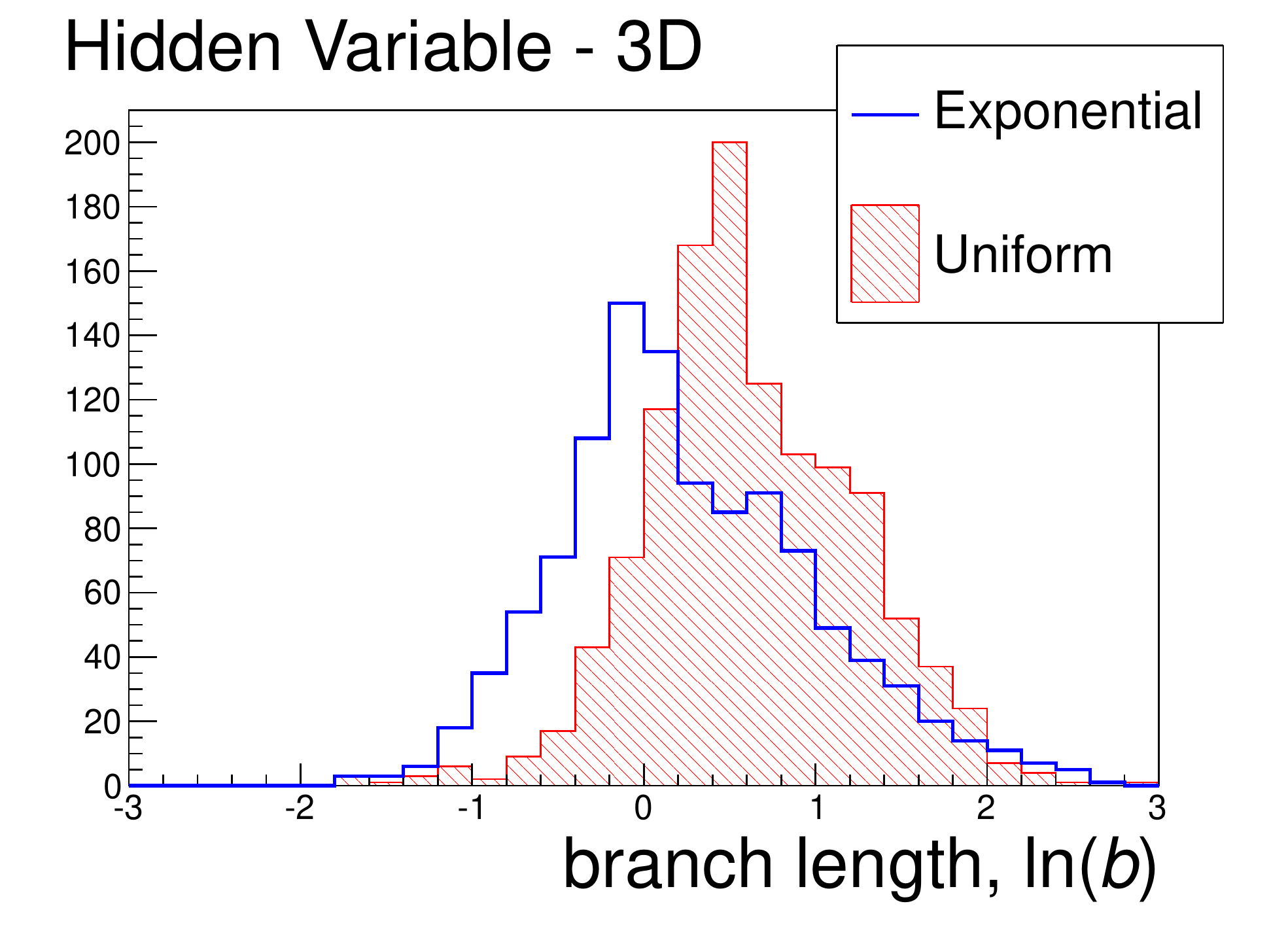}
\caption{The distributions of $l$ (top left), $\ln(\bar{l})$ (top right), $d$ (bottom left), and $\ln(b)$ (bottom right) for the uniform and exponential trees in Figs.~\ref{fig: ex4trees2d} and \ref{fig: ex4treesXZ3d}.}
\label{fig: ex4indiv3d}

\vskip 30pt

\includegraphics[width=0.495\textwidth]{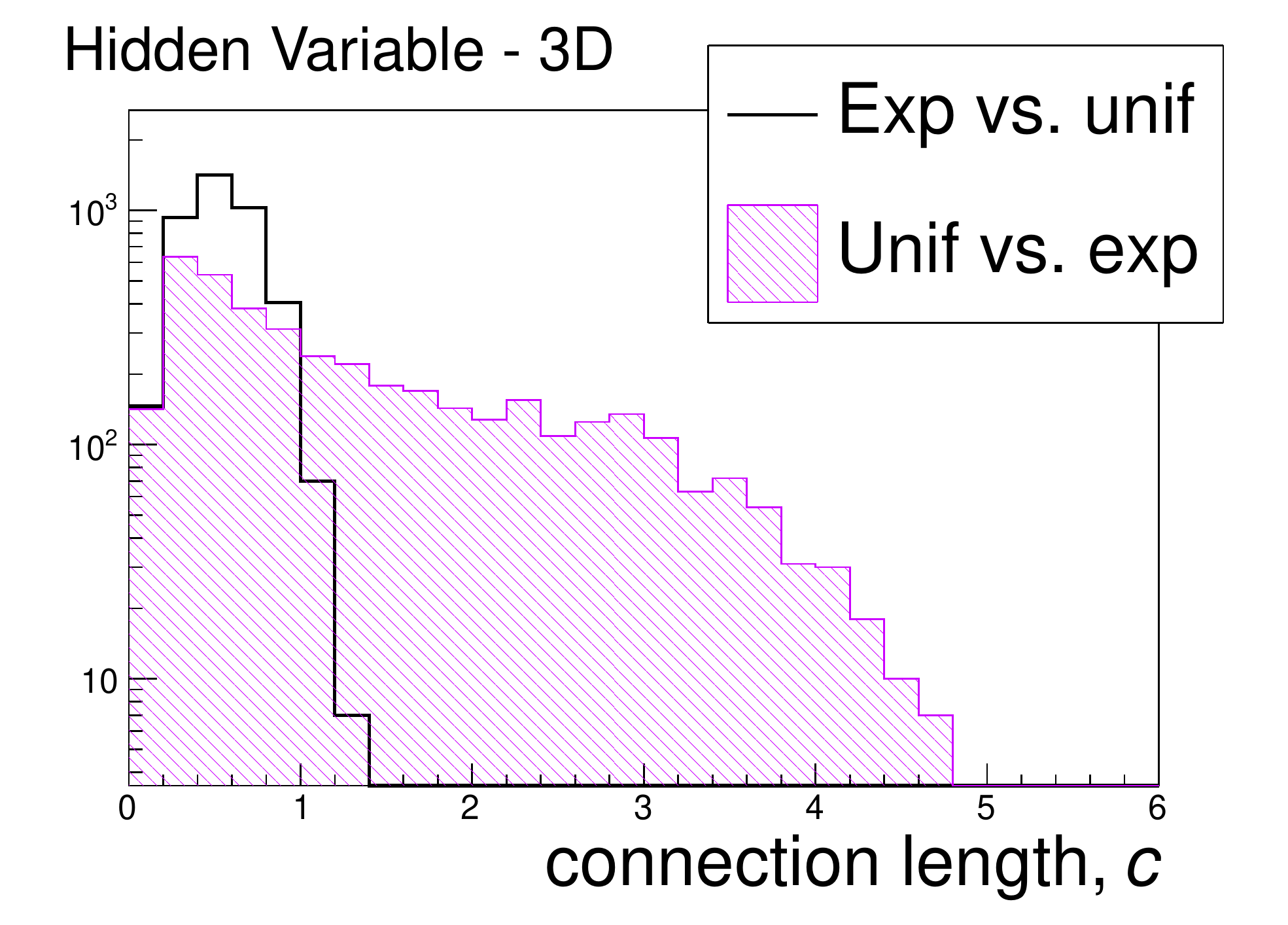}
\includegraphics[width=0.495\textwidth]{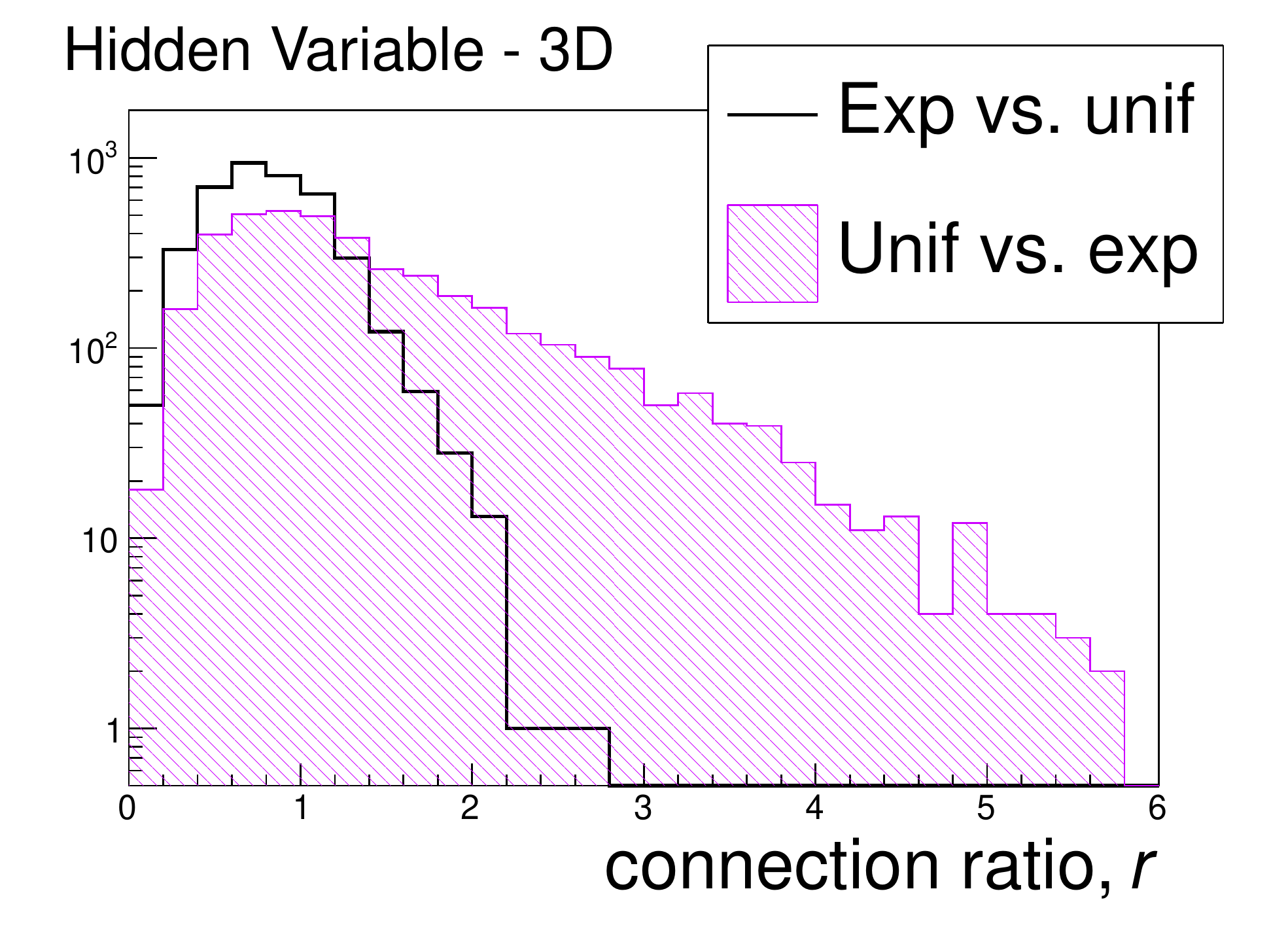}
\caption{The distributions of $c$ (left) and $r$ (right) for the uniform and exponential trees.}
\label{fig: ex4comp3d}
\end{figure}

The distributions of $l$ in figure~\ref{fig: ex4indiv3d} differ not only in mean but also in shape.  The long tails on the exponential tree distribution of $\ln(\bar{l})$ show the abundance of edges that are longer and shorter than the average.  The distributions of $d$ are very similar to those in figure~\ref{fig: ex4indiv2d}, but they may indicate slightly more of a filamentary structure in the exponential tree.  In spite of its more filamentary structure, the exponential tree in fact has shorter branches, as is evident in figure~\ref{fig: ex4indiv3d}.  The $c$ and $r$ distributions in figure~\ref{fig: ex4comp3d} show events present in the uniform tree that one would not expect to find in the exponential tree.

This fourth artificial example illustrates the promise of MST analyses: structure that may be obscure or unknown to the observer can be uncovered on a statistical basis.  While the pictures in figure~\ref{fig: ex4trees3d} show no differences, figure~\ref{fig: ex4indiv3d} and especially figure~\ref{fig: ex4comp3d} show major differences.

\section{Collider data}
In an attempt to provide a quasi-realistic example, we simulated basic LHC processes that lead to final states with two energetic, opposite-charge muons using the \Pythia\ event generator~\cite{Sjostrand}.  These processes are: direct Drell-Yan production, di-boson (WW, WZ, and ZZ) production, top quark pair production, and Z~$\rightarrow \tau^+ \tau^-$ followed by $\tau \rightarrow \mu\nu\nu$.  Such events are of general interest both for precision measurements and also for searches for new physics.

We defined two samples of MC events to compare.  The first sample (\DYonly) contains only Drell-Yan production of lepton pairs.  The other sample (\All) contains all the most important processes: Drell-Yan, top quark production, di-boson production, and Z~$\rightarrow \tau^+ \tau^-$.  These processes contributed to the second sample in accordance with their cross sections and the relevant branching ratios.  In order to approximate an event sample of interest, we required that the missing transverse momentum be greater than 50 GeV.\footnote{The so-called missing transverse momentum is the magnitude of the vector sum of the transverse momenta of all undetected particles.  For SM processes, neutrinos are the main contributors.  It is possible, however, that new particles such as dark matter particles could also contribute.}  Both \DYonly\ and \All\ samples contained nearly 6000 events each.  For the purposes of this study, we used the events to define vertices in the $(\MLL, \qT)$ plane, where $\MLL$ is the invariant mass and $\qT$ is the transverse momentum of the lepton pair.

\begin{figure} \centering
\includegraphics[width=0.495\textwidth]{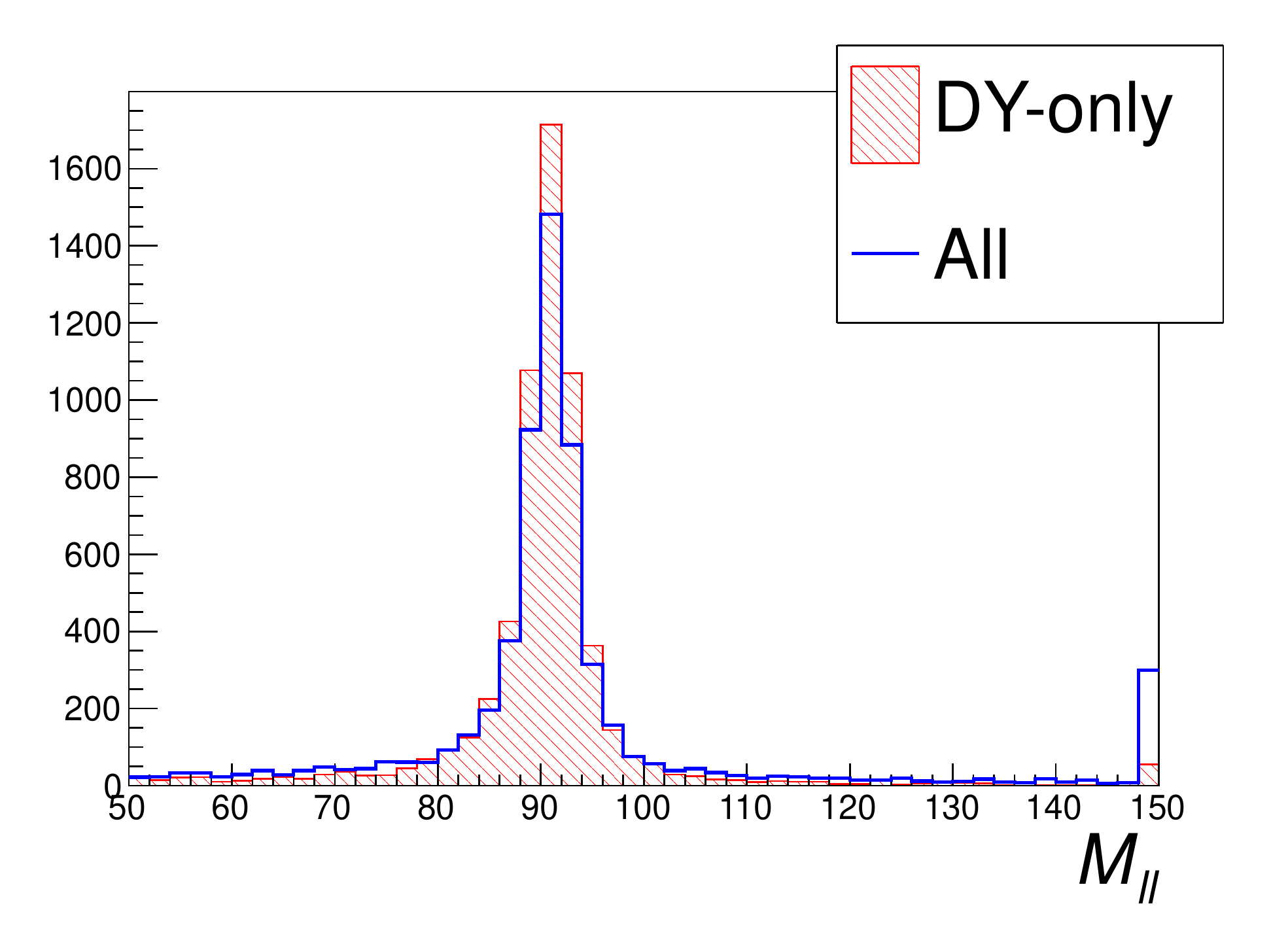}
\includegraphics[width=0.495\textwidth]{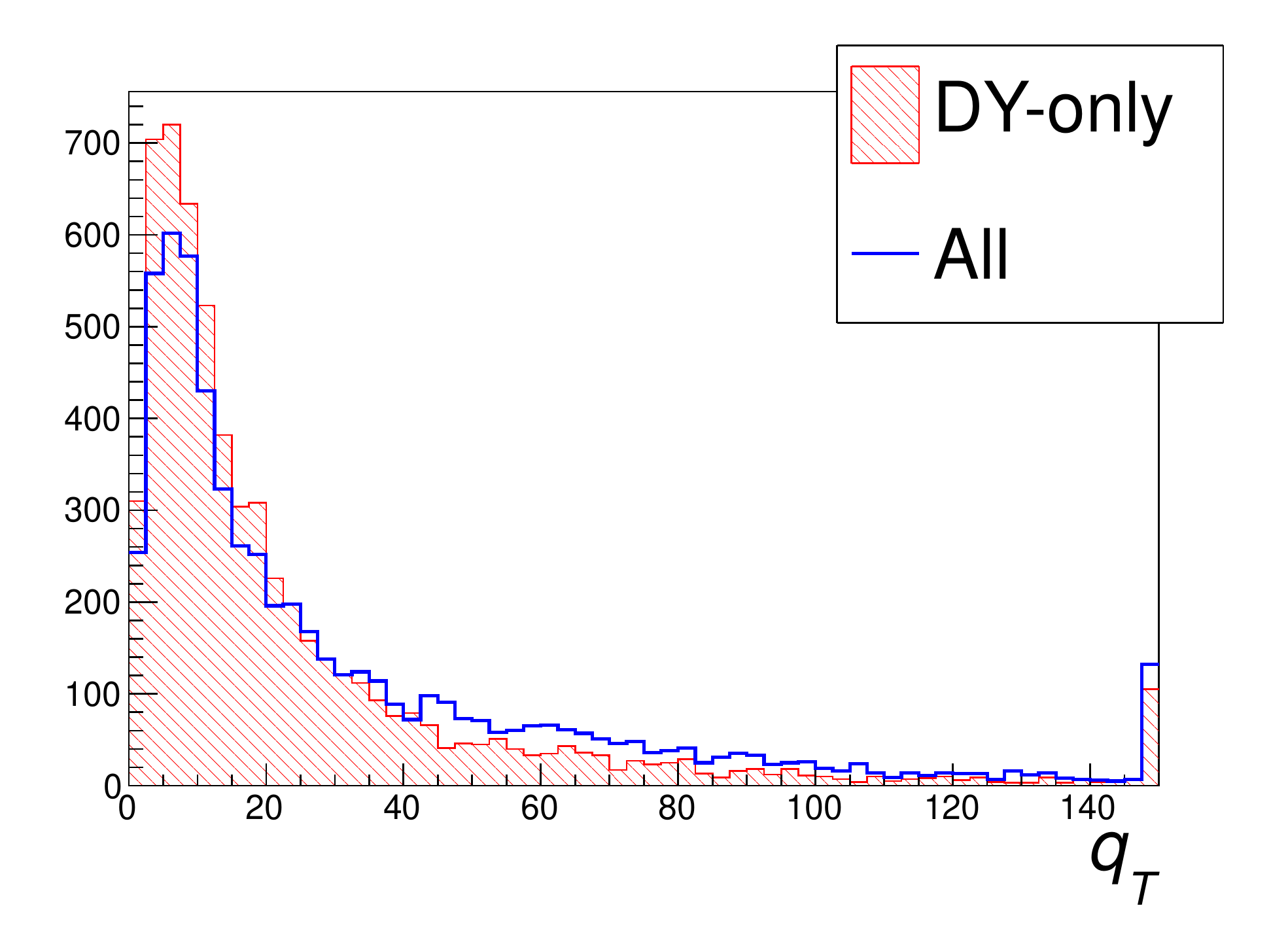}
\caption{The distributions of $\MLL$ (left) and $\qT$ (right) for the \DYonly\ and \All\ trees.  The last bin for each is an overflow bin.}
\label{fig: dyhists}
\end{figure}

\clearpage
The distributions of $\MLL$ and $\qT$ are displayed in figure~\ref{fig: dyhists}.  This is, essentially, the view of the data taken when differential cross sections $d\sigma / d\MLL$ and $d\sigma / d\qT$ are measured.  Since both the \DYonly\ and the \All\ histograms have the same number of entries, we can only draw conclusions about differences in shape.  Figure~\ref{fig: dyhists} shows that those differences are small but not insignificant.

\subsection{Complete plane analysis}
\begin{figure} \centering
\includegraphics[width=0.495\textwidth]{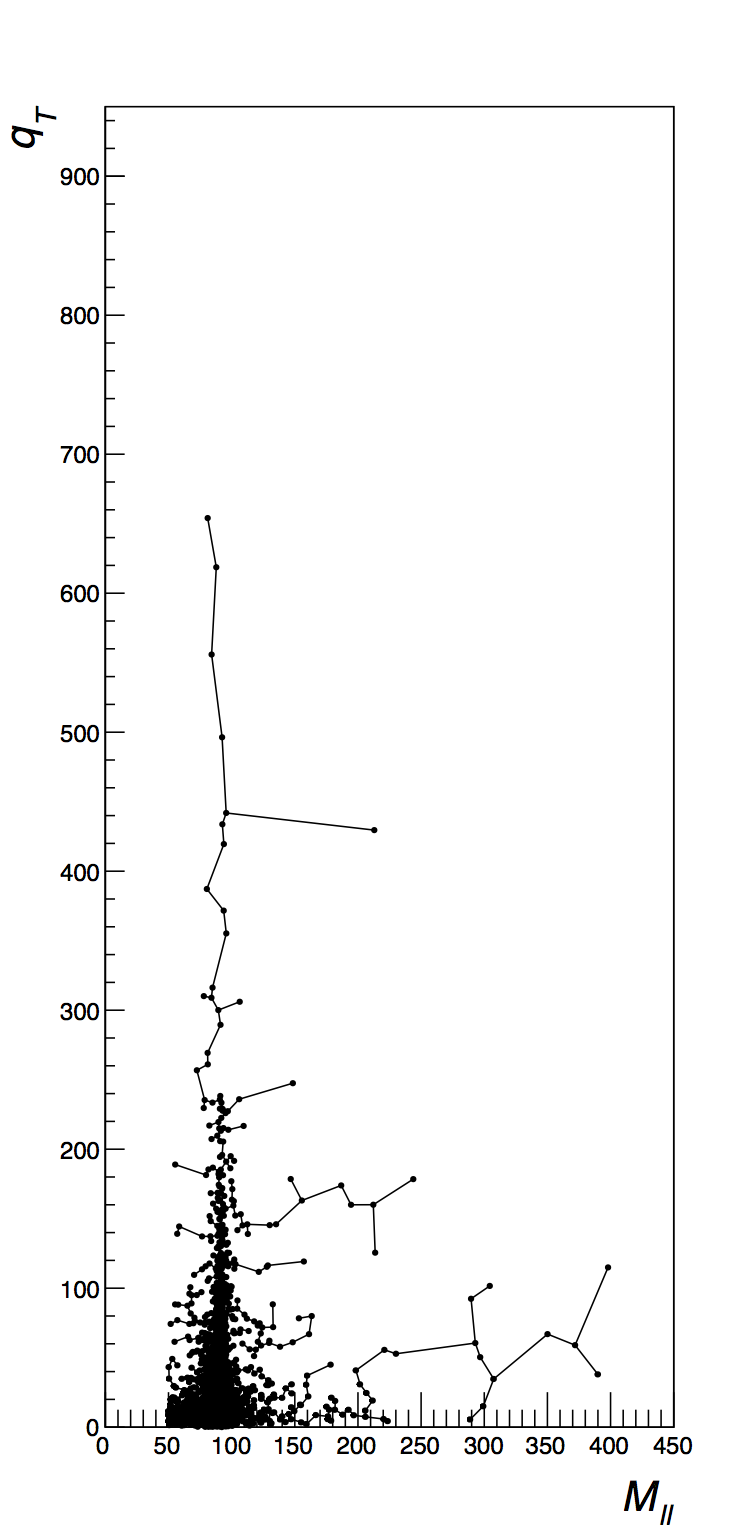}
\includegraphics[width=0.495\textwidth]{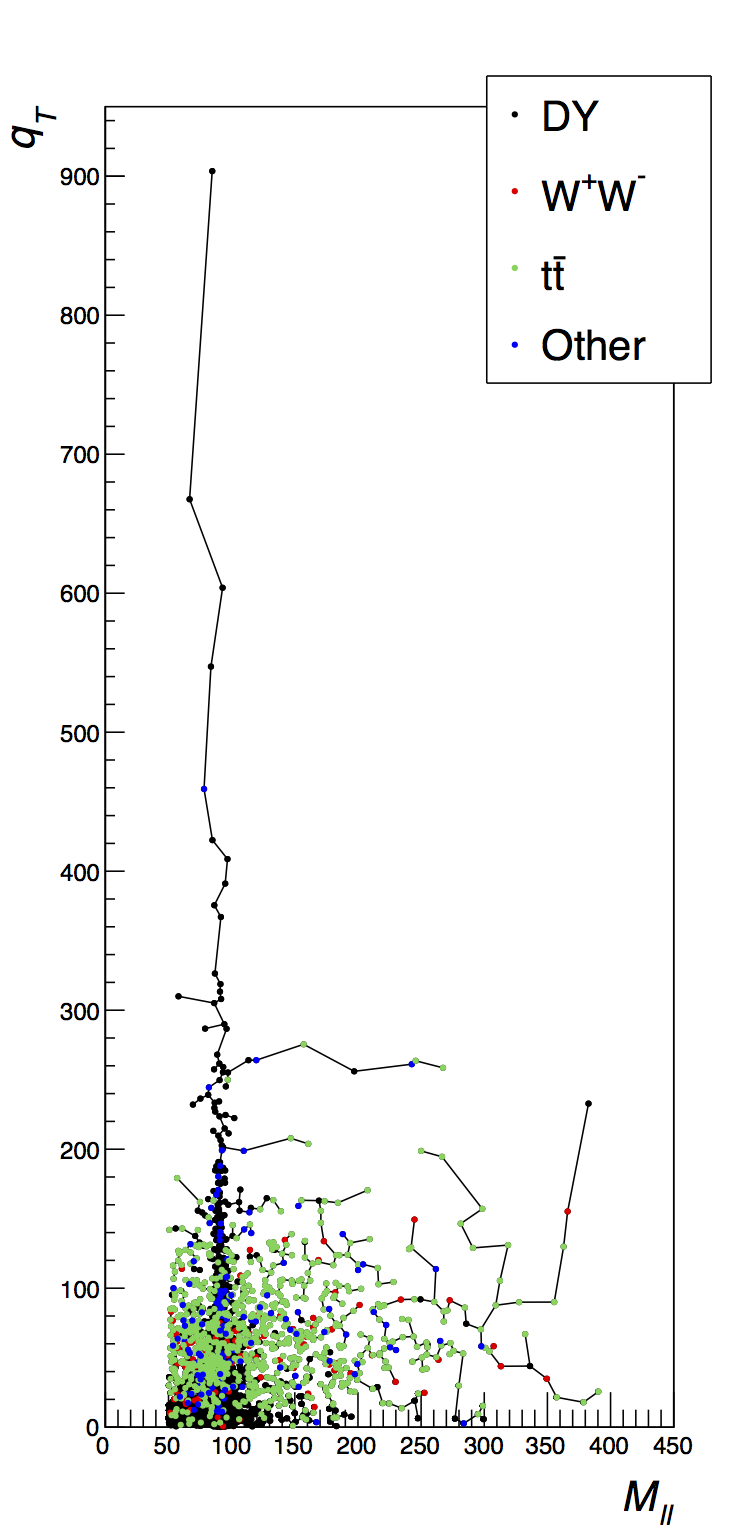}
\caption{The \DYonly\ (left) and \All\ (right) trees.  There are nearly 6000 events in each tree.  ``Other'' includes WZ, ZZ, and Z~$\rightarrow \tau^+ \tau^-$.}
\label{fig: dy1trees}
\end{figure}

Figure~\ref{fig: dy1trees} shows the MSTs for the two samples.  The \DYonly\ and \All\ trees in figure~\ref{fig: dy1trees} are visually different.  The vertices in figure~\ref{fig: dy1trees} (right) are colored according to the processes.  The tree is constructed ignoring this information.

\begin{figure} \centering
\includegraphics[width=0.495\textwidth]{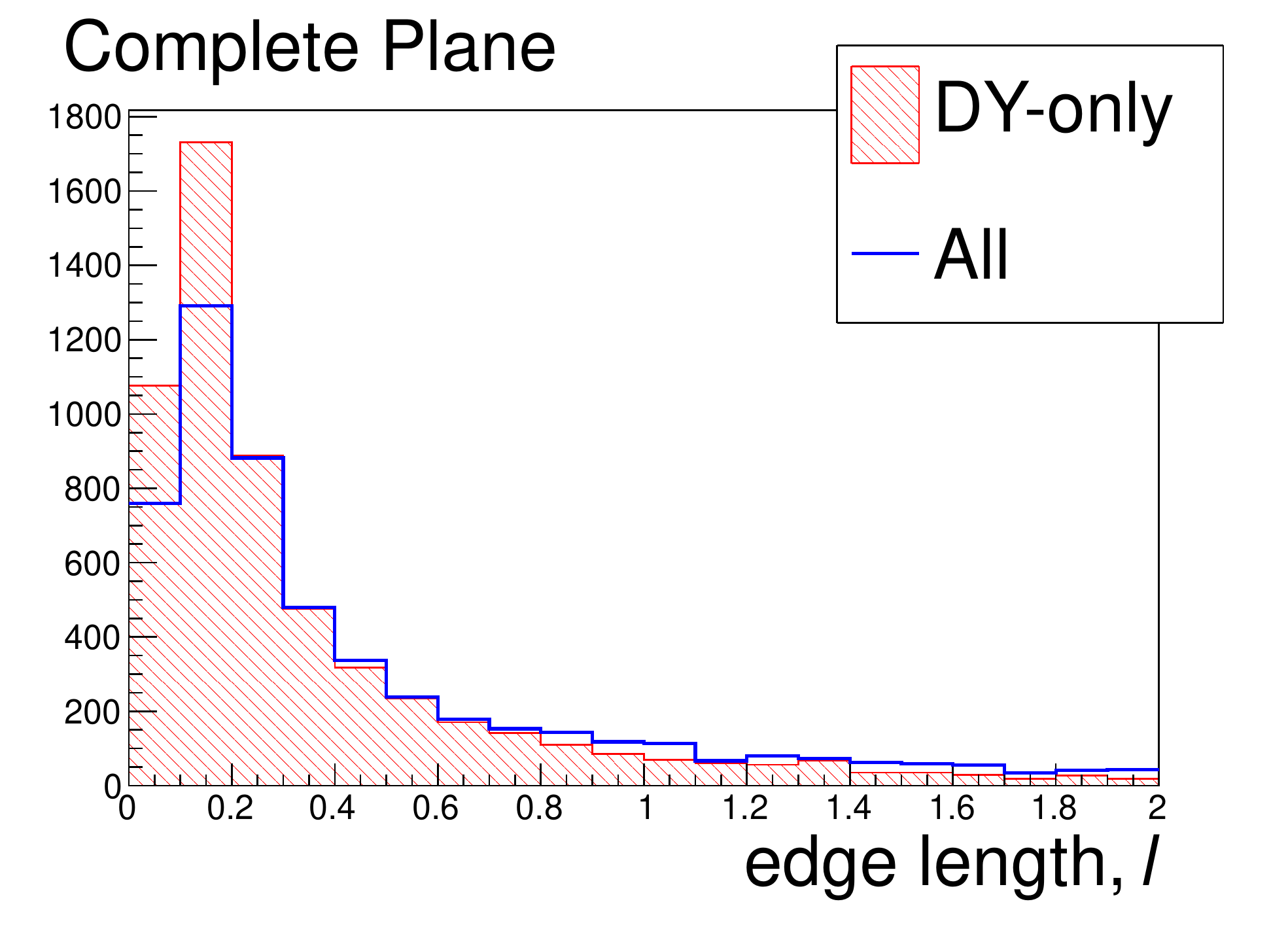}
\includegraphics[width=0.495\textwidth]{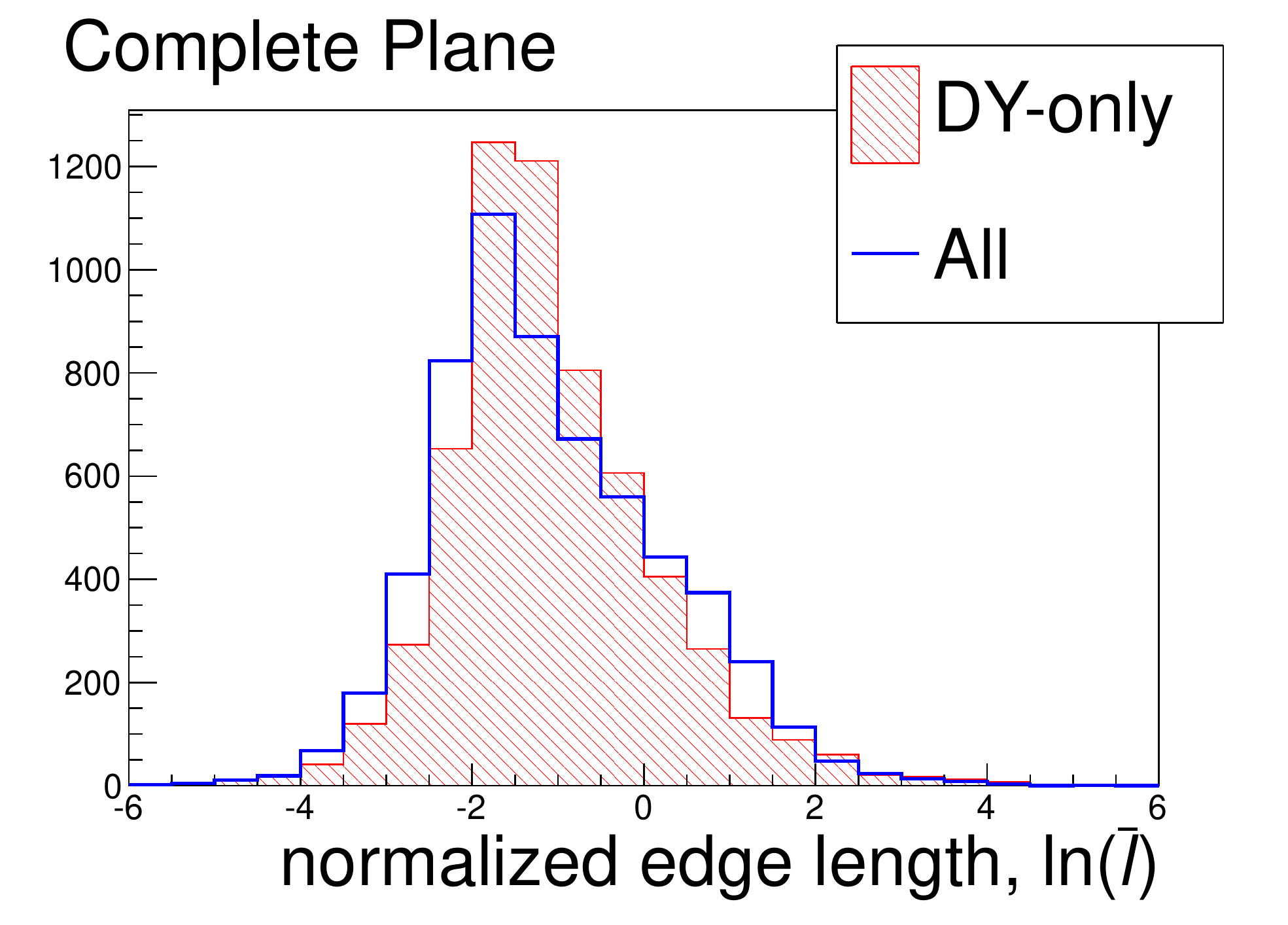}
\includegraphics[width=0.495\textwidth]{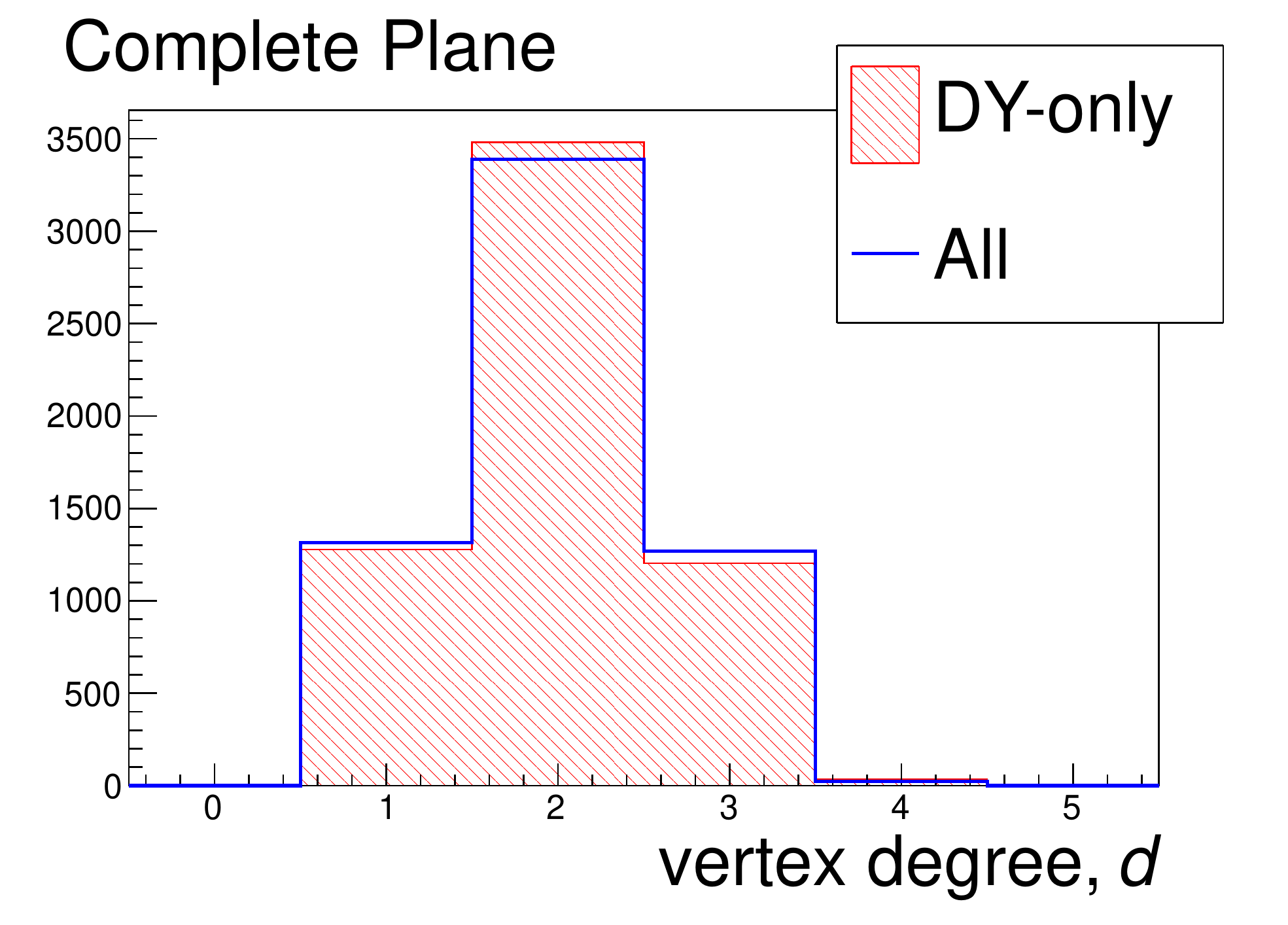}
\includegraphics[width=0.495\textwidth]{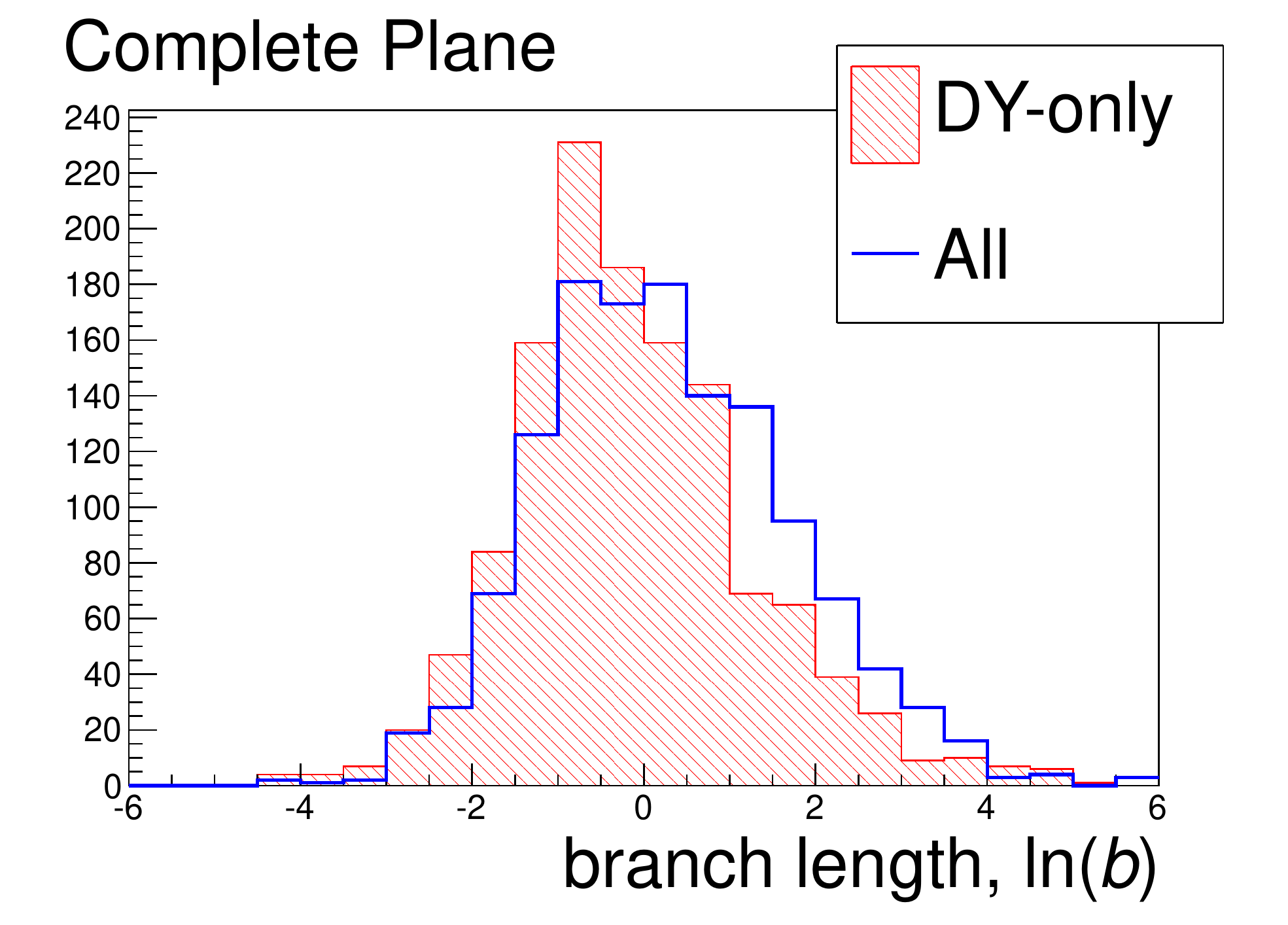}
\caption{The distributions of $l$ (top left), $\ln(\bar{l})$ (top right), $d$ (bottom left), and $\ln(b)$ (bottom right) for the \DYonly\ and \All\ trees in figure~\ref{fig: dy1trees}.}
\label{fig: dy1indiv}

\vskip 30pt

\includegraphics[width=0.495\textwidth]{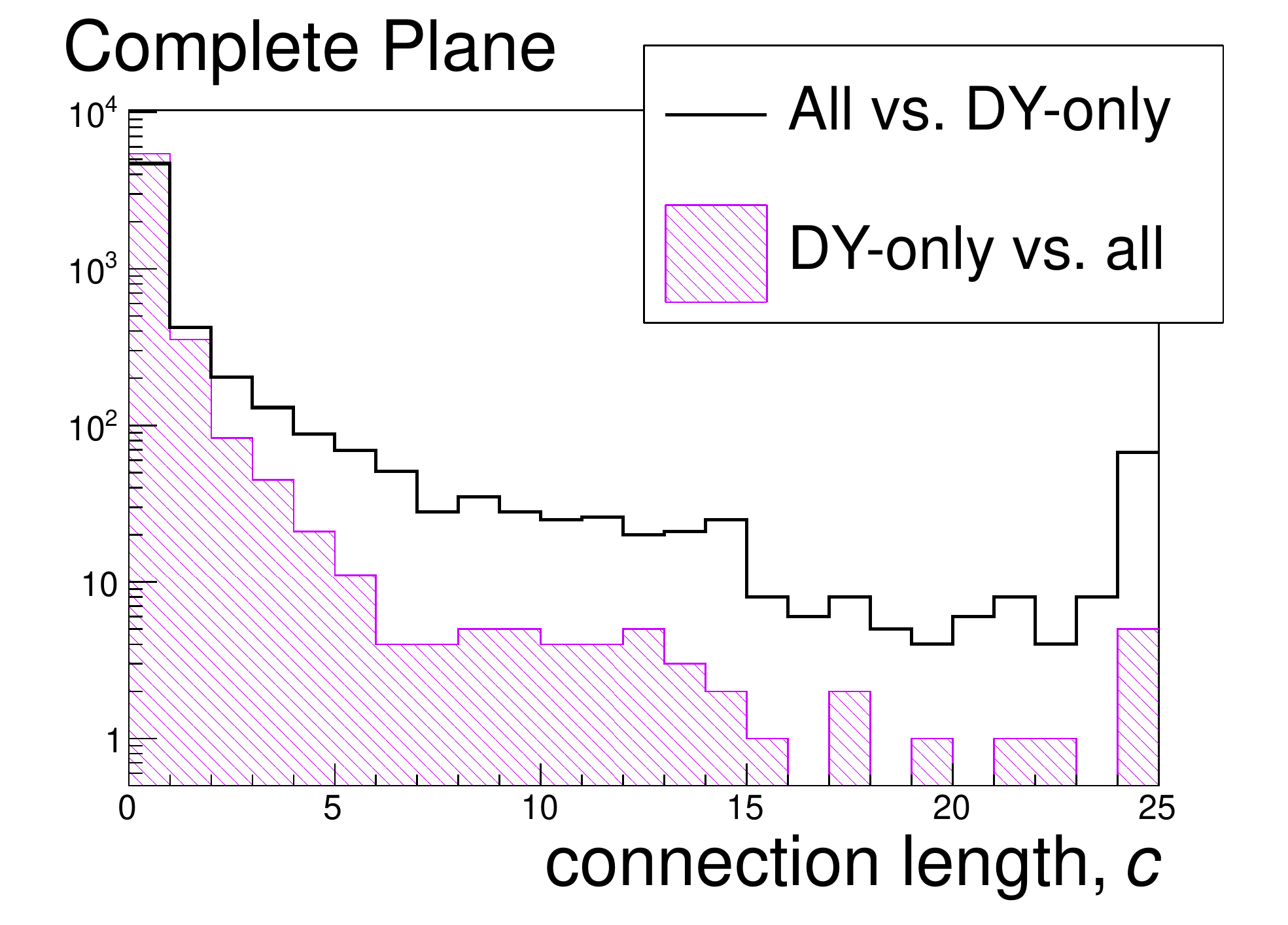}
\includegraphics[width=0.495\textwidth]{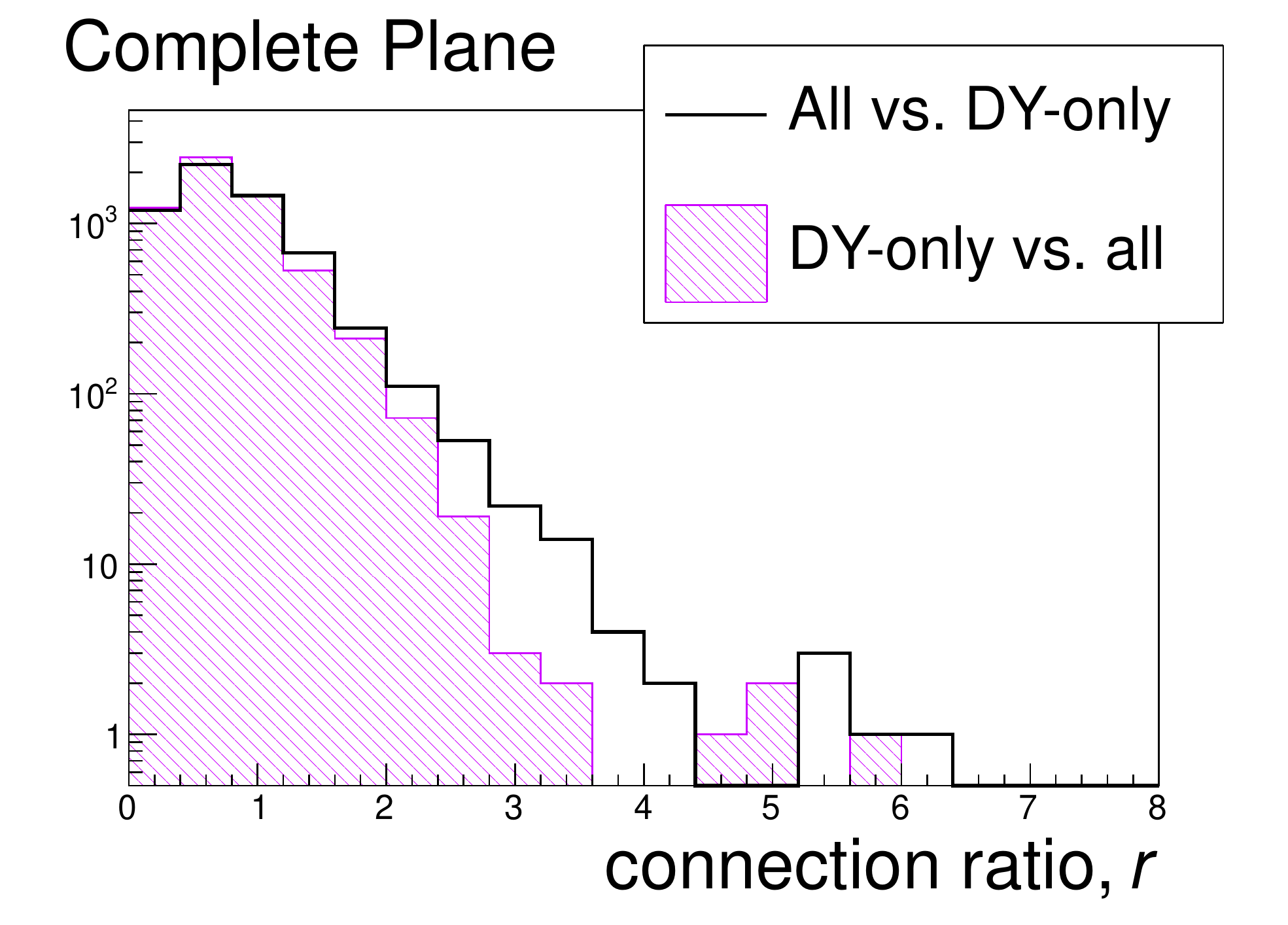}
\caption{(left): The distributions of $c$ for the \DYonly\ and \All\ trees.  The last bin is an overflow bin.  (right): The distributions of $r$ for the \DYonly\ and \All\ trees.}
\label{fig: dy1comp}
\end{figure}

In figure~\ref{fig: dy1indiv} we see that the distributions of $l$ and $\ln(\bar{l})$ are not as sharply peaked for the \All\ tree, reflecting its more sparsely distributed events.  The distributions of $d$ are indistinguishable for the two trees.  The \All\ tree has slightly longer branches on average, however, which are located in the region populated by non-DY events.  These non-DY events are even more apparent in figure~\ref{fig: dy1comp}, where the long tails of the $c$ and $r$ distributions for the \All\ tree are visible.

\subsection{Suppression of concentrated DY region}
The previous example shows some of the differences we expect, but we can magnify them using our physical insight to focus on the interesting areas of the $(\MLL, \qT)$ plane.  Using the same samples as before, we now apply weights such that events in the rectangle $50 < \MLL < 90,\ 0 < \qT < 100$~GeV have weight zero and all other events have weight one.  Effectively, the number of vertices is reduced from 6000 to 3739 for \DYonly\ and 3819 for \All.  Each edge is given a weight equal to the product of its vertices' weights.  This does not affect the structure of the trees, but reduces the number of events included in each histogram.  In particular, it censors a portion of the tree that physicists may consider uninteresting.

\begin{figure} \centering
\includegraphics[width=0.495\textwidth]{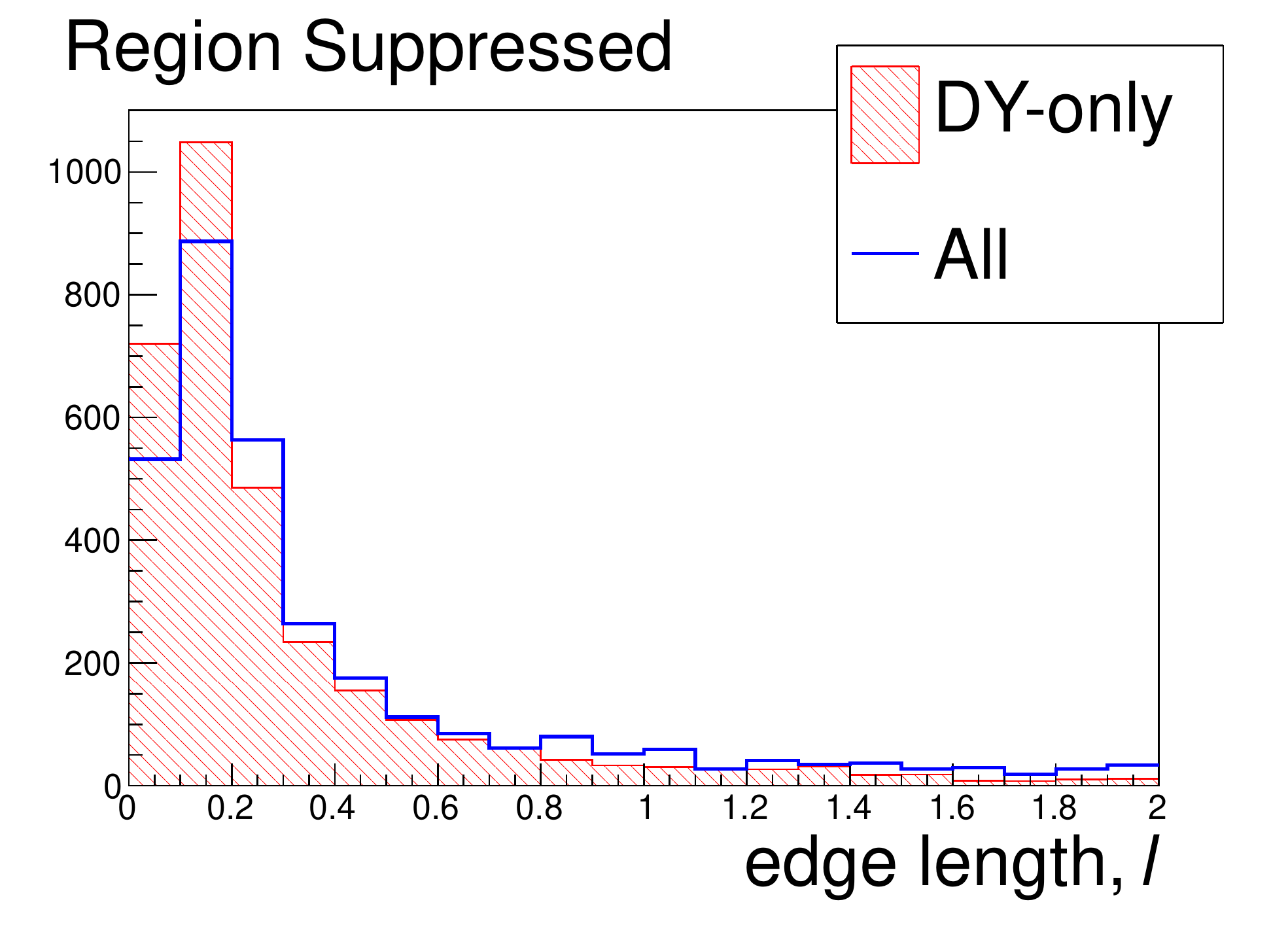}
\includegraphics[width=0.495\textwidth]{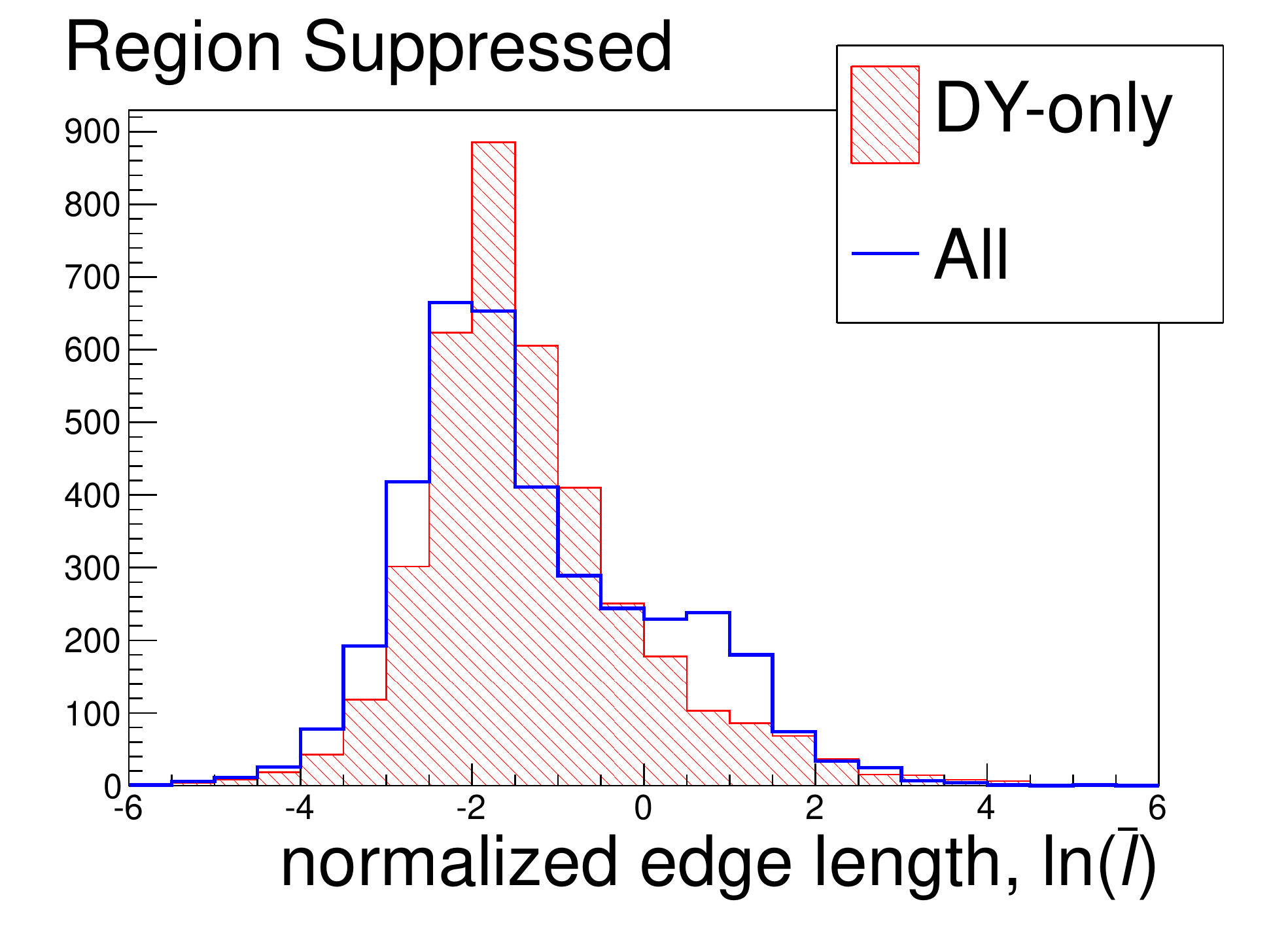}
\includegraphics[width=0.495\textwidth]{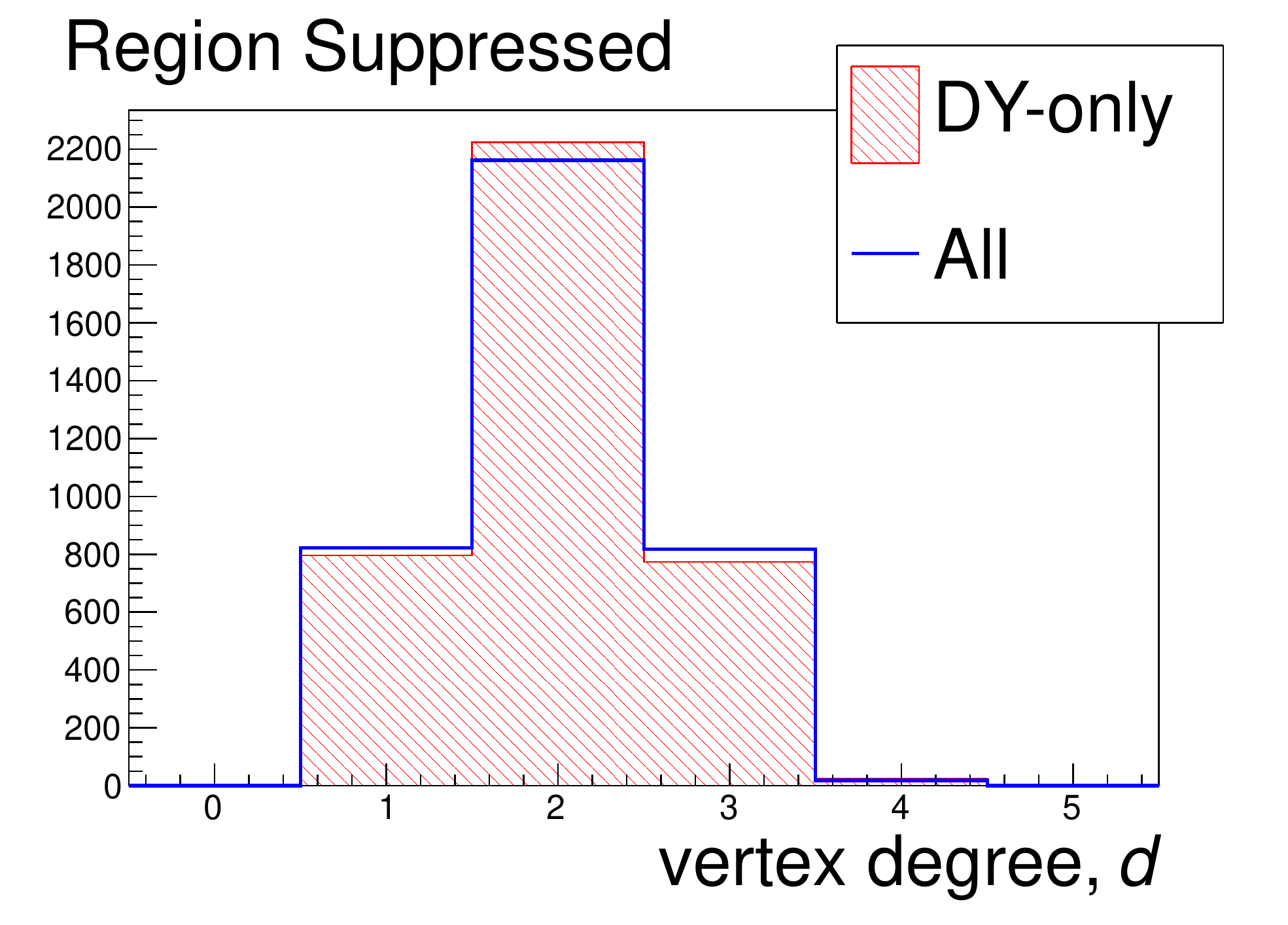}
\includegraphics[width=0.495\textwidth]{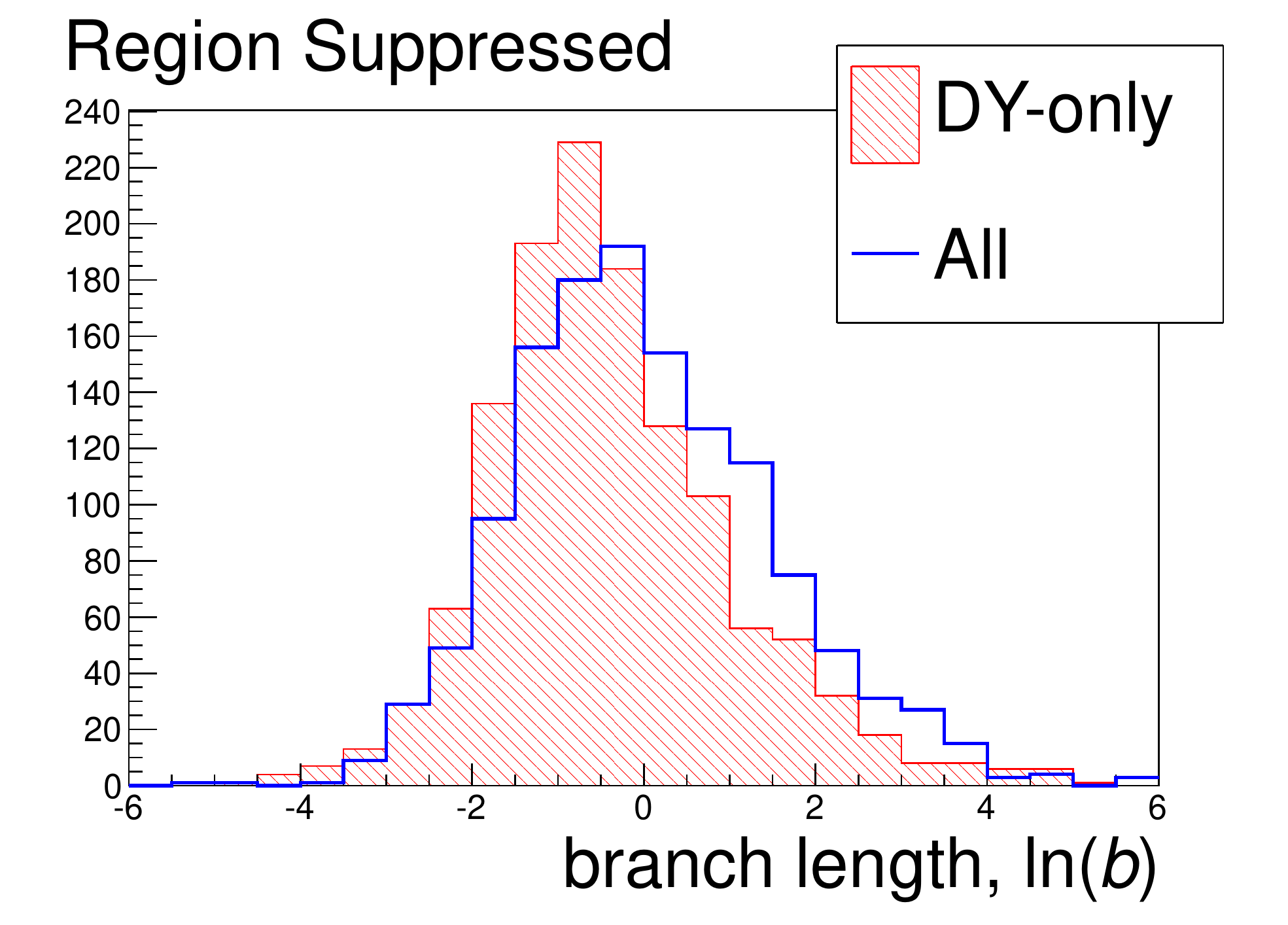}
\caption{The distributions of $l$ (top left), $\ln(\bar{l})$ (top right), $d$ (bottom left), and $\ln(b)$ (bottom right) for the \DYonly\ and \All\ trees with events in the region $50 < \MLL < 90,\ 0 < \qT < 100$~GeV suppressed.}
\label{fig: dy2indiv}

\vskip 20pt

\includegraphics[width=0.495\textwidth]{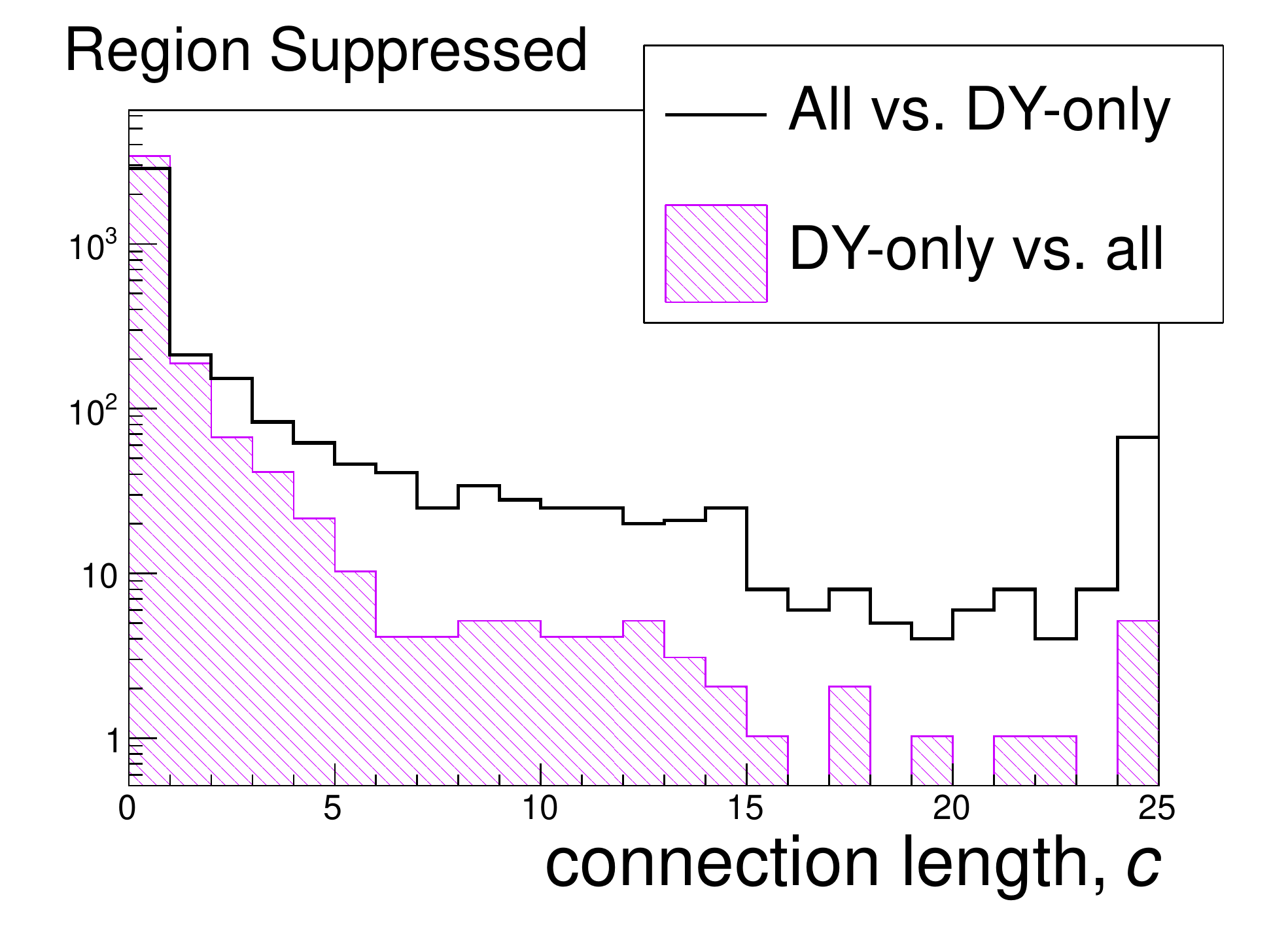}
\includegraphics[width=0.495\textwidth]{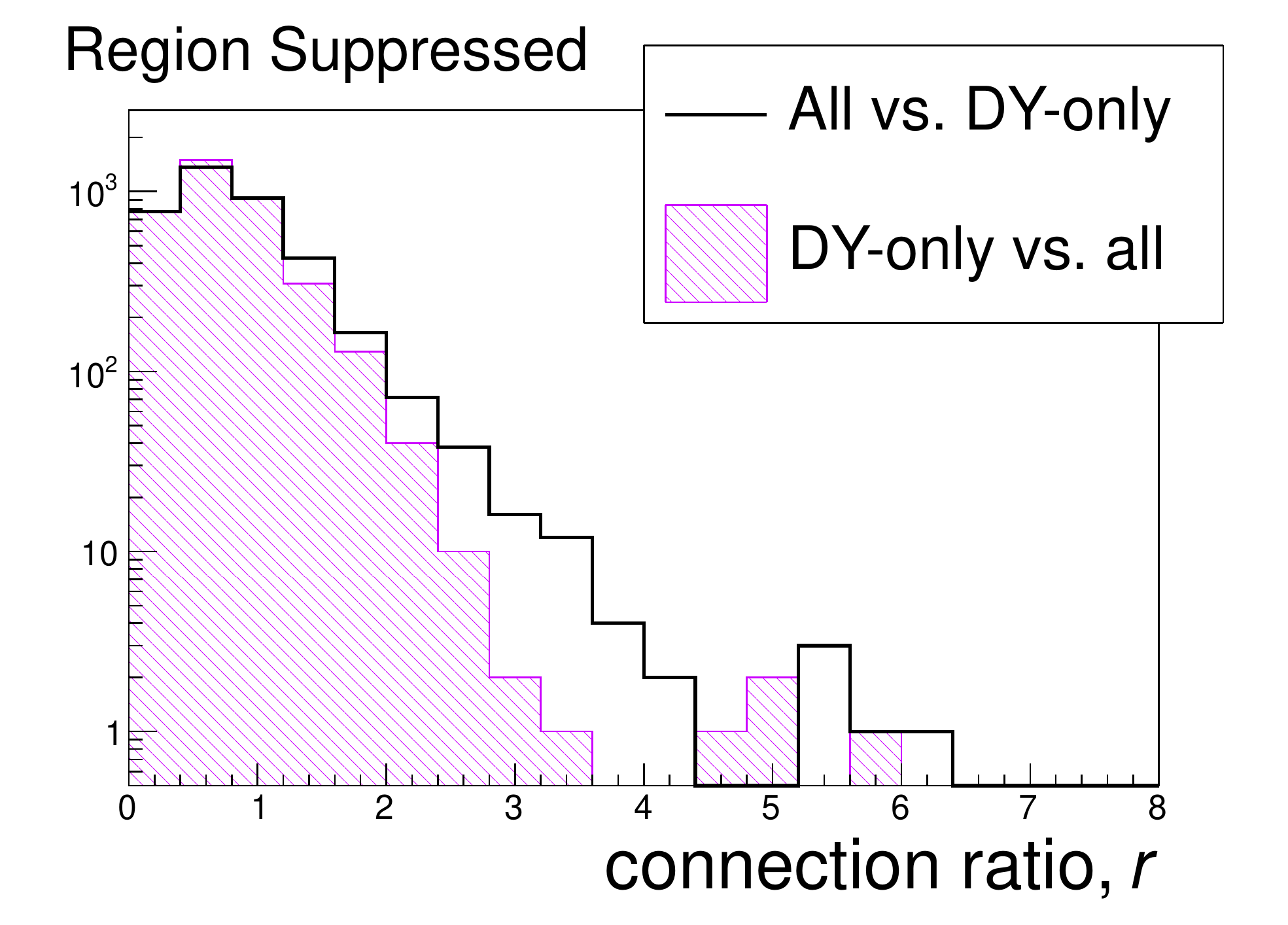}
\caption{(left): The distributions of $c$ for the \DYonly\ and \All\ trees with events in the region $50 < \MLL < 90,\ 0 < \qT < 100$~GeV suppressed.  The last bin is an overflow bin.  (right): The distributions of $r$ for the same trees.}
\label{fig: dy2comp}
\end{figure}

For the $l$ and $\ln(\bar{l})$ distributions in figure~\ref{fig: dy2indiv}, the size of the \DYonly\ histogram has been normalized to the size of the \All\ histogram so that we can focus on differences in shape.  For $\ln(\bar{l})$, the \All\ histogram has grown a shoulder for $\ln(\bar{l}) \approx 0.5$.  The \DYonly\ histogram for the $d$ distribution has also been normalized to the size of the corresponding \All\ histogram.  For $\ln(b)$, however, the \DYonly\ histogram was normalized using the factor from the $\ln(\bar{l})$ distribution because it is normal for trees of comparable size to have a different number of branches, though they must have the same number of edges.  The \All\ $\ln(b)$ distribution has a gentler peak than it did before the weights were applied.  Interestingly, after the weighting, the distributions of $c$ and $r$ in figure~\ref{fig: dy2comp}  are nearly identical in shape to those in figure~\ref{fig: dy1comp}.

\subsection{Suppression of DY peak}
We can try to further increase the differences between the two samples by focusing on the high-mass region only.  To do so, we give weight zero to all events with $\MLL < 100$~GeV.

\begin{figure} \centering
\includegraphics[width=0.495\textwidth]{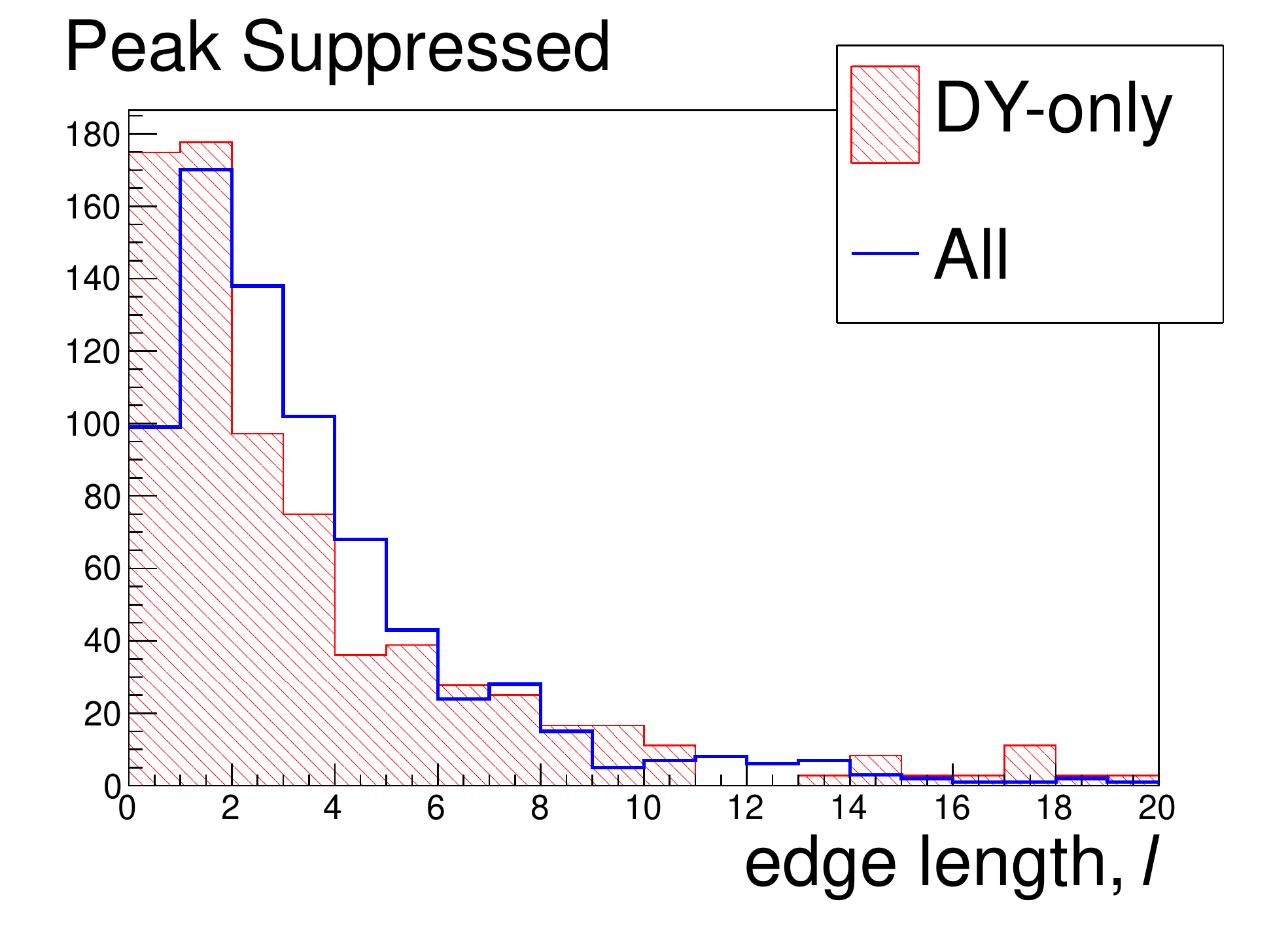}
\includegraphics[width=0.495\textwidth]{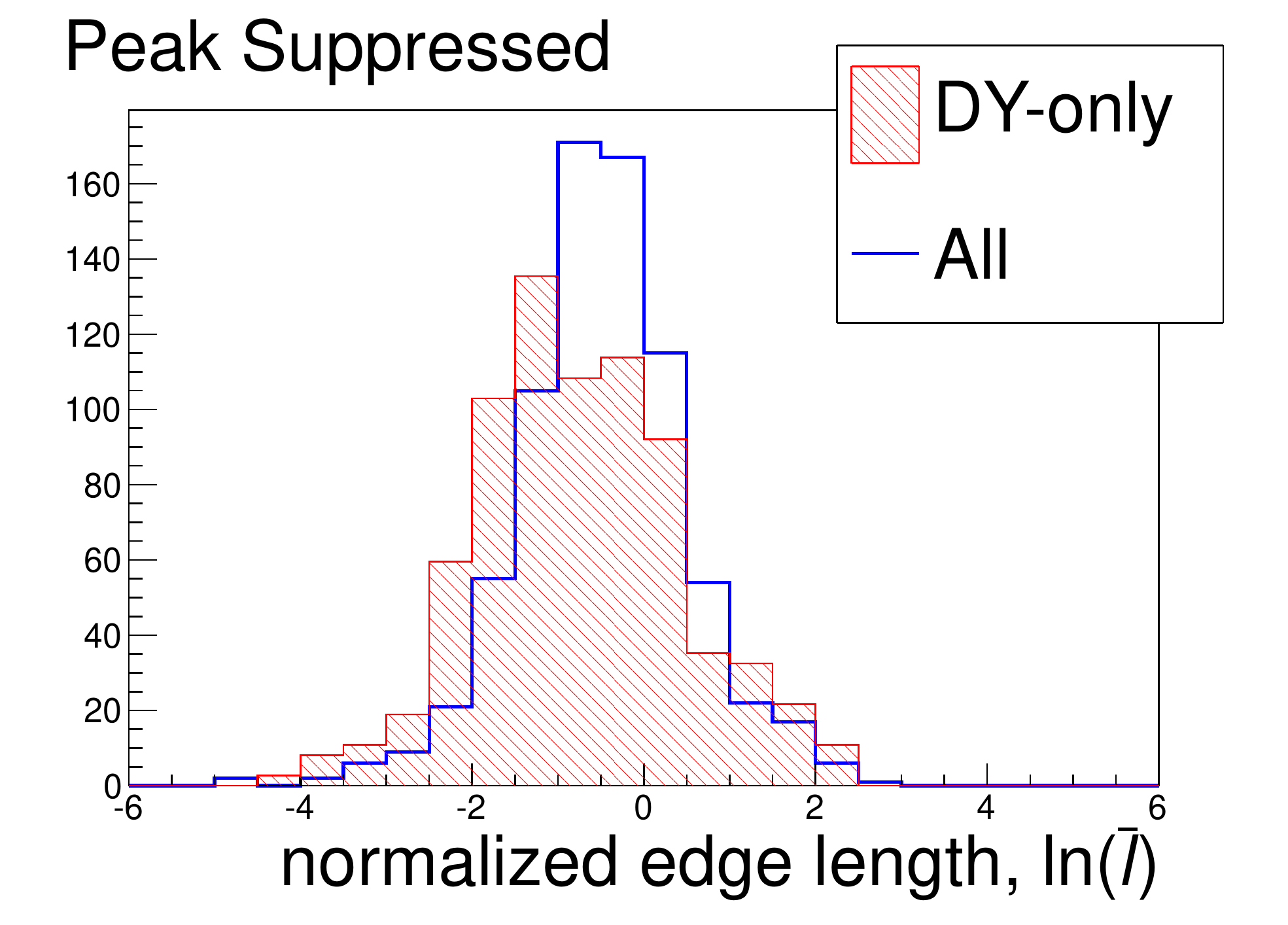}
\includegraphics[width=0.495\textwidth]{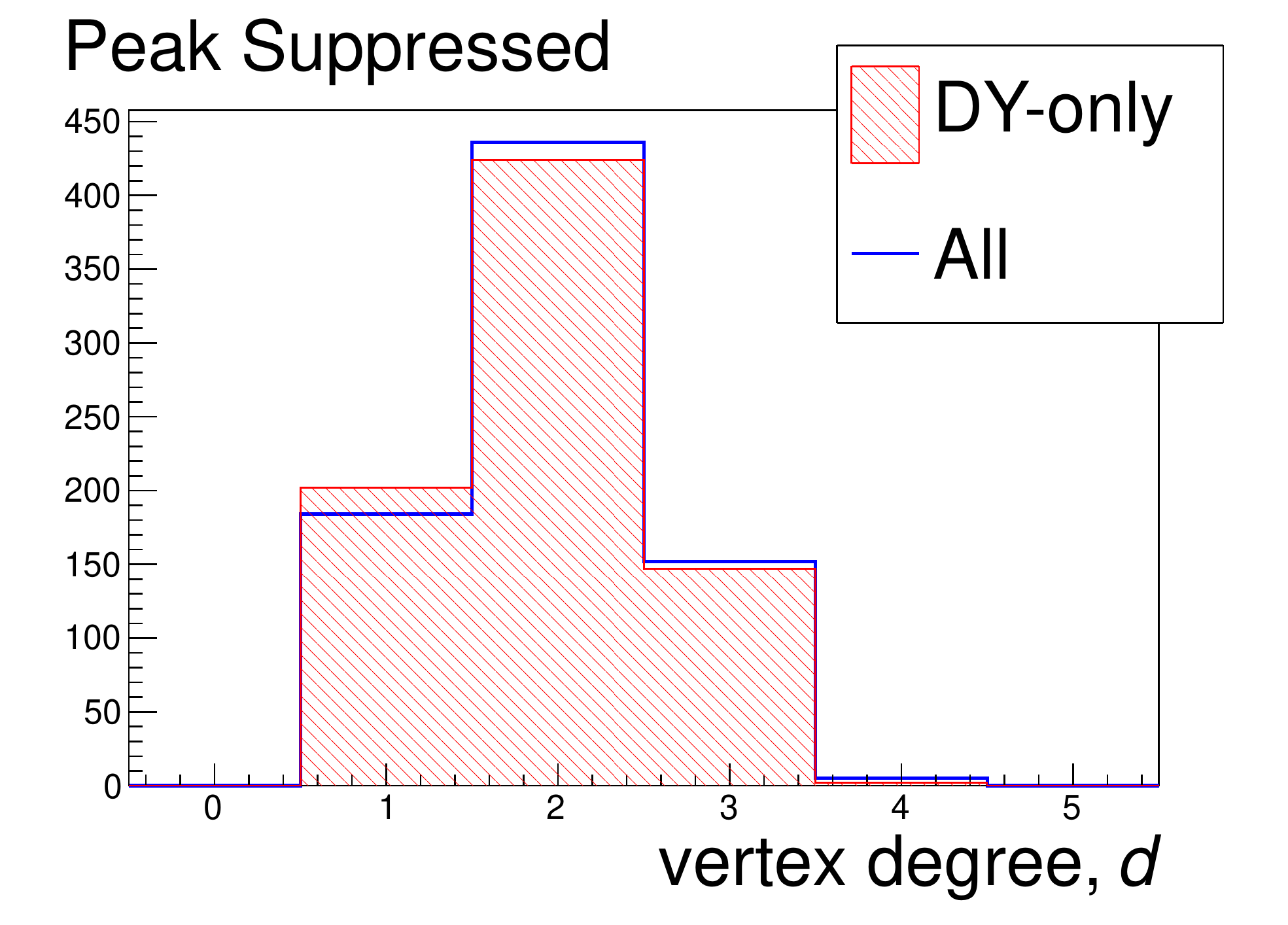}
\includegraphics[width=0.495\textwidth]{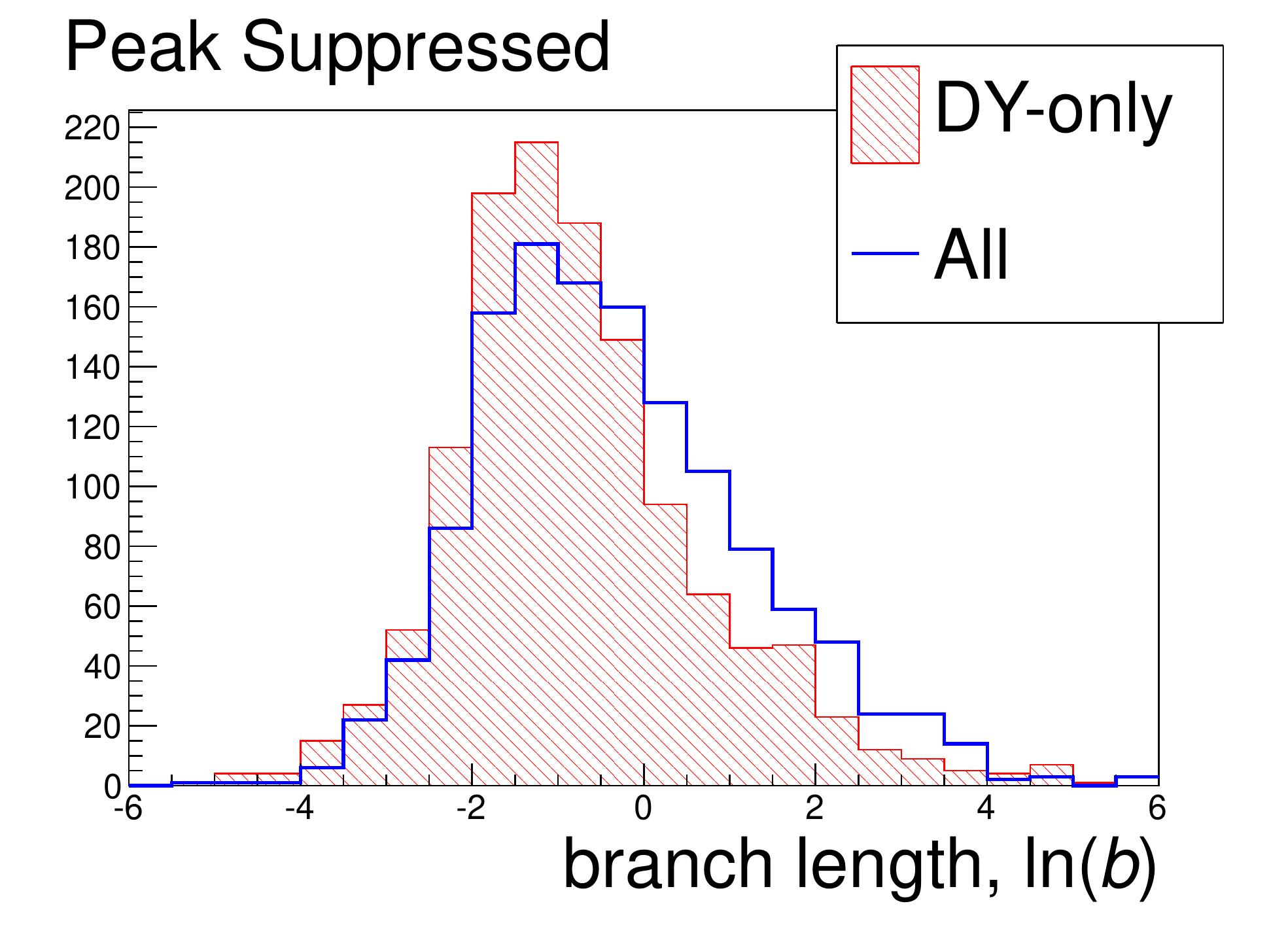}
\caption{The distributions of $l$ (top left), $\ln(\bar{l})$ (top right), $d$ (bottom left), and $\ln(b)$ (bottom right) for the \DYonly\ and \All\ trees with events in the region $\MLL < 100$~GeV suppressed.}
\label{fig: dy3indiv}

\vskip 30pt

\includegraphics[width=0.495\textwidth]{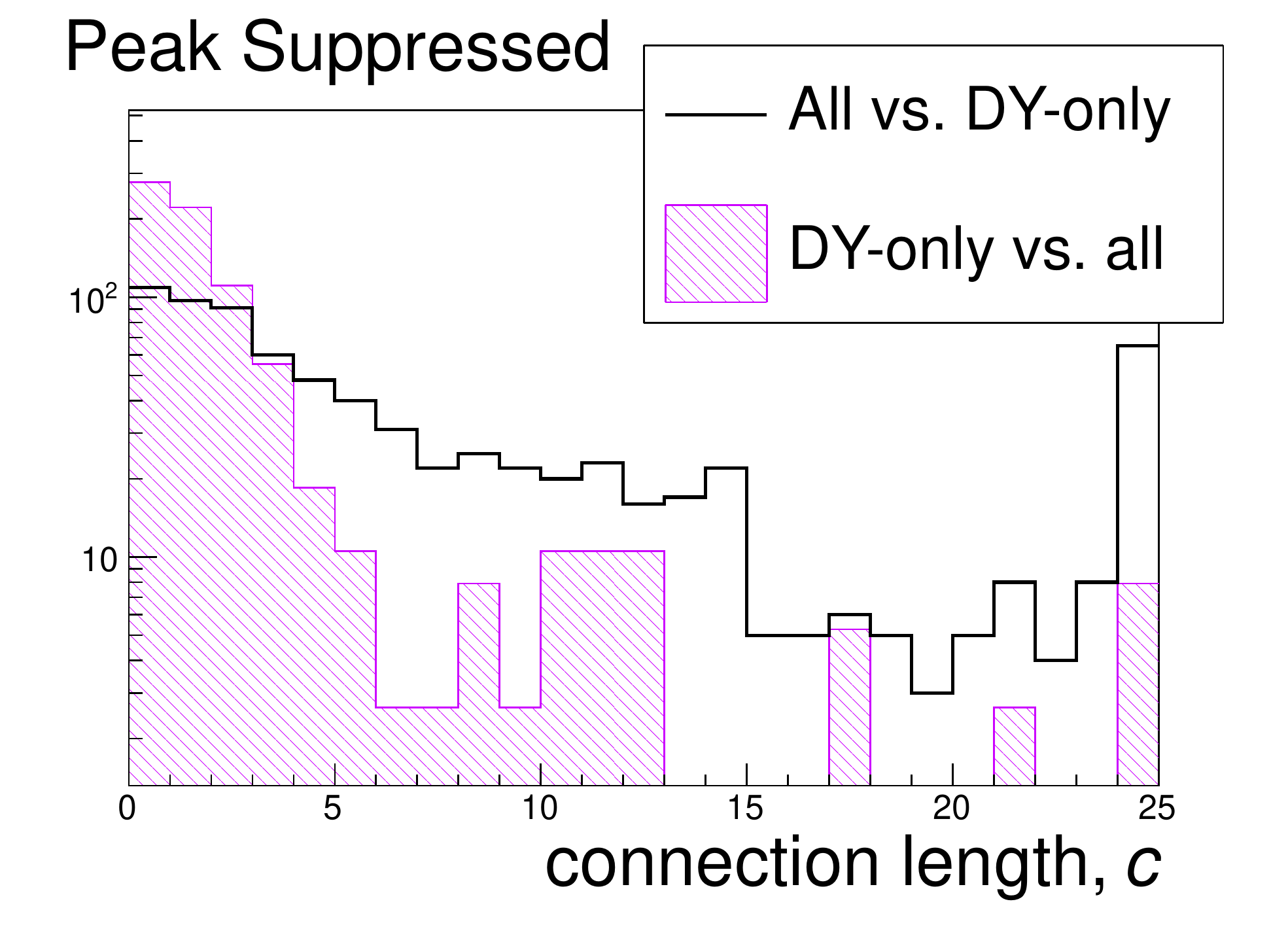}
\includegraphics[width=0.495\textwidth]{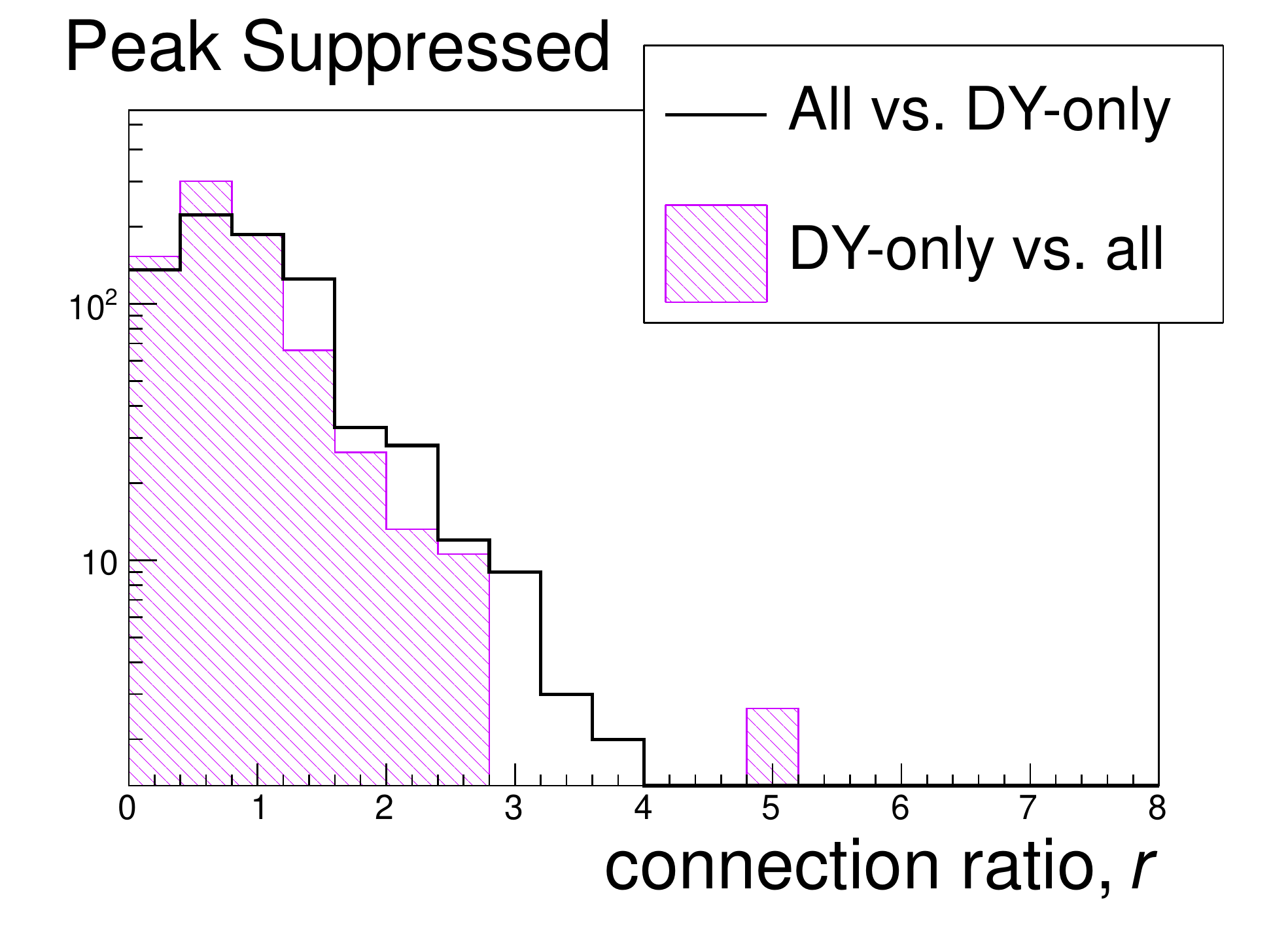}
\caption{(left): The distributions of $c$ for the \DYonly\ and \All\ trees with events in the region $\MLL < 100$~GeV suppressed.  The last bin is an overflow bin.  (right): The distributions of $r$ for the same trees.}
\label{fig: dy3comp}
\end{figure}

This weighting scheme removes most of the short edges from both trees, so both $l$ distributions are much wider as seen in figure~\ref{fig: dy3comp}.  The removal of so many DY events broadens the \DYonly\ $\ln(\bar{l})$ distribution, and the \All\ distribution moves upward.  We see that the weighted \DYonly\ tree has shorter edges on average, as expected for a tree with more concentrated vertices.  The difference in the $d$ distributions is modest, but they suggest a more filamentary structure in the \DYonly\ tree.  The slightly broader distribution of $\ln(b)$ for the \All\ tree shows its longer branches.  Finally, there is a dramatic difference in the shapes of the histograms of $c$ in figure~\ref{fig: dy3comp}, indicating the presence of many non-DY events in the \All\ sample.  The $r$ distribution also shows these events, although not as dramatically.

\subsection{Incorporating MST information in collider data analysis}
The MST information explored here can be used to improve the separation of events from different processes.  As an example, we consider the following task: given the kinematic region with two muons with $\MLL > 100$~GeV as in the previous subsection, can we estimate the fraction of events coming from $\ttbar$ production?

\begin{figure} \centering
\includegraphics[width=0.48\textwidth, trim={0 1cm 0 1cm}, clip]{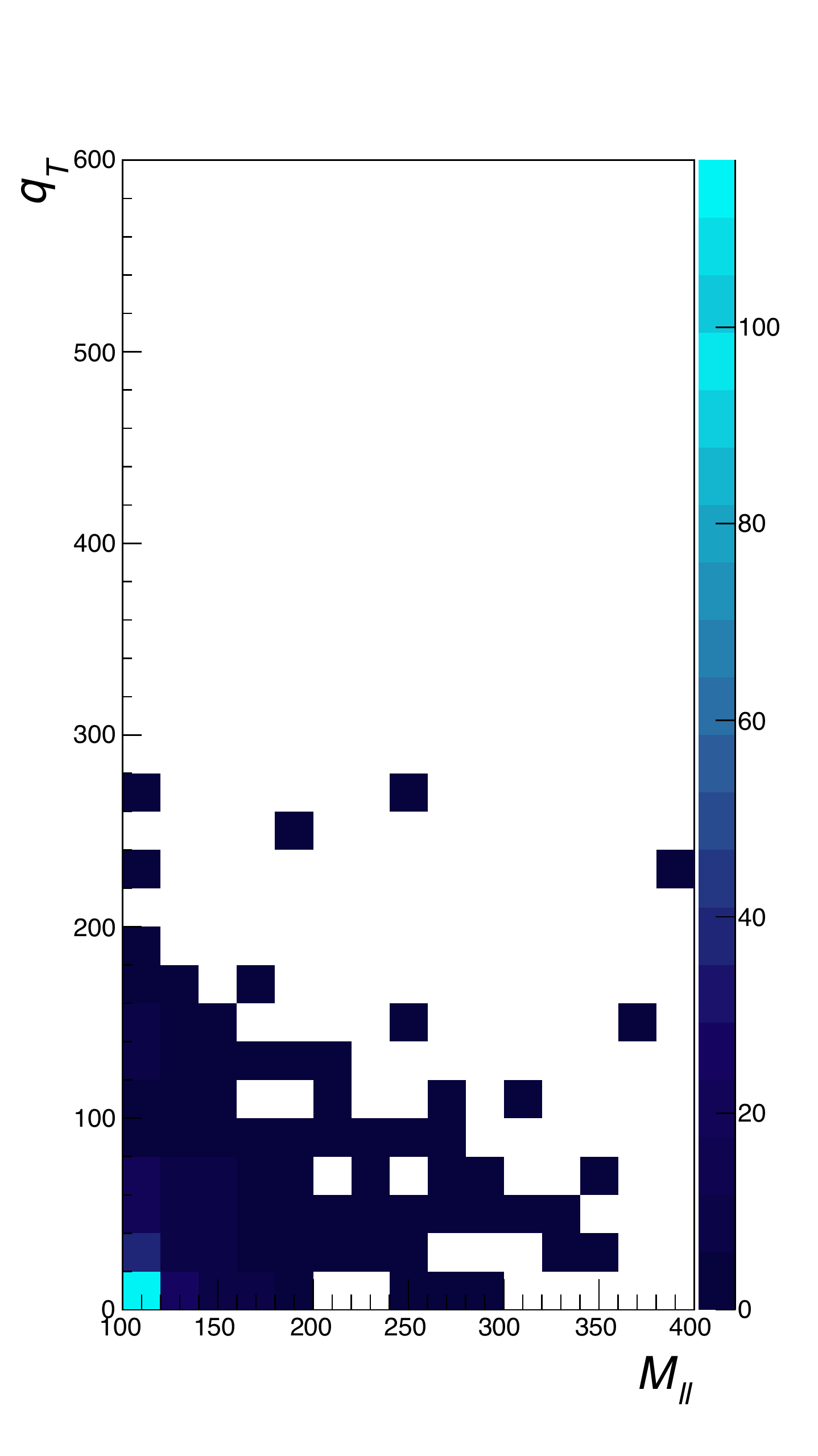}
\includegraphics[width=0.48\textwidth, trim={0 1cm 0 1cm}, clip]{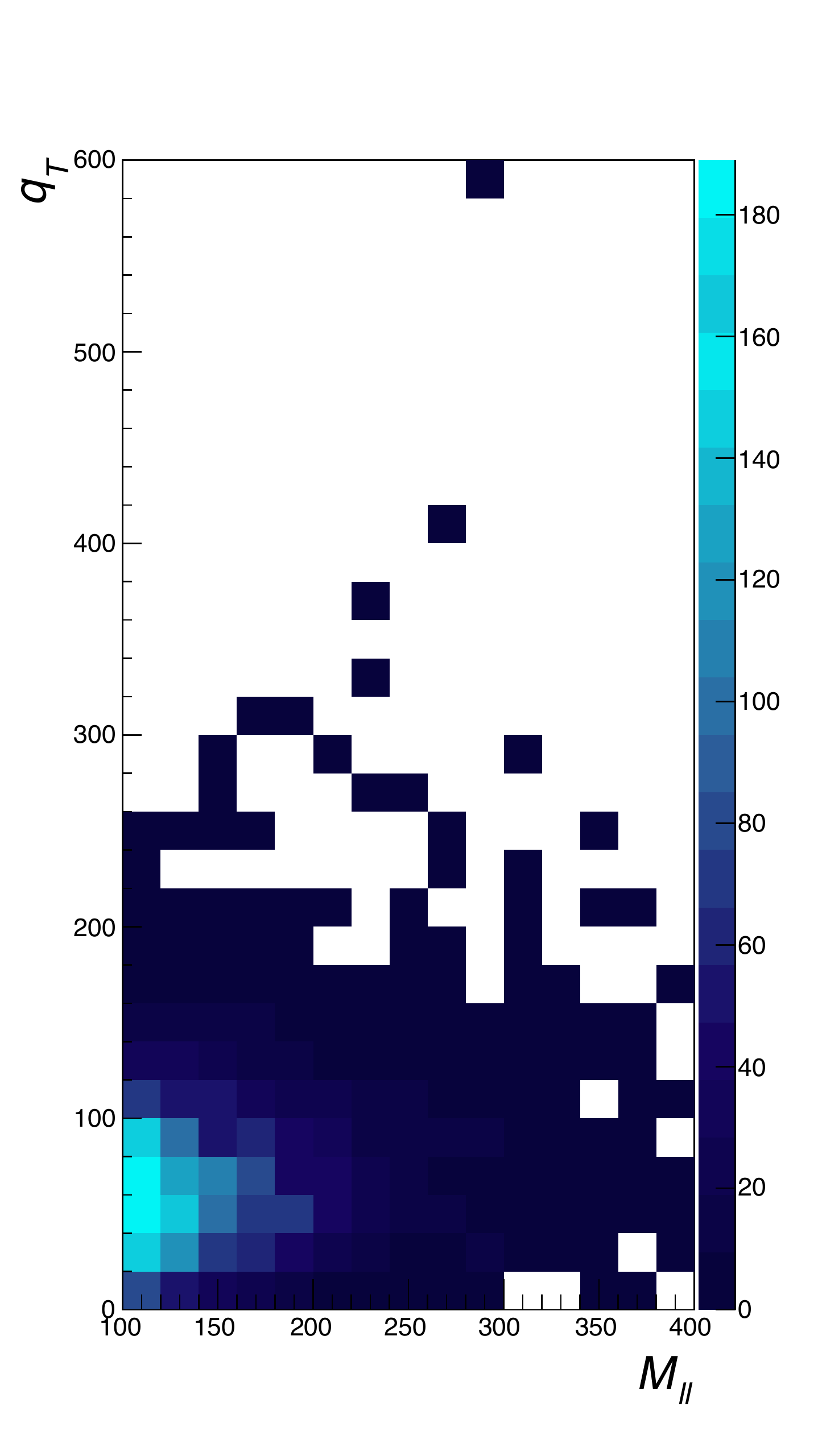}
\caption[.]{\label{fig:qTMLLplane}
The distribution of events in the $(q_T,\MLL)$ plane, for
Drell-Yan and diboson events~(left), and for $\ttbar$ events~(right).}
\end{figure}

A standard approach to this problem might be to take the $(q_T, \MLL)$ plane and perform a binned maximum likelihood fit.  As shown in figure~\ref{fig:qTMLLplane}, the distribution of $\ttbar$ and Drell-Yan plus di-boson events is rather different.  A set of six bins was defined to take advantage of the different distribution of events in this kinematic plane, and a negative log-likelihood function was defined:
\begin{equation}
\label{eq:NLL0}
 Q(\alpha) = -2 \ln L = -2 \sum_{j=1}^{N} \Bigl[ (1-\alpha) B_j + \alpha S_j \Bigr]
\end{equation}
where $N$ is the number of bins, $B_j$ is the pdf for the Drell-Yan plus di-boson  events, and $S_j$ is the corresponding pdf for the $\ttbar$ events.  The one free parameter,~$\alpha$, corresponds to the fraction of selected events that come from the $\ttbar$ process.  The two pdfs satisfy normalization conditions $\sum_{j=1}^N B_j = \sum_{j=1}^N S_j = 1$.  Consequently, the function $Q(\alpha)$ profits only from the difference in shape for the background (Drell-Yan plus di-boson) and signal~($\ttbar$).

\begin{figure} \centering
\includegraphics[width=0.6\textwidth]{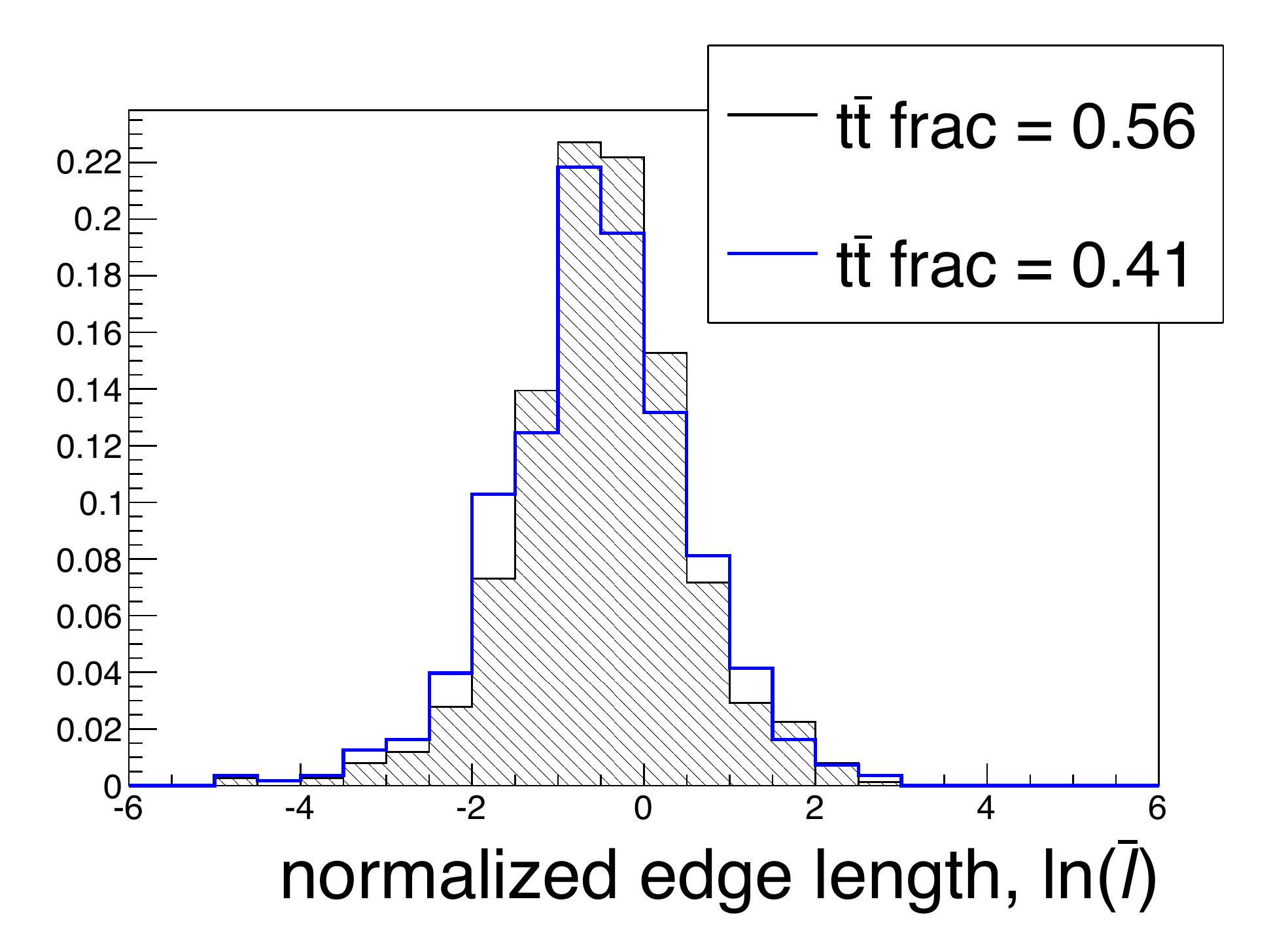}
\caption[.]{\label{fig:lnlhists}
Comparison of the distributions of the logarithm of the normalized
edge length,  $\ln(\bar{l})$, built from two trees with two different
fractions of $\ttbar$.}
\end{figure}

The difference in the distribution of background and signal events visible in figure~\ref{fig:qTMLLplane} leads to a weak constraint on~$\alpha$.  In order to demonstrate the utility of MST quantities, we select the logarithm of the normalized edge length, $\ln(\bar{l})$, that has shown some discriminating power in the toy examples of the previous section.  This single quantity is sensitive to the fraction of events that come from the signal process ($\ttbar)$, as shown in figure~\ref{fig:lnlhists}.  The maximum differentiation afforded by this quantity comes from the shape of the distribution in figure~\ref{fig:lnlhists}, but to be conservative, only the \emph{mean} of the distribution is used.  

The mean $\mu_l$ of the distribution of the logarithm of the normalized edge length varies linearly with the $\ttbar$ fraction.  We parametrized this linear relation of $\alpha$ with $\mu_l$ using simulations in which we varied the $\ttbar$ fraction directly.  For each value of $\alpha$, a new tree was built, the distribution of $\ln(\overline{l})$ was formed, and $\mu_l$ was computed.

In order to incorporate the new MST-related information, an additional $\chi^2$-like term was added to the negative log-likelihood function, expanding the expression in eq.~(\ref{eq:NLL0}):
\begin{equation}
\label{eq:NLL1}
 Q(\alpha) = -2 \sum_{j=1}^{N} \Bigl[ (1-\alpha) B_j + \alpha S_j \Bigr]
 - \frac{ (\mu_{\mathrm{obs}} - \mu_l(\alpha) )^2}{ \sigma^2_l }
\end{equation}
where $\mu_{\mathrm{obs}}$ is the default value for $\mu_l$ and $\mu_l(\alpha)$ is the linear function of $\mu_l$ in terms of $\alpha$.  The variance $\sigma^2_l$ is determined by the distributions as
in figure~\ref{fig:lnlhists}.

\begin{figure} \centering
\includegraphics[width=0.6\textwidth]{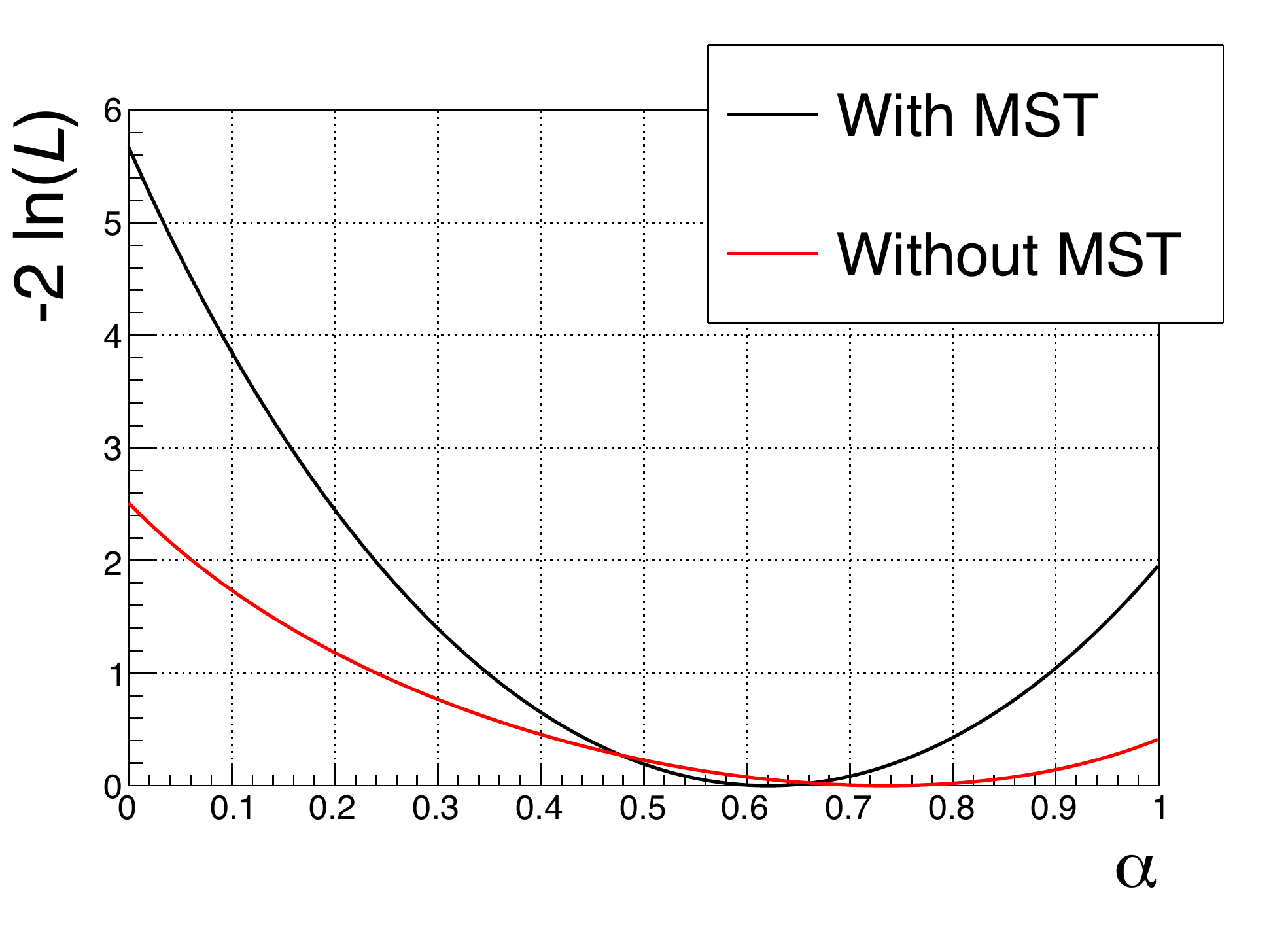}
\caption[.]{\label{fig:alphacurves}
Comparison of $Q(\alpha)$ for the binned likelihood function alone (red curve),
Eq.~(\ref{eq:NLL0}), and for the binned likelihood plus MST information,
Eq.~(\ref{eq:NLL1}) (black curve).}
\end{figure}

The inclusion of this single MST information improves the constraint on the signal fraction~$\alpha$.   Figure~\ref{fig:alphacurves} shows the curves for $Q(\alpha)$ for the binned likelihood alone, as in 
eq.~(\ref{eq:NLL0}), and also for the binned likelihood augmented by the MST information, eq.~(\ref{eq:NLL1}).  On the basis of these curves, the standard deviation for~$\alpha$ decreases from $\sigma_\alpha \approx 0.5$ to $\sigma_\alpha \approx 0.3$.  This is a substantial gain, and illustrates the potential of MST quantities to enhance the statistical power of data collected by particle physics experiments.

\section{Conclusions}
We have presented an early exploration of the use of minimal spanning trees~(MST) in particle physics.   We briefly defined what an MST is and recalled how they were used in cosmological studies.  We devised a number of simple examples that demonstrate how MST quantities can differentiate structures in a phase or feature space.

We hope that MSTs can play a useful role in particle physics.  We devised a simple illustration for events with a pair of oppositely-charged muons and significant missing transverse energy; the distributions of tree-based quantities show that samples with different compositions can be distinguished.  To the extent that distributions of tree-based quantities are statistically independent of distributions of kinematic and other quantities, discriminating power in, for example, a multi-variate analysis will improve: the tree-based quantities can be considered as new additions to the feature space utilized in multi-variate analyses.  While a distribution of a tree-based quantity for the entire tree will not be very discriminating (because the information about which kinematic region is interesting has been discarded), distributions for targeted regions of phase or feature space can be interesting, as our study of di-muon plus missing transverse energy events shows.

This paper is not at all exhaustive of what can be studied with MSTs.  For example, we did not use any pruning (deletion of short branches) which is common in other applications.   Also, our collider example utilized only two kinematic quantities but many more are available.  Finally, we have not tried to use tree-based quantities in machine learning or multi-variate analyses of the type we presented in ref.~\cite{Rainbolt}.   Clearly, opportunities for further development of MSTs in particle physics remain.

\section*{Acknowledgments}
The authors would like to thank S.~Bhattacharya and B.~Pollack for useful comments on the manuscript.  The authors gratefully acknowledge the support provided by the Department of Energy under award number DE-SC0010143.

\bibliographystyle{unsrt}

\end{document}